\documentclass[reqno, 11pt]{amsart}

\usepackage{amsfonts,amssymb,amsbsy,amsmath,amsthm,enumerate}
\usepackage{txfonts}
\usepackage{eucal}
\usepackage{dsfont}
\usepackage{mathrsfs}

\usepackage{diagrams}

\usepackage[dvips]{graphicx}
\usepackage{color}

\newtheorem{theorem}{Theorem}[section]
\newtheorem{proposition}[theorem]{Proposition}
\newtheorem{lemma}[theorem]{Lemma}
\newtheorem{corollary}[theorem]{Corollary}

\newtheorem{definition}[theorem]{Definition}
\newtheorem{remark}[theorem]{Remark}

\newtheorem{example}[theorem]{Example}

\numberwithin{equation}{section}

\numberwithin{figure}{section}
\numberwithin{table}{section}

\newcommand\beq{\begin{equation}}
\newcommand{\bea}{\begin{eqnarray}}
\newcommand{\eea}{\end{eqnarray}}
\newcommand{\beas}{\begin{eqnarray*}}
\newcommand{\eeas}{\end{eqnarray*}}
\newcommand{\beql}{\begin{equation} \label}
\newcommand{\eeq}{\end{equation}}

\newcommand{\R}{\mathbb R}
\newcommand{\N}{\mathbb N}
\newcommand{\C}{\mathbb C}                           

\newcommand{\Z}{\mathbb Z}
\newcommand{\T}{\mathbb T}

\newcommand{\s}[1]{\CMcal{#1}}
\newcommand{\f}[1]{\mathcal{#1}}                  
\newcommand{\bb}[1]{\mathscr{#1}}
\newcommand{\rr}[1]{\mathfrak{#1}}
\newcommand{\n}[1]{\mathds {#1}}

\newcommand{\bra}[1]{\langle #1|}
\newcommand{\ket}[1]{|#1\rangle}
\newcommand{\ketbra}[2]{|#1\rangle\langle#2|}

\newcommand{\expo}[1]{{\rm e}^{#1}}                 
\newcommand{\dd}{\,{\rm d}}
\newcommand{\ii}{\,{\rm i}\,}
\newcommand{\ncint}{\mathrel{{\ooalign{$\int$\cr\kern+.07em\raise.15ex\hbox{$\pmb{\scriptstyle-}$}\cr}}}}           \newcommand{\ncpartial}{\mathrel{{\ooalign{$\partial$\cr\kern+.29em\raise.79ex\hbox{$\pmb{\scriptstyle-}$}\cr}}}}

\newcommand{\virg}[1]{\lq\lq#1\rq\rq}                
\newcommand{\ie}{{\sl i.\,e.\,}}
\newcommand{\eg}{{\sl e.\,g.\,}}
\newcommand{\cf}{{\sl cf.\,}}
\newcommand{\etc}{{\sl etc.\,}}

\newcommand{\red}{\textcolor[rgb]{0.50,0.00,0.00}}       

\begin{document}

\title[Differential geometric invariants for time-reversal symmetric Bloch-bundles]{
Differential geometric invariants for\\ time-reversal symmetric Bloch-bundles: the \virg{Real} case}


\author[G. De~Nittis]{Giuseppe De Nittis}
\address[De~Nittis]{Facultad de Matem\'aticas,
Pontificia Universidad Cat\'olica,
Santiago, Chile}
\email{gidenittis@mat.puc.cl}

\author[K. Gomi]{Kiyonori Gomi}
\address[Gomi]{Department of Mathematical Sciences, Shinshu University,  Nagano, Japan}
\email{kgomi@math.shinshu-u.ac.jp}




\begin{abstract}
\vspace{-4mm}
Topological quantum systems subjected to an even (resp. odd) time-reversal symmetry can be classified by looking at the related \virg{Real} (resp. \virg{Quaternionic}) Bloch-bundles. If from one side the topological classification of these \emph{time-reversal} vector bundle theories has been completely described in \cite{denittis-gomi-14} for the \virg{Real} case and in \cite{denittis-gomi-14-gen} for the \virg{Quaternionic} case, from the other side it seems that a classification in terms of  
differential geometric invariants is still missing in the literature. With this article (and its companion \cite{denittis-gomi-15}) we want to cover this gap. More precisely, we extend in an equivariant way the theory of connections on principal bundles and vector bundles endowed with a  time-reversal symmetry. In the  \virg{Real} case we generalize the  Chern-Weil theory and we show that the assignment of
a \virg{Real} connection, along with  the related \emph{differential}  Chern class and its \emph{holonomy}, suffices for the classification of  \virg{Real} vector bundles in low dimensions.
\end{abstract}


\maketitle

\vspace{-5mm}
\tableofcontents

\section{Introduction}\label{sect:intro}

Families of operators $H(x)$ which depend  continuously on \emph{parameters} $x\in X$ are ubiquitous in mathematical physics. Systems of this type usually show peculiar properties that are of topological nature, hence extremely stable under perturbations. 
The recent enormous  interest for this new  \virg{paradigm} of \emph{topological} quantum states is justified by the fact that states of this type can be realized in certain condensed matter systems called \emph{topological insulators} (see \eg \cite{hasan-kane-10}
for a modern review on the subject). Dealing with the business  of topological quantum states one has to face two types of problems: The first concerns the classification of all possible topological phases while the second regards the way in which different phases can be distinguished. The first problem can be usually faced in a certain generality by using quite abstract tools like 
{
\emph{$K$-theories} \cite{kitaev-09,freed-moore-13,thiang-15} 
or \emph{equivariant homotopy} techniques \cite{kennedy-guggenheim-15,kennedy-zirnbauer-16} 
or \emph{cohomology theories} \cite{denittis-gomi-14,denittis-gomi-14-gen}
\footnote{{It is interesting to point out that
 these three approaches do not always give the same results depending on the number of valence/conduction bands (a.k. the \emph{stable rank} in the mathematical jargon). The $K$-theory is the least fine of these techniques being totally insensitive to the stable rank (by construction). However, the $K$-theory has the enormous advantage of being easily computable. On the opposite side, the homotopy theory is the most precise of these approaches but presents the unfortunate disadvantage of being not (algorithmically) computable in general. The technique based on  cohomology is seated in the center: It is algorithmically computable and at the same time sufficiently precise to be sensible to the stable rank.}
}.}
Instead, an adequate answer to the second problem necessitates  the use of less sophisticated and more manageable objects like \emph{differential forms} and \emph{connections}. This is the reason why differential geometry  offers a privileged framework for the study of topological phases of quantum systems. 
{The aim of this paper is twofold: first, to provide a coherent and self-contained review of the differential geometric tools  commonly used for the study of topological phases of systems when symmetries are absent;
second,  to develop  \emph{new} geometric differential  tools capable to classify 
topological phases for quantum systems subjected to   \emph{time-reversal symmetry}.}

\medskip

Before exposing the main contributions of this paper, we formalize in a precise way the type of systems  of interest  and we recall how the notion of topological phase emerges.
\begin{definition}[Topological quantum systems\footnote{The setting described in Definition \ref{def:tsq} can be generalized to unbounded operator-valued maps $x\mapsto H(x)$ by requiring 
the continuity  of the \emph{resolvent map} $x\mapsto R_z(x):=\big(H(x)-z\n{1}\big)^{-1}\in\bb{K}(\s{H})$. 
 Another possible generalization consists in replacing the norm-topology with the open-compact topology as in \cite[Appendix D]{freed-moore-13}. However, these kind of generalizations have no particular consequences  for the purposes of this work.}]\label{def:tsq}
Let $X$ be a compact Hausdorff space,
$\s{H}$ a separable Hilbert space and $\bb{K}(\s{H})$ the algebra of compact operators on $\s{H}$.
A \emph{topological quantum system} is a self-adjoint map
\begin{equation}\label{eq:tqsA1}
X\;\ni\;x\; \longmapsto\; H(x)=H(x)^*\;\in\;\bb{K}(\s{H})
\end{equation}
continuous with respect to the norm-topology of $\bb{K}(\s{H})$. Let $\sigma(H(x))=\{\lambda_j(x)\ |\ j\in\s{I}\subseteq\Z\}\subset\R$,  be the sequence of eigenvalues of $H(x)$ ordered  according to  $\ldots\lambda_{-2}(x)\leqslant\lambda_{-1}(x)<0\leqslant\lambda_1(x)\leqslant\lambda_2(x)\leqslant \ldots$. The map $x\mapsto \lambda_j(x)$ (which is continuous by standard perturbative arguments \cite{kato-95}) is called \emph{$j$-th energy band}. 
An \emph{isolated family} of  bands is a (finite) collection $\{\lambda_{j_1}(\cdot),\ldots,\lambda_{j_m}(\cdot)\}$
of energy bands such that 
\begin{equation}\label{eq:tqsA2}
\min_{x\in X}\ {\rm dist}\left(
\bigcup_{s=1}^m\{\lambda_{j_s}\}\;,\; \bigcup_{j\in\s{I}\setminus\{j_1,\ldots,j_m\}}\{\lambda_{j}\}
\right)\;=\;C_g\;>0.
\end{equation}
Inequality \eqref{eq:tqsA2} is usually called \emph{gap condition}.
\end{definition}

A standard construction associates to each  topological quantum system of type 
\eqref{eq:tqsA1} with an {isolated family} of $m$ bands a complex vector bundle $\bb{E}\to X$ of rank $m$ which, 
to some extent, can be named \emph{Bloch-bundle}\footnote{This name is justified by the fact that 
the \emph{Bloch theory} for electrons in a crystal  (see \eg \cite{ashcroft-mermin-76}) provides 
some of the most interesting
examples of topological quantum system.}. 
The first step of this construction is the realization of a continuous map of rank $m$ spectral projections $X\ni x\mapsto P(x)$
associated with the isolated family of  bands by means of the \emph{Riesz-Dunford integral}; the second step
 turns out to be a concrete application of the \emph{Serre-Swan Theorem} \cite[Theorem 2.10]{gracia-varilly-figueroa-01} which relates vector bundles and continuous family of projections.
By borrowing  the accepted terminology in the field of topological insulators we will refer to systems of type described in Definition \ref{def:tsq} as  \emph{topological quantum systems
 in class {\bf A}} (see \eg \cite{schnyder-ryu-furusaki-ludwig-08}). In this case the (Cartan) label {\bf A} expresses  the absence of symmetries and extra structures.
Since topological quantum systems
 in class {\bf A}
  lead to complex vector bundles (without any other extra structure) they are topologically classified by the set ${\rm Vec}_\C^m(X)$ of  isomorphism classes  of rank $m$ complex vector bundles over the base space $X$.  The theory of classification of complex vector bundles is a classical, well-studied, problem in topology. Under rather general assumptions for the base space $X$, the set ${\rm Vec}_\C^m\big(X\big)$ can be classified by using homotopy theory techniques and, for dimension $d\leqslant 4$ (and for all $m$) a complete description can be done in terms of cohomology groups and Chern classes \cite{peterson-59}  (see \eg \cite[Section 3]{denittis-gomi-14} for a summary of these standard results). Even more relevant is the fact that when $X$ has a manifold structure the Chern classes can be easily described in terms of certain differential forms built from any connection: 
This is the core of the celebrated \emph{Chern-Weil theory} (\cf Appendix \ref{sect:ChernÐWeil} and references therein).
The relevance of this result is huge both for the mathematical
description and the physical interpretation of the topological phases. Indeed, the use of the differential calculus makes the computation of the de Rham cohomology of a manifold much more accessible than the singular cohomology of a topological space. Moreover, the description of the topological phases in terms of differential forms is a fundamental step to associate topological protected states of the system with different values of suitable observables. The study example in this context is provided by the \emph{quantum Hall effect}
where inequivalent physical states characterized by different values of the Hall conductance are  distinguished by Chern numbers through the \emph{Kubo-Chern formula} \cite{thouless-kohmoto-nightingale-nijs-82,bellissard-elst-schulz-baldes-94,denittis-landi-12}.

\medskip

Topological quantum systems  described in Definition \ref{def:tsq} occur naturally
 in the study of $d$-dimensio\-nal 
 systems of independent quantum  particles that are invariant under continuous translations (\eg free electrons) or 
subjected to a periodic potential (\eg electrons interacting with a crystal). In the first case, after a Fourier transform, one obtains a fibered operator
parametrized by the points of a sphere $\n{S}^d:=\{k\in\R^{d+1}\ |\ \|k\|=1\}$ 
{that results as the one-point compactification of the momentum space
by the identification on physical grounds of the infinite momenta 
(notice that the finiteness of these \virg{extended} momenta is compensated by  the introduction of an extra dimension)
}. In the second case the Bloch-Floquet transform \cite{kuchment-93} provides a fibered operator
parametrized by the points of the \emph{Brillouin} torus $\n{T}^d:=\n{S}^1\times\ldots\times\n{S}^1$.
A detailed description of these two relevant examples,  together with a precise construction of the related Bloch-bundles, can be found 
in \cite[Section 2]{denittis-gomi-14}. 
Although $X=\n{S}^d$ and $X=\n{T}^d$ are the most common situations occurring in condensed matter, through this work we will consider  the  general case of a compact manifold $X$. The physical motivation behind this choice relies 
on the observation that
operator-valued maps like \eqref{eq:tqsA1} are not exclusively models for condensed matter systems.
For instance, they can be used as models for quantum Hamiltonians $H(x)$ perturbed by external (electromagnetic, strains, $\ldots$) fields,  $x$  taking values in the \emph{parameter space} $X$; see \eg the rich monographs \cite{bohm-mostafazadeh-koizumi-niu-zwanziger-03} and \cite{chruscinski-jamiolkowski-04} {or the recent paper
\cite{gat-robbins-15}}. An example of a  topological quantum system associated with the physics of the harmonic oscillator and not related with any condensed matter system is discussed in Appendix \ref{sect:model_HO}.

\medskip

The phenomenology of a topological quantum system  can be  enriched by the presence of symmetries. In this work we are mainly interested in the effects induced by symmetries of \emph{time-reversal} type. 

\begin{definition}[Time-reversal symmetric topological quantum systems]\label{def:tsqAI}
Let $x\mapsto H(x)$ be a topological quantum system as described in Definition \ref{def:tsq}. Assume that $X$ is endowed with an \emph{involution} $\tau:X\to X$ (\cf Section \ref{sect:connect_TR_VB}) and $\s{H}$ with a \emph{complex conjugation} (\ie an anti-linear involution) $C:\s{H}\to\s{H}$.
We say that $x\mapsto H(x)$ is a topological quantum system with a \emph{time-reversal symmetry}
 if there exists a continuous unitary-valued map $x\mapsto J(x)$ on $X$ such that
\begin{equation}\label{eq:tqsA3}
J(x)^*\; H(\tau(x))\; J(x)\;=\;C\;H(x)\;C\;,\qquad\quad C\;J(\tau(x))\;C\;=\; \pm J(x)^*\;.
\end{equation}
If the second equation in   \eqref{eq:tqsA3} is realized with the sign $+$ (resp. $-$) we say that the
{time reversal symmetry} is \emph{even} (resp. \emph{odd}). Topological quantum systems with an  even time-reversal symmetry (+TR)
are said  to be of class {\bf AI}. Similarly,
systems with an  odd time-reversal symmetry (-TR)
are said to be of class {\bf AII}\footnote{Also in this case the Cartan labels {\bf AI} and {\bf AII}  are borrowed from the standard terminology of topological insulators \cite{schnyder-ryu-furusaki-ludwig-08}.}. 
\end{definition}
{
Before discussing the geometric implications of Definition \ref{def:tsqAI}
let us say something about the underlying physical motivations. On a physical ground a Hamiltonian  $H$ possesses a symmetry of time-reversal type if there is  an \emph{anti-unitary} operator  $\Theta:=JC$ such that $\Theta^2=\pm\n{1}$  which commutes with $H$. The passage to a parameter dependent  description as in Definition \ref{def:tsqAI} requires that  both $H$ and $\Theta$ have to be simultaneously \virg{fiberable} operators. For instance in the case of electronic systems this means that both $H$ and $\Theta$ have to be periodic operators, hence
simultaneously decomposable over the Brillouin zone via a Bloch-Floquet transform.
In many of the most relevant physical applications the operator $\Theta$ acts only at  level of internal (spinorial) degrees of freedom and not on the  spatial degrees of freedom. In  these cases the dependence of $\Theta$ (and hence of $J$) by the quasi-momentum is trivial and the 
unitary-valued map $x\mapsto J(x)$ in  \eqref{eq:tqsA3} turns out to be just a constant map $J$. On the other hand this is not the most general possible situation and it is possible to design models where the basic symmetries need to be \emph{twisted} by a phase which depends on the position (\eg in the spirit of the Zak's magnetic translations). For this reason we chose to consider in this work, at least as a mathematical interest, the more general situation described in Definition \ref{def:tsqAI}. We point out that the effect of a non-constant map $J$ can be not totally harmless as shown by Proposition \ref{prop:equivGB_conn}.
}

\medskip

Also topological quantum systems with time-reversal symmetry lead to complex vector bundles. However, the presence of the symmetry endows
 the  vector bundles   with an extra structure:
 In the case of a +TR symmetry one obtains \emph{\virg{Real} vector bundles} (or simply \emph{$\rr{R}$-bundles})
 while in presence of a -TR symmetry  one ends up to \emph{\virg{Quaternionic} vector bundles} (or  \emph{$\rr{Q}$-bundles} for short). \virg{Real} vector bundles have been invented 
 by M.~F. Atiyah in \cite{atiyah-66} while J. L. Dupont introduced for the first time the notion of \virg{Quaternionic} vector bundles  in \cite{dupont-69}\footnote{In this work  J. L. Dupont called this new category of vector bundles
 \emph{symplectic}. However, the  name of {symplectic} vector bundle  is also used in the literature for vector bundles with a quaternionic fiber (see \eg \cite{husemoller-94}). For this reason we chose to use the name  \virg{Quaternionic}  in order to avoid confusion.
}. Both $\rr{R}$-bundles and $\rr{Q}$-bundles are complex vector bundles over an \emph{involutive space} $(X,\tau)$ endowed with an involutive automorphism of the total space which covers the involution $\tau$ and restricts to an \emph{anti-linear} map between conjugate  fibers. More precisely:
 
\begin{definition}[$\rr{R}$-bundles and $\rr{Q}$-bundles]\label{def_vec_R&QB}
Let $(X,\tau)$ be a topological space endowed with a continuous involution $\tau$ and $\pi:\bb{E}\to X$ a complex vector bundle of rank $m$. We say that $\bb{E}$ has a \virg{Real} structure if there is a topological homeomorphism $\Theta:\bb{E}\to \bb{E}$ such that:
\begin{enumerate}
\item[\upshape{(Eq.)}] The bundle projection $\pi$ is \emph{equivariant} in the sense that $\pi\circ\Theta=\tau\circ\pi$;\vspace{1.2mm}
\item[\upshape{(A.L.)}] The homeomorphism $\Theta$ acts \emph{anti-linearly} at each fiber, \ie $\Theta(\lambda p)=\bar{\lambda}\Theta(p)$ for all $p\in\bb{E}$ and $\lambda\in\C$ (here $\bar{\lambda}$ stands for the complex conjugate of $\lambda$);
\vspace{1.2mm}
\item[{\upshape ($\rr{R}$)}] $\Theta^2(p)=+p$ for all $p\in\bb{E}$.
\end{enumerate}
We say that $\bb{E}$ has a \virg{Quaternionic} structure if {\upshape ($\rr{R}$)} is replaced by
\begin{enumerate}
\item[{\upshape ($\rr{Q}$)}] $\Theta^2(p)=-p$ for all $p\in\bb{E}$.
\end{enumerate}
\end{definition}

\noindent    
$\rr{R}$-bundles and $\rr{Q}$-bundles define  new categories of locally trivial fibered objects with their own morphisms. 
A vector bundle \emph{morphism} $f:\bb{E}\to \bb{E}'$ between two vector bundles  (over the same base space)
is a  continuous map  which is \emph{fiber preserving} in the sense that  $\pi=\pi'\circ f$
 and which restricts to a \emph{linear} map on each fiber. Similarly an $\rr{R}$-morphism (resp. $\rr{Q}$-morphism) $f:(\bb{E},\Theta)\to(\bb{E}',\Theta')$ between two  $\rr{R}$-bundles (resp. $\rr{Q}$-bundles)
over the same involutive space $(X,\tau)$ 
  is a vector bundle morphism  commuting with the involutions, \ie $f\circ\Theta\;=\;\Theta'\circ f$. The set of equivalence classes of isomorphic $\rr{R}$-bundles of  rank $m$ over $(X,\tau)$ 
  is denoted by ${\rm Vec}_{\rr{R}}^m(X,\tau)$. Similarly the set of equivalence classes  of rank $m$ $\rr{Q}$-bundles is denoted by ${\rm Vec}_{\rr{Q}}^m(X,\tau)$. Topological quantum systems of type {\bf AI}  and {\bf AII} are classified by  
  ${\rm Vec}_{\rr{R}}^m(X,\tau)$ and ${\rm Vec}_{\rr{Q}}^m(X,\tau)$, respectively.
  
  \medskip

Topological quantum systems of type {\bf AI}  and {\bf AII} have been completely classified \red{up to dimension four} in \cite{denittis-gomi-14,denittis-gomi-14-gen} in terms of  proper characteristic classes belonging to suitable equivariant cohomology theories.
 The major aim of the present work is to connect these \virg{abstract}  classifying invariants  with more \virg{concrete} objects that can be defined at a differential geometric level  in the case in which $(X,\tau)$ carries a suitable manifold structure (see properties H.1 - H.4 below). There are many interesting physical situations covered by this framework. For instance, in the case of   free electrons with a time-reversal symmetry (independently if odd or even) the involutive space is the \emph{TR sphere} $\tilde{\n{S}}^d\equiv({\n{S}}^d,\tau)$ with involution $\tau(k_0,k_1,\ldots,k_d)=(k_0,-k_1,\ldots,-k_d)$ for all $(k_0,k_1,\ldots,k_d)\in{\n{S}}^d$. {This particular form of the involution encodes the physical information that only the zero momentum and the infinite momentum are left unchanged by the time inversion while all the others are reversed (see \cite[Section 2]{denittis-gomi-14} for more details).} 
  Similarly, in the case of electrons in a crystal subjected to a time-reversal symmetry one has that the Brillouin zone is the involutive \emph{TR torus} $\tilde{\n{T}}^d\equiv({\n{T}}^d,\tau)$ where the involution $\tau$ is   induced by the product structure $\tilde{\n{T}}^d=\tilde{\n{S}}^1\times\ldots\times \tilde{\n{S}}^1$.
  
        \medskip
  
Mathematically, this paper consists of two major parts each of which contains new relevant results. The {\bf first part}, which essentially coincides with Section \ref{sect:connect_TR_VB}, provides a coherent construction of the theory of connections for bundles endowed with a time-reversal structure. {Since this construction amounts to a natural, but non trivial, extension of the usual theory of connections to the categories of \virg{Real} and \virg{Quaternionic} vector bundles, we found appropriate and pedagogical to present the new theory of time-reversal connections in parallel with a systematic review of the classical theory.} Here the main goals are:
\begin{itemize}
\item[-] The definition and the description of \emph{time-reversal} (\virg{Real} and \virg{Quaternionic}) \emph{principal $\n{U}(m)$-bundles}. This includes the content of Section \ref{sect:equiv_princ_bund}.\vspace{1.3 mm}
\item[-]The definition of the notion of  \emph{equivariant connection} on principal bundles endowed with  time-reversal (\virg{Real} and \virg{Quaternionic})  structures and its local description (Section \ref{sect:equiv_sect}). The extension of the notion of  {equivariant connection} to {time-reversal} vector bundles (Section \ref{sect:equiv_sect_VB}). The identification of the condition for the equivariance of the \emph{Grassmann-Berry connection} (Section \ref{sect:bott-chern-berry}).
\vspace{1.3 mm} 

\item[-] The study of the \emph{holonomy}  (Section \ref{sect:equiv_holonm}) and the \emph{curvature} (Section \ref{sect:equiv_curv}) defined from  a time-reversal equivariant connection. \vspace{1.3 mm}

\end{itemize}
 In the {\bf second part}, which agrees with  Section \ref{sect:R-bundles}, we use the notion of time-reversal equivariant connection in order to provide a classification of \virg{Real} vector bundles based on differential geometric invariants. More precisely:
\begin{itemize}
\item[-] We prove that the \emph{topological} \virg{Real} Chern classes, which are elements of the equivariant Borel cohomology $H_{\Z_2}^\bullet(X,\Z(1))$ (Section \ref{subsec:borel_cohom} and \ref{sect:eq_chern_class}) and which are sufficient to classify
 \virg{Real} vector bundles in low dimensions \cite{denittis-gomi-14}, can be described as even degree differential forms of a graded de Rham complex: this is the  \virg{Real} analog of the  Chern-Weil theory (Section \ref{sect:equiv_ChernÐWeil}). \vspace{1.3 mm}

\item[-] The classifying cohomology $H_{\Z_2}^2(X,\Z(1))$ generally possess a non-vanishing torsion part (even when the homology of $X$ is torsion free) but the differential representatives of the   \virg{Real} Chern classes can detect only  the free part of $H_{\Z_2}^2(X,\Z(1))$. However, when $H_{\Z_2}^2(X,\Z(1))$ is pure torsion
 we can exploit the holonomy associated to a \emph{flat \virg{Real} connection} in order to classify \virg{Real} vector bundles (Section \ref{sect:non-triv}, \ref{sect:real_flat_connections} and \ref{sect:non-triv_low-dim}). These results can be summarized in the following:
\end{itemize}
 
 \medskip
\noindent 
{\bf Classification principle} (\cf Theorem \ref{theo:class_princ}){\bf.} \emph{Let $(X,\tau)$ be an involutive manifold 
such that ${\rm dim}(X)\leqslant 3$
(plus certain technical assumptions). 
Then
\begin{enumerate}
\item[(i)] If $H^2_{\Z_2}(X,\Z(1))$ has no torsion then 
\virg{Real} vector bundles over $(X,\tau)$ are classified by the \emph{differential} \virg{Real} Chern class associated to any  \virg{Real} connection;\vspace{1.3mm}
\item[(ii)]  If $H^2_{\Z_2}(X,\Z(1))$ is pure torsion then 
\virg{Real} vector bundles over $(X,\tau)$ are distinguished by the holonomy  with respect to a flat \virg{Real} connection (modulo \emph{gauge} transformations).
\end{enumerate}}
\vspace{1.3 mm}

  The paper is completed by a series of appendices which have the role of
  {reviewing}
  the basic facts in the theory of connections on principal bundles (Appendix \ref{sect:remind_connect}),  the Chern-Weil theory (Appendix \ref{sect:ChernÐWeil}) and the theory of \emph{spectral sequences} (Appendix \ref{spectral}) which is a ubiquitous tool for computation in cohomology. 
 The intention of including these three appendices is to make this paper self-consistent {and accessible to
possible non expert readers}. A last Appendix \ref{sec:new_cohom} contains just a few technical computations.
 
 \medskip

The natural  {\bf third part} of our analysis, consisting of the differential geometric classification of  \virg{Quaternionic} vector bundles,
will appear in a separated incoming article \cite{denittis-gomi-15}. We decided to  split the full content in two parts since the \virg{Quaternionic} case is more involved than the \virg{Real} case and  necessitates the introduction of certain technical notions which cannot be covered in a few pages.

\medskip

\medskip

\noindent
{\bf Acknowledgements.} 
{GD's research is supported
 by
the  grant \emph{Iniciaci\'{o}n en Investigaci\'{o}n 2015} - $\text{N}^{\text{o}}$ 11150143 funded  by FONDECYT.	 KG's research is supported by 
the JSPS KAKENHI Grant Number 15K04871.}
GD wants to thank the organizers of the workshop \emph{Topological Phases of Quantum Matter} at the Erwin Schr\"{o}dinger Institute  (Wien) where part of this work was done. 
{The authors would like to thank the anonymous reviewer for the useful comments and suggestions which help us to improve the quality of our paper.}
\medskip

\section{Theory of connections for time-reversal symmetric bundles}\label{sect:connect_TR_VB}

Throughout this paper, we deal with topological spaces $X$ endowed with homeomorphisms $\tau:X\to X$ such that $\tau^2={\rm Id}_X$. They are usually called
 \emph{involutions}. We assume, when not stated otherwise, that:\vspace{1.3 mm}
\begin{itemize}
\item[(H.1)] $X$ is compact, Hausdorff,  path-connected and without boundary;\vspace{1.3 mm}
\item[(H.2)] $X$ is a \emph{smooth} manifold of finite dimension $d$;\vspace{1.3 mm}
\item[(H.3)] The {involution} $\tau:X\to X$ is smooth in the sense that it
preserves the  manifold structure.\vspace{1.3 mm}
\end{itemize}
Assumptions (H.1) and (H.2) together say that $X$ is a \emph{closed} manifold. Moreover, (H.2) and (H.3)  imply (see \eg \cite[Theorem 3.6]{may-96})
\begin{itemize}
\vspace{1.3 mm}
\item[(H.4)] The involutive space $(X,\tau)$ admits the structure of a $\Z_2$-CW-complex. \vspace{1.3 mm}
\end{itemize}
For the sake of completeness, let us recall that an involutive space $(X,\tau)$ has the structure of a $\Z_2$-CW-complex if it admits a skeleton decomposition given by gluing cells of different dimensions which carry a $\Z_2$-action. 
For a precise definition of the notion of 
$\Z_2$-CW-complex, the reader can refer to  \cite[Section 4.5]{denittis-gomi-14} or \cite{may-96,matumoto-71,allday-puppe-93}.

\medskip

We assume that the reader is familiar with the most basic  facts concerning the theory of vector bundles and we will refer to the standard monographs
\cite{milnor-stasheff-74,hirsch-76,husemoller-94}
when necessary. For what concerns the theory of \virg{Real} and \virg{Quaternionic} vector bundles we refer to \cite{denittis-gomi-14} and \cite{denittis-gomi-14-gen}, respectively. Let us recall that  vector bundles are usually defined in the \emph{topological category} meaning that  all the maps involved in the various definitions are continuous functions between topological spaces. However, if the base space $X$ has an additional smooth (resp. $C^r$-differentiable) 
manifold structure one can define vector bundles in the \emph{smooth category} (resp. $C^r$-\emph{category})
by requiring that all  spaces involved in the definitions carry a smooth (resp. $C^r$-differentiable) 
manifold structure and maps are smooth (resp. $C^r$) functions. More details about this question are available in \cite[Chapt. 4]{hirsch-76} but here we need to  recall  a fundamental fact:
\begin{theorem}[Topological vs. smooth category]\label{theo:top-smooth}
Every topological (complex, real or quaternionic) vector bundle  $\bb{E}\to X$ over a smooth manifold $X$ has a compatible smooth bundle structure which is unique  up to smooth isomorphisms. The same is true if the space $X$ is endowed with a smooth involution $\tau$ and $\bb{E}$ carries a \virg{Real} or \virg{Quaternionic} structure. In particular one has
$$
{^{\rm top}{\rm Vec}^m_{\n{F}}}(X)\;\simeq\;{^{C^r}{\rm Vec}^m_{\n{F}}}(X)\;,\qquad\quad\n{F}\;=\;\R\;,\;\C\;,\n{H}\;,\qquad\ \ r\in\N\cup\{\infty\}\;,
$$
 namely the set of equivalence classes of topological rank $m$ (complex, real or quaternionic) vector bundles agrees with the set of equivalence classes of $C^r$ rank $m$ (complex, real or quaternionic) vector bundles. Similarly, one has that
 $$
{^{\rm top}{\rm Vec}^m_{\rr{T}}}(X,\tau)\;\simeq\;{^{C^r}{\rm Vec}^m_{\rr{T}}}(X,\tau)\;,\qquad\quad\rr{T}\;=\;\rr{R}\;,\;\rr{Q}\;,
$$
where $\rr{R}$ and $\rr{Q}$ stand for the \virg{Real} or \virg{Quaternionic} structure, respectively. 
\end{theorem}

\noindent
Theorem \ref{theo:top-smooth} can be understood as a particular manifestation
 of the extremely general \emph{Oka-Grauert principle} \cite{grauert-58} which is valid also in the \emph{equivartiant} category \cite{heinzner-kutzschebauch-95}.
A direct proof of Theorem \ref{theo:top-smooth} in the case of complex, real or quaternionic vector bundles can be found in \cite[Chapt. 4, Theorem 3.5]{hirsch-76}.
The key argument of the proof is that continuous classification maps can be deformed to smooth classification maps. Since also \virg{Real} or \virg{Quaternionic} vector bundles admit a homotopy classification (see \cite[Theorem 4.13]{denittis-gomi-14} and \cite[Theorem 2.13]{denittis-gomi-14-gen}, respectively) a similar argument applies also for these categories. The importance of Theorem \ref{theo:top-smooth} resides in the fact that it allows us to
 switch from the topological to the $C^r$ or smooth category, and vice versa, in an arbitrary manner. In the rest of this work we will often make use of this freedom. 

\subsection{Time-reversal principal bundles: \virg{Real} and \virg{Quaternionic} structures}\label{sect:equiv_princ_bund}
The study of vector bundles is equivalent,  both topologically and geometrically, to the study of \emph{principal bundles}. On the other side, as discussed in Appendix \ref{sect:remind_connect} (and references therein), the theory of connections is more naturally developed on the domain of principal bundles. The aim of this section is to adapt the \virg{Real} and \virg{Quaternionic} structures to principal bundles. This is, indeed, the first step toward a consistent definition of    \virg{Real} and \virg{Quaternionic} connections.
Before proceeding, let us recall that principal bundles are related to vector bundles through the \emph{structure group}.
Since both \virg{Real} and \virg{Quaternionic} vector bundles over a compact base space admit an equivariant Hermitian metric (see \cite[Remark 4.11]{denittis-gomi-14} and \cite[Proposition 2.10]{denittis-gomi-14-gen}, respectively), it turns out that the relevant structure group is the \emph{unitary group} $\n{U}(m)$ together with its Lie algebra $\rr{u}(m)$ of anti-Hermitian matrices.

\begin{definition}[$\rr{R}$ and $\rr{Q}$-principal bundles]\label{def_princ_R&QB}
Let $(X,\tau)$ be a topological space endowed with a continuous involution $\tau$ and $\pi:\bb{P}\to X$ a 
principal $\n{U}(m)$-bundle (see Appendix \ref{sect:remind_connect} for the definition).
 We say that $\bb{P}$ has a \virg{Real} structure if there is a topological homeomorphism $\hat{\Theta}:\bb{P}\to \bb{P}$ such that:
\begin{enumerate}
\item[\upshape{(Eq.)}] The bundle projection $\pi$ is \emph{equivariant} in the sense that $\pi\circ\hat{\Theta}=\tau\circ\pi$;\vspace{1.2mm}
\item[\upshape{(Inv.)}] $\hat{\Theta}$ is an \emph{involution}, \ie $\hat{\Theta}^2(p)=p$ for all $p\in\bb{P}$;
\vspace{1.2mm}
\item[{\upshape ($\hat{\rr{R}}$)}] The right $\n{U}(m)$-action on fibers and the homeomorphism $\hat{\Theta}$ fulfill the condition
$$
\hat{\Theta}(R_u(p))\;=\;R_{\overline{u}}(\hat{\Theta}(p))\;,\qquad\quad \forall\ p\in\bb{P}\;,\ \ \ \forall\ u\in \n{U}(m)
$$
where $R_u(p)=p\cdot u$ denotes the right $\n{U}(m)$-action and $\overline{u}$ is the complex conjugate of $u$.
\end{enumerate}
We say that $\bb{P}$ has a \virg{Quaternionic} structure if 
the structure group $\n{U}(2m)$ has even rank and 
{\upshape ($\hat{\rr{R}}$)} is replaced by
\begin{enumerate}
\item[{\upshape ($\hat{\rr{Q}}$)}] The right $\n{U}(2m)$-action on fibers and the homeomorphism $\hat{\Theta}$ fulfills the condition
$$
\hat{\Theta}(R_u(p))\;=\;R_{\sigma(u)}(\hat{\Theta}(p))\;,\qquad\quad \forall\ p\in\bb{P}\;,\ \ \ \forall\ u\in \n{U}(2m)
$$
where $\sigma:\n{U}(2m)\to\n{U}(2m)$ is the involution given by $\sigma(u):=Q\cdot\overline{u}\cdot Q^{-1}=-Q\cdot\overline{u}\cdot Q$ with
\beql{eq:Q-mat}
Q
\;:=\;
\left(
\begin{array}{rr|rr|rr}
0 & -1 &        &        &   &    \\
1 &  0 &        &        &   &    \\
\hline
  &    & \ddots &        &   &    \\
  &    &        & \ddots &   &    \\
\hline
  &    &        &        & 0 & -1 \\
  &    &        &        & 1 &  0
\end{array}
\right)\;
\eeq
a  symplectic matrix. 
\end{enumerate}
\end{definition} 

\noindent
We will often refer to \virg{Real} and \virg{Quaternionic} principal bundles with the abbreviations $\rr{R}$-principal bundles
and $\rr{Q}$-principal bundles, respectively. 
\begin{remark}\label{rk:quat_str}{\upshape
Let us point out  that both in  the \virg{Real} and in the \virg{Quaternionic} case the map $\hat{\Theta}$ is subjected to the 
involutive property (Inv.). This means that both $\rr{R}$- and $\rr{Q}$-principal bundles are examples of \emph{$\Z_2$-equivariant} 
principal bundles (indeed
properties (Eq.) and (Inv.) define these objects). This is  an important difference with respect to the case of  \virg{Quaternionic} vector bundles 
for which the total space is endowed with an \emph{anti}-involution (\cf property ($\rr{Q}$) in Definition \ref{def_vec_R&QB}). Let us also recall that in the case of involutive base spaces with fixed points a \virg{Quaternionic}  structure necessarily requires a structure group of even dimension \cite[Proposition 2.2]{denittis-gomi-14-gen}.
}\hfill $\blacktriangleleft$
\end{remark}

Morphisms (and isomorphisms) of  $\rr{R}$- and $\rr{Q}$-principal bundles are defined in a natural way: If $(\bb{P},\hat{\Theta})$ and $(\bb{P}',\hat{\Theta}')$ are two of such principal bundles over the same involutive space $(X,\tau)$ then an $\rr{R}$- or $\rr{Q}$-morphism  is a principal bundle morphism $f:\bb{P}\to\bb{P}'$
(as described in Appendix \ref{sect:remind_connect}) such that $f\circ\hat{\Theta}=\hat{\Theta}'\circ f$. We will use the symbols  
${\rm Prin}^{\rr{R}}_{\n{U}(m)}(X,\tau)$ and ${\rm Prin}^{\rr{Q}}_{\n{U}(2m)}(X,\tau)$ for the sets of equivalence classes of \virg{Real} and \virg{Quaternonic} principal bundles over  $(X,\tau)$, respectively. An $\rr{R}$-principal bundle over $(X,\tau)$
is called \emph{trivial} if it is isomorphic to the {product} $\rr{R}$-principal bundle $X\times\n{U}(m)$ with {trivial $\rr{R}$-structure} $\hat{\Theta}_0:(x,u)\mapsto(\tau(x),\overline{u})$. In much the same way, a \emph{trivial} $\rr{Q}$-principal bundle is isomorphic to the
{product} $\rr{Q}$-principal bundle  $X\times\n{U}(2m)$ endowed with the  {trivial $\rr{Q}$-structure} $\hat{\Theta}_0:(x,u)\mapsto(\tau(x),\sigma(u))$. 

\medskip

The usual equivalence (of categories) between  principal bundles  and complex vector bundles which leads to isomorphisms of type \eqref{eq:isoPV1} extends also to $\rr{R}$- and $\rr{Q}$-principal bundles. 
\begin{proposition}\label{prop:equi_top_Prin_QRVB}
Let ${\rm Prin}^{\rr{R}}_{\n{U}(m)}(X,\tau)$ and ${\rm Prin}^{\rr{Q}}_{\n{U}(2m)}(X,\tau)$ be the sets of equivalence classes of $\rr{R}$- and $\rr{Q}$-principal bundles over the involutive space $(X,\tau)$. Then, there are isomorphisms
$$
{\rm Prin}^{\rr{R}}_{\n{U}(m)}(X,\tau)\;\simeq\ {^{\rm top}{\rm Vec}^m_{\rr{R}}}(X,\tau)\;,\qquad\quad {\rm Prin}^{\rr{Q}}_{\n{U}(2m)}(X,\tau)\;\simeq\ {^{\rm top}{\rm Vec}^{2m}_{\rr{Q}}}(X,\tau).
$$ 
\end{proposition}

\proof
We start with the (more interesting) \virg{Quaternionic} case. Let $(\bb{P},\hat{\Theta})$ be a $\rr{Q}$-principal bundle over $(X,\tau)$ with bundle projection $\pi:\bb{P}\to X$.
 The \emph{associated bundle} construction (\cf Appendix \ref{sect:remind_connect}) provides a
complex vector bundle $\pi':\bb{E}\to X$ with total space $\bb{E}:=\bb{P}\times\C^{2m}/\n{U}(2m)$ where the equivalence relation is $(p,{\rm v})\sim(p\cdot u, u^{-1} {\rm v})$ for some $u\in \n{U}(2m)$ and the new bundle projection $\pi'=\pi\circ {\rm pr}_1$ is the composition of $\pi$ with the  canonical (first component) projection ${\rm pr}_1:\bb{P}\times\C^{2m}\to \bb{P}$.
 This vector bundle inherits a $\rr{Q}$-structure $\Theta:\bb{E}\to \bb{E}$ given by $\Theta([(p,{\rm v})]):=[\hat{\Theta}(p),Q\overline{\rm v}]$. 
This definition is compatible 
  with the equivalence relation. Indeed,   $\hat{\Theta}(p\cdot u)=\hat{\Theta}(p)\cdot \sigma(u)$ by definition, and 
$$
Q\overline{(u^{-1}\rm v)}\;=\;-\;\left(Q\cdot\overline{u^{-1}}\cdot Q^2\right)\overline{\rm v}\;=\;-\left(\left(Q\cdot\overline{u}\cdot Q\right)^{-1}\cdot Q\right)\overline{\rm v}\;=\;\left(\sigma(u)^{-1}\cdot Q\right)\overline{\rm v}\;.
$$
With this definition, the three properties (Eq.), (A.L.) and ($\rr{Q}$) in Definition \ref{def_vec_R&QB} are easily verifiable.
Moreover, the standard Hermitian metric on $\bb{E}$ given by  $\rr{m}\big([(p,{\rm v})],[(p,{\rm w})]\big):=\overline{\rm v}\cdot {\rm w}$ is evidently $\Theta$-equivariant in the sense that $\rr{m}\circ\Theta=\overline{\rr{m}}$.

Conversely, let $(\bb{E},\Theta)$ be a rank $2m$ $\rr{Q}$-vector bundle over $(X,\tau)$ with a bundle projection $\pi':\bb{E}\to X$ and a
$\Theta$-equivariant Hermitian metric $\rr{m}$ (which exists in view of \cite[Proposition 2.10]{denittis-gomi-14-gen}).
Consider the associated \emph{orthonormal frame} bundle $\bb{P}:=\bb{F}(\bb{E})$. As explained in Appendix \ref{sect:remind_connect}, this is a principal 
$\n{U}(2m)$-bundle over $X$ and each point $p\in \bb{P}_x$ can be identified with a choice of an $\rr{m}$-orthonormal basis 
of $\bb{E}_x$, \ie $p\equiv\{{\rm v}_1,\ldots,{\rm v}_{2m}\}$. The frame bundle $\bb{P}$ can be endowed with the involution $\hat{\Theta}:\bb{P}\to \bb{P}$ given by $\hat{\Theta}:\{{\rm v}_1,\ldots,{\rm v}_{2m}\}\mapsto \{\Theta({\rm v}_2),-\Theta({\rm v}_1),\ldots,\Theta({\rm v}_{2m}), -\Theta({\rm v}_{2m-1})\}$. With this definition the conditions (Eq.) and (Inv.) of Definition \ref{def_princ_R&QB} are   trivially satisfied. In order to check  condition ($\hat{\rr{Q}}$)
we notice that for each $u\in\n{U}(2m)$ one has  $R_u(p)\equiv\{{\rm v}'_1,\ldots,{\rm v}'_{2m}\}$ with 
${\rm v}'_i:=\sum_{j=1}^{2m}u_{ji}{\rm v}_j$. Since $\Theta({\rm v}'_i):=\sum_{j=1}^{2m}\overline{u_{ji}}\;\Theta({\rm v}_j)$ by anti-linearity, one concludes after a straightforward calculation the expected relation $\hat{\Theta}(R_u(p))=R_{Q\cdot \bar{u}\cdot Q^{-1}}(\hat{\Theta}(p))$.

Finally, the isomorphism ${\rm Prin}^{\rr{Q}}_{\n{U}(2m)}(X,\tau)\simeq\ {^{\rm top}{\rm Vec}^{2m}_{\rr{Q}}}(X,\tau)$ (and the equivalence of categories) can be proved as in the complex case by showing that the  \emph{associated bundle construction}  and  the \emph{frame bundle construction}  are, up to $\rr{Q}$-isomorphisms, mutually inverse mappings between $\rr{Q}$-principal bundles and $\rr{Q}$-vector bundles.

The proof for the \virg{Real} case follows exactly the same line as in the \virg{Quaternionic} case. The only difference in this case (it is in fact a  simplification!)  is that the matrix $Q$ has to be consistently replaced by the identity matrix.
\qed

\medskip

\begin{example}[TR principal bundles with product structure]\label{ex:PB_prod}{\upshape
Let $\bb{E}\to X$ be a  \virg{Real} or \virg{Quaternionic} vector bundle over the involutive space $(X,\tau)$ obtained from 
a continuous map of
rank $m$ spectral projections $X\ni x\mapsto P(x)$ associated with a time-reversal symmetric topological quantum system in the sense of Definition \ref{def:tsqAI}. More specifically  this means that the projections are subjected to the transformation relation
$J(x)^* \cdot  P(x) \cdot  J(x) = \overline{P(\tau(x))}$ where $x\mapsto J(x)$ is a continuous map with values in the unitary operators on a fixed Hilbert space $\s{H}$ such that $J(\tau(x)) \cdot\overline{J(x)}=\pm\n{1}$.
In many interesting cases the Hilbert space $\s{H}$ has finite dimension $N$ (or one can reduce to consider a similar situation
by an argument similar to \cite[Chapter 3, Theorem 5.5]{husemoller-94} when $X$ is a finite dimensional CW-complex). In this case  the vector bundle $\bb{E}\to X$ can be seen as a subbundle of a trivial bundle $X\times\C^N\to X$ with time-reversal structure  $\Theta:(x,{\rm v})=(\tau(x),J(x)\bar{\rm v})$. The proof of Proposition \ref{prop:equi_top_Prin_QRVB} shows that also the associated principal bundle  $\bb{F}(X\times \C^N)=X\times \n{U}(N)$  inherits a time-reversal structure which is given by $\hat{\Theta}:(x,u)=(\tau(x),J(x)\cdot \overline{u})$ in the \virg{Real} case or by $\hat{\Theta}:(x,u)=(\tau(x),J(x) \cdot \overline{u}\cdot Q)$ in the \virg{Quaternionic} case (if $N$ is even).
}\hfill $\blacktriangleleft$
\end{example}

\begin{remark}[Equivariant local trivializations]\label{rk:loc_triv}{\upshape
\virg{Real} and \virg{Quaternionic} principal bundles are special types of principal $\n{U}(m)$-bundles and so they admit local trivializations by definition (\cf Appendix \ref{sect:remind_connect}). Less obvious is that these objects are also locally trivial in the category of principal bundles over spaces with involution. However, we can deduce immediately from Proposition \ref{prop:equi_top_Prin_QRVB} that $\rr{R}$- and $\rr{Q}$-principal bundles admit \emph{equivariant} local trivializations. With this we mean that for each given $\rr{R}$- or $\rr{Q}$-principal bundle $(\bb{P},\hat{\Theta})$
over the involutive space $(X,\tau)$ there exists an open  cover $\{\f{U}_\alpha\}$  of $\tau$-\emph{invariant} sets $\tau(\f{U}_\alpha)=\f{U}_\alpha$
and  local trivializations $h_\alpha:\bb{P}|_{\f{U}_\alpha}\to\f{U}_\alpha\times\n{U}(m)$ \emph{equivariant} in the sense that $h_\alpha\circ \hat{\Theta}=\hat{\Theta}_0\circ h_\alpha$ where $\hat{\Theta}_0$ is the trivial structure on the product principal bundle $\f{U}_\alpha\times\n{U}(m)$. As a matter of fact Proposition \ref{prop:equi_top_Prin_QRVB} states that each  $\rr{R}$- or $\rr{Q}$-principal bundle can be seen as the frame bundle of an associated $\rr{R}$- or $\rr{Q}$-vector bundle. These latter are locally trivial objects (\cf \cite[Proposition 4.10]{denittis-gomi-14} and \cite[Proposition 2.9]{denittis-gomi-14-gen}) and, as explained in Appendix \ref{sect:remind_connect},  local trivializations of the underlying vector bundle automatically provide local trivializations for the principal bundle. The equivariance of the local trivializations $h_\alpha$ automatically implies the $\tau$-equivariance of the  transition functions $\varphi_{\beta,\alpha}:\f{U}_\alpha\cap \f{U}_\beta\to\n{U}(m)$. More precisely,  one has that $\varphi_{\beta,\alpha}(\tau(x))=\overline{\varphi_{\beta,\alpha}(x)}$ in the  \virg{Real} case and $\varphi_{\beta,\alpha}(\tau(x))=\sigma(\varphi_{\beta,\alpha}(x))$ in the  \virg{Quaternionic} case. 
}\hfill $\blacktriangleleft$
\end{remark}

Let us finish this section by commenting on the case of an involutive base space $(X,\tau)$ with a smooth structure in the sense of  assumptions (H.1)-(H.3).
In this situation one can extend Definition \eqref{def_princ_R&QB} to the \emph{smooth category} simply by asking  the appropriate regularity for all the maps that enter into the definition. Moreover, Proposition \ref{prop:equi_top_Prin_QRVB} automatically extends to the smooth  category. Combining this fact with Theorem \ref{theo:top-smooth} one immediately obtains:

\begin{corollary}\label{cor:equi_categ}
Let $(X,\tau)$ be an involutive space which verifies  assumptions (H.1)-(H.3). Then
$$
\begin{aligned}
&{\rm Prin}^{\rr{R}}_{\n{U}(m)}(X,\tau)\;&\simeq&\; {^{C^r}{\rm Vec}^m_{\rr{R}}}(X,\tau)\;&\simeq&\; {^{\rm top}{\rm Vec}^m_{\rr{R}}}(X,\tau)\\
&{\rm Prin}^{\rr{Q}}_{\n{U}(2m)}(X,\tau)\;&\simeq&\; {^{C^r}{\rm Vec}^{2m}_{\rr{Q}}}(X,\tau)\;&\simeq&\; {^{\rm top}{\rm Vec}^{2m}_{\rr{Q}}}(X,\tau)
\end{aligned}
$$ 
where the principal bundles are understood in the $C^r$-category with $r\in\N\cup\{\infty\}$.
\end{corollary}
\noindent
As a consequence   we have the freedom to use smooth structures on $\rr{R}$- and $\rr{Q}$-principal bundles in order to explore the  classification of topological quantum system subjected to a TR symmetry.

\subsection{Time-reversal equivariant connections}\label{sect:equiv_sect}
In this section $(X,\tau)$ will be an involutive space that verifies conditions (H.1) - (H.3) and we consider  principal $\n{U}(m)$-bundles in the smooth category $\pi:\bb{P}\to X$ endowed with a \virg{Real} or \virg{Quaternionic} structure $\hat{\Theta}:\bb{P}\to \bb{P}$ as in Definition   \ref{def_princ_R&QB}.
We use the symbol $\omega\in\Omega^1(\bb{P},\rr{u}(m))$ for  \emph{connection} 1-forms associated to  given  horizontal distributions $p\mapsto H_p$ of $\bb{P}$ (for a reminder of the theory of connections see Appendix \ref{sect:remind_connect}). We observe that the Lie algebra $\rr{u}(m)$ has two natural involutions: a \emph{real} involution $\rr{u}(m)\ni\xi\mapsto \overline{\xi}\in\rr{u}(m)$ and a \emph{quaternionic} involution $\rr{u}(2m)\ni \xi\mapsto \sigma(\xi):=-Q\cdot\overline{\xi}\cdot Q\in \rr{u}(2m)$. Here $\xi\in \rr{u}(m)$ is any anti-Hermitian matrix of size $m$ and the matrix $Q$ has been defined in \eqref{eq:Q-mat}. Finally, given a $k$-form $\phi \in\Omega^k(\bb{P}, \bb{A})$ with value in some structure $\bb{A}$ (module, ring, algebra, group, \etc) and a smooth map $f: \bb{P}\to \bb{P}$ we denote with $f^*\phi:=\phi\circ f_\ast$ the \emph{pull-back} of  $\phi$ with respect to the map $f$ (and $f_\ast:T\bb{P}\to T\bb{P}$ is the \emph{push forward} of vector fields).
Given a $\rr{u}(m)$-valued $k$-form $\phi \in\Omega^k(\bb{P}, \rr{u}(m))$ we define the \emph{complex conjugate} form $\overline{\phi}$ pointwise, \ie 
$\overline{\phi}_p({\rm w}^1_p,\ldots,{\rm w}^k_p):=\overline{{\phi}_p({\rm w}^1_p,\ldots,{\rm w}^k_p)}$ for every $k$-tuple $\{{\rm w}^1_p,\ldots,{\rm w}^k_p\}\subset T_p\bb{P}$ of tangent vectors at $p\in\bb{P}$. It follows that $f^*\overline{\phi}=\overline{f^*\phi}$ for every smooth map $f:\bb{P}\to\bb{P}$. Similarly, if $\phi \in\Omega^k(\bb{P}, \rr{u}(2m))$ we define $\sigma(\phi)$ pointwise by
$\sigma(\phi)_p({\rm w}^1_p,\ldots,{\rm w}^k_p):=-Q\cdot\overline{{\phi}_p({\rm w}^1_p,\ldots,{\rm w}^k_p)}\cdot Q$. Hence, one has that $\sigma(f^*\phi)=f^*\sigma(\phi)$.

\begin{lemma}\label{lemma:invol_connect}
Let $(X,\tau)$ be an involutive space that verifies conditions (H.1) - (H.3) and $\pi:\bb{P}\to X$ a smooth principal $\n{U}(m)$-bundle over $X$
endowed with a \virg{Real}  structure $\hat{\Theta}:\bb{P}\to \bb{P}$ as in Definition   \ref{def_princ_R&QB}.
If  $\omega\in\Omega^1(\bb{P},\rr{u}(m))$ is a {connection} 1-form then also  $\hat{\Theta}^\ast\overline{\omega}$ is a {connection} 1-form.
Similarly, if $\hat{\Theta}:\bb{P}\to \bb{P}$ is a \virg{Quaternionic}  structure and $\omega\in\Omega^1(\bb{P},\rr{u}(2m))$ is a {connection} 1-form then also  $\hat{\Theta}^\ast\sigma(\omega)$ is a {connection} 1-form.
\end{lemma}
\proof
We start with the \virg{Real} case. In view of Proposition \ref{prop_commect} we need only to verify that $\hat{\Theta}^\ast\overline{\omega}$ verifies properties (a') and (b') that characterize a {connection} 1-form. For (a') let $\xi_p^*:= V_p(\xi)\in  T_p\bb{P}$ be the \emph{vertical} vector 
 at the point $p\in\bb{P}$ associated with $\xi\in \rr{u}(m)$ by \eqref{eq:defi_vert_vec}. Then
 $$
 \big(\hat{\Theta}^\ast\overline{\omega}\big)_p(\xi_p^*)\;=\;  \overline{\omega}_{\hat{\Theta}(p)}(\hat{\Theta}_\ast\xi_p^*)
 \;=\;  \overline{\omega}_{\hat{\Theta}(p)}\big(\overline{\xi}_{\hat{\Theta}(p)}^*\big)\;=\;\overline{\overline{\xi}}\;=\;{\xi}
 $$
where we used the equality $\hat{\Theta}_\ast\xi_p^*=\overline{\xi}_{\hat{\Theta}(p)}^*$ which can be checked from the definition \eqref{eq:defi_vert_vec}. The arbitrariness in the choice of $\xi\in \rr{u}(m)$  implies  $\hat{\Theta}^\ast\overline{\omega}\circ V_p={\rm Id}_{\rr{u}(m)}$. For  property (b'), given a tangent vector ${\rm w}_p\in T_p\bb{P}$ at the point $p\in\bb{P}$
and a $u\in\n{U}(m)$ one has  that
$$
\begin{aligned}
\big(R^*_u(\hat{\Theta}^\ast\overline{\omega})\big)_p({\rm w}_p)\;&=\;\big(\hat{\Theta}^\ast\overline{\omega}\big)_{R_u(p)}\big((R_u)_*{\rm w}_p\big) \;=\;\overline{\omega}_{(\hat{\Theta}\circ R_u)(p)}\big((\hat{\Theta}\circ R_u)_*{\rm w}_p\big)\\
&=\;\overline{\omega}_{(R_{\overline{u}}\circ\hat{\Theta})(p)}\big((R_{\overline{u}}\circ\hat{\Theta})_*{\rm w}_p\big)\;=\;\overline{(R_{\overline{u}}^*\omega)}_{\hat{\Theta}(p)}\big(\hat{\Theta}_*{\rm w}_p\big)\\
&=\;\overline{\overline{u}^{-1}\cdot(\omega)_{\hat{\Theta}(p)}\big(\hat{\Theta}_*{\rm w}_p\big)\cdot \overline{u}}\;=\;{u}^{-1}\cdot
(\hat{\Theta}^\ast\overline{\omega})_p({\rm w}_p)
\cdot u
\end{aligned}
$$
which proves that $R^*_u(\hat{\Theta}^\ast\overline{\omega})={\rm Ad}(u^{-1})\circ (\hat{\Theta}^\ast\overline{\omega})$.

For the \virg{Quaternionic} case the proof follows exactly along the same line. The only difference consists in the proper equality $\hat{\Theta}_\ast\xi_p^*=\sigma(\xi)_{\hat{\Theta}(p)}^*$ in the proof of (a').
\qed

\medskip

\noindent
With these premises we are now in position to give the following definitions.

\begin{definition}[\virg{Real} and \virg{Quaternionic} equivariant connections]\label{def:R&Q_connec}
Let $(X,\tau)$ be an involutive space that verifies conditions (H.1) - (H.3) and $\pi:\bb{P}\to X$ a smooth principal $\n{U}(m)$-bundle over $X$
endowed with a \virg{Real} or a \virg{Quaternionic} structure $\hat{\Theta}:\bb{P}\to \bb{P}$ as in Definition   \ref{def_princ_R&QB}.
We say that a {connection} 1-form $\omega\in\Omega^1(\bb{P},\rr{u}(m))$ is \emph{equivariant} if $\hat{\Theta}^\ast\overline{\omega}=\omega$
in the \virg{Real} case or $\hat{\Theta}^\ast\sigma(\omega)=\omega$ in the \virg{Quaternionic} case. 
Equivariant connections in the \virg{Real} case are called \virg{Real} connections (or $\rr{R}$-connections). Similarly, we call \virg{Quaternionic} connections (or $\rr{Q}$-connections) the equivariant connections in the \virg{Quaternionic} case. 
\end{definition}

\begin{remark}\label{rk:equiv_cond_alternativ}{\upshape
Let $\omega$ be a {connection} 1-form for the (smooth)
 principal $\n{U}(m)$-bundle $\pi:\bb{P}\to X$. 
 Proposition \ref{prop_commect} assures that both $\overline{\omega}$ in the \virg{Real} case or $\sigma(\omega)$ in the \virg{Quaternionic} case
 are {connection} 1-forms for $\bb{P}$.
Moreover, a standard result states that  for every (smooth) map $\hat{\Theta}:\bb{P}\to \bb{P}$ also $\hat{\Theta}^*\omega$ is a {connection} 1-form for $\bb{P}$. This means that the equivariance condition can be equivalently written as $\hat{\Theta}^*\omega=\overline{\omega}$ in the \virg{Real} case and $\hat{\Theta}^*\omega=\sigma(\omega)$ in the \virg{Quaternionic} case.
}\hfill $\blacktriangleleft$
\end{remark}

\begin{proposition}[Existence of equivariant connections]
Let $(X,\tau)$ be an involutive space that verifies conditions (H.1) - (H.3) and $\pi:\bb{P}\to X$ a smooth principal $\n{U}(m)$-bundle over $X$
endowed with a \virg{Real} or \virg{Quaternionic} structure $\hat{\Theta}:\bb{P}\to \bb{P}$. Then there exists at least one  equivariant (\virg{Real} or \virg{Quaternionic}, respectively)  connection  on $\bb{P}$.
\end{proposition}
\proof
Theorem \ref{teo:exist_connect} assures that $\bb{P}$ has at least one connection 1-form $\omega$ and Lemma \ref{lemma:invol_connect} shows that also $\omega'=\hat{\Theta}^*\overline{\omega}$ in the \virg{Real} case or $\omega'=\hat{\Theta}^*\sigma(\omega)$ in the \virg{Quaternionic} case 
are connections. Since convex combinations of connections are still connections we can define the \emph{average} connection $\omega_{\rm eq}:=\frac{1}{2}(\omega+\omega')$ which is equivariant by construction.
\qed

\medskip

\noindent
A  major consequence of the this result is that it shows that the spaces of equivariant connections are non empty. 
We denote with $\rr{A}_{\rr{R}}(\bb{P})\subset \Omega^1(\bb{P},\rr{u}(m))$ the space of $\rr{R}$-connections on the \virg{Real} principal bundle $(\bb{P},\hat{\Theta})$. Similarly,  $\rr{A}_{\rr{Q}}(\bb{P})\subset \Omega^1(\bb{P},\rr{u}(2m))$ will denote the space of $\rr{Q}$-connections on the \virg{Quaternionic} principal bundle $(\bb{P},\hat{\Theta})$. Let us  introduce the sets of equivariant 1-forms
\begin{equation}\label{eq:set_equi_form}
\begin{aligned}
\Omega^1_{\rr{R}}(\bb{P},\rr{u}(m))\;&:=\left\{\omega\in \Omega^1(\bb{P},\rr{u}(m))\ |\ \hat{\Theta}^*\overline{\omega}=\omega\right\}\;\\
\Omega^1_{\rr{Q}}(\bb{P},\rr{u}(2m))\;&:=\left\{\omega\in \Omega^1(\bb{P},\rr{u}(2m))\ |\ \hat{\Theta}^*\sigma(\omega)=\omega\right\}\;.
\end{aligned}
\end{equation}
We recall that a 1-form is called \emph{horizontal} if it vanishes on vertical vectors. The set of $\rr{u}(m)$-valued 1-form on $\bb{P}$ which are {horizontal} and which verify the transformation rule (b') in Proposition \ref{prop_commect} is denoted with $\Omega^1_{\rm hor}(\bb{P},\rr{u}(m),{\rm Ad})$. Accordingly, we can define the sets
$$
\begin{aligned}
\Omega^1_{\rr{R},{\rm hor}}(\bb{P},\rr{u}(m),{\rm Ad})\;&:=\; \Omega^1_{\rm hor}(\bb{P},\rr{u}(m),{\rm Ad})\cap\; \Omega^1_{\rr{R}}(\bb{P},\rr{u}(m))\\
\Omega^1_{\rr{Q},{\rm hor}}(\bb{P},\rr{u}(2m),{\rm Ad})\;&:=\; \Omega^1_{\rm hor}(\bb{P},\rr{u}(2m),{\rm Ad})\cap\; \Omega^1_{\rr{Q}}(\bb{P},\rr{u}(2m))\;.\\
\end{aligned}
$$
\begin{proposition}
The sets $\rr{A}_{\rr{R}}(\bb{P})$ and $\rr{A}_{\rr{Q}}(\bb{P})$ are closed with respect convex combinations with real coefficients.
Moreover, they are affine spaces modelled on the vector spaces $\Omega^1_{\rr{R},{\rm hor}}(\bb{P},\rr{u}(m),{\rm Ad})$ and $\Omega^1_{\rr{Q},{\rm hor}}(\bb{P},\rr{u}(2m),{\rm Ad})$, respectively.
\end{proposition}
\proof
The proof follows directly from Remark \ref{rk:convex_connect} and Corollary \ref{cor:aaf_prin}.
\qed

\medskip

Connection 1-forms of a principal $\n{U}(m)$-bundles can be described in terms of collections of local 1-forms on the base space subjected to suitable gluing rules (see Theorem \ref{theo:loc_connect} and equation \ref{eq:conn_trans_rul_bis}). This fact extends to the categories of \virg{Real} and \virg{Quaternionic} principal bundles, provided that an extra equivariance condition is added.  Let $\pi:\bb{P}\to X$ be an $\rr{R}$ or $\rr{Q}$-principal bundle over the involutive space $(X,\tau)$ and consider an \emph{equivariant}  local trivialization $\{\f{U}_\alpha,h_\alpha\}$ in the sense of 
Remark \ref{rk:loc_triv} with related transition functions $\{\varphi_{\beta,\alpha}\}$. On each open set  $\f{U}_\alpha\subset X$ we can define a local (smooth) section  $\rr{s}_\alpha(x):=h_\alpha^{-1}(x,\n{1})$ with $\n{1}\in\n{U}(m)$ the identity matrix. Due to the equivariance of the maps $h_\alpha$ the sections  $\rr{s}_\alpha$ inherit the equivariance property $\rr{s}_\alpha\circ\tau=\hat{\Theta}\circ \rr{s}_\alpha$. Let $\omega$ be an equivariant connection 1-form for $\bb{P}$. We can use the equivariant sections $\rr{s}_\alpha$ to construct the pull-backs $\s{A}_\alpha:=(\rr{s}_\alpha)^*\omega$. By definition, each $\s{A}_\alpha$ is a local 1-form with values in the Lie algebra $\rr{u}(m)$, \ie an element of $\Omega^1(\f{U}_\alpha,\rr{u}(m))$. Moreover, a simple calculation shows $\tau^*\s{A}_\alpha=(\rr{s}_\alpha\circ\tau)^*\omega=(\hat{\Theta}\circ \rr{s}_\alpha)^*\omega=\rr{s}_\alpha^*(\hat{\Theta}^*\omega)$. This implies that $\tau^*\overline{\s{A}_\alpha}=\s{A}_\alpha$ if $\omega$ is a \virg{Real} connection and $\tau^*\sigma(\s{A}_\alpha)=\s{A}_\alpha$ if $\omega$ is a \virg{Quaternionic} connection. In analogy with \eqref{eq:set_equi_form}
we introduce the sets of 1-forms
\begin{equation}\label{eq:set_equi_form_base}
\begin{aligned}
\Omega^1_{\rr{R}}(\f{U},\rr{u}(m))\;&:=\left\{\s{A}\in \Omega^1(\f{U},\rr{u}(m))\ |\ \tau^*\overline{\s{A}}=\s{A}\right\}\;\\
\Omega^1_{\rr{Q}}(\f{U},\rr{u}(2m))\;&:=\left\{\s{A}\in \Omega^1(\f{U},\rr{u}(2m))\ |\ \tau^*\sigma(\s{A})=\s{A}\right\}\;.
\end{aligned}
\end{equation}
where $\f{U}\subset X$ is some $\tau$-invariant open set. We are now in position to state the following result:
\begin{theorem}[Local description of equivariant connections]\label{theo:loc_connect_equiv}
Let $(X,\tau)$ be an involutive space that verifies conditions (H.1) - (H.3) and $\pi:\bb{P}\to X$ a smooth principal $\n{U}(m)$-bundle over $X$
endowed with a \virg{Real} or \virg{Quaternionic} structure $\hat{\Theta}:\bb{P}\to \bb{P}$. Let $\{\f{U}_\alpha,h_\alpha\}$ be a set of equivariant  local trivializations (in the sense of Remark \ref{rk:loc_triv}) with related equivariant transition functions $\{\varphi_{\beta,\alpha}\}$.
There is a one-to-one correspondence between equivariant connections $\omega\in \rr{A}_{\rr{T}}(\bb{P})$ and collections of equivariant 1-forms $\{\s{A}_\alpha\in\Omega^1_{\rr{T}}(\f{U}_\alpha,\rr{u}(m))\}$ which verify the transformation rules
\begin{equation}\label{eq:conn_trans_rul-equiv}
\s{A}_\alpha\;=\; \varphi_{\beta,\alpha}^{-1}\;\cdot\; \s{A}_\beta\;\cdot\; \varphi_{\beta,\alpha}\;+\;\varphi_{\beta,\alpha}^{-1}\;\cdot\;{\rm d }\varphi_{\beta,\alpha}\;,\qquad\text{on}\ \ \f{U}_\alpha\cap\f{U}_\beta\;. 
\end{equation}
Here, the subscript $\rr{T}$ is used 
as a label for both the \virg{Real} structure $\rr{R}$ and the \virg{Quaternionic} structure $\rr{Q}$. Moreover, in the \virg{Quaternionic} case the rank $m$ has to be considered even.
\end{theorem}
\proof
The proof follows exactly as in the non-equivariant case (\cf Theorem \ref{theo:loc_connect} and related references). Moreover, 
as a consequence of the equivariance of the transition functions $\varphi_{\beta,\alpha}$ (\cf Remark \ref{rk:loc_triv}) one can easily check  the compatibility between the transformation rule \eqref{eq:conn_trans_rul-equiv} and  the equivariance condition for 1-forms in $\Omega^1_{\rr{T}}(\f{U}_\alpha,\rr{u}(m))$.
\qed

\medskip

\begin{corollary}\label{corol:uniq_R-connection}
Let $X$ be a manifold with  trivial $\Z_2$-action. Then any \virg{Real}
$\n{U}(1)$-principal bundle over $X$ admits only one \virg{Real} connection.
\end{corollary}
\proof
Let $\omega$ be a connection 1-form identified by family of local 1-forms $\{\s{A}_\alpha\}$. 
Since the action of $\tau$ is trivial  the first of \eqref{eq:set_equi_form_base} leads to $\s{A}_\alpha=\overline{\s{A}_\alpha}$. However,
this implies $\s{A}_\alpha=0$ since $\s{A}_\alpha$ must take value in $\rr{u}(1)=\ii\R$. Then the unique \virg{Real} connection
$\omega$ is specified by the local conditions $(\sigma_\alpha)^*\omega=0$ where $\sigma_\alpha:\f{U}_\alpha\to \bb{P}$ are the canonical sections defined by the \virg{Real} trivialization.
\qed

\medskip
\medskip

We conclude this section by discussing a simple but interesting (at least in low dimension) situation in which equivariant connections can be constructed explicitly.

\begin{example}[Equivariant connections on TR bundles with product structure]\label{ex:low_dim}{\upshape
Let $\pi:X\times\n{U}(m)\to X$ be the product principal $\n{U}(m)$-bundle
over an involutive space $(X,\tau)$ which verifies conditions (H.1) - (H.3). As already discussed in Example \ref{ex:PB_prod}
each map $J:X\to\n{U}(m)$ which verifies the equivariance condition $J(\tau(x))\cdot\overline{J(x)}=\n{1}$ endows the product bundle with the \virg{Real} structure $\hat{\Theta}:(x,u)\mapsto(\tau(x),J(x)\cdot\overline{u})$. Similarly, if one replaces $m$ with $2m$, one can endow the product bundle
with a \virg{Quaternionic} structure $\hat{\Theta}:(x,u)\mapsto(\tau(x),J(x)\cdot\overline{u}\cdot Q)$ provided that the map $J:X\to\n{U}(2m)$
verifies the equivariance condition $J(\tau(x))\cdot\overline{J(x)}=-\n{1}$. By exploiting  the product structure of $X\times\n{U}(m)$ one can try to describe the equivariant connections with the help of the  \emph{Maurer-Cartan}  1-form $\theta\in \Omega^1_{\rm left}(\n{U}(m),\rr{u}(m))$ given by  $\theta_u=u^{-1}\cdot{\rm d}u$ for all $u\in\n{U}(2m)$. The pull-back $\omega_{\rm flat}:=({\rm pr}_2)^*\theta$ under the natural projection ${\rm pr}_2:X\times\n{U}(m)\to X$ defines a connection 1-form $\omega_{\rm flat}\in\rr{A}(X\times\n{U}(m))$ called canonical \emph{flat} connection (\cf Appendix \ref{sect:remind_connect}). Moreover, the inclusion $\imath:X\to X\times\n{U}(m)$ given by $\imath(x)=(x,\n{1})$ provides the pull-back $\imath^*:\Omega^1(X\times \n{U}(m),\rr{u}(m))\to
\Omega^1(X,\rr{u}(m))$
which is well known  to restrict to  a one-to-one correspondence $\rr{A}(X\times \n{U}(m))\simeq \Omega^1(X,\rr{u}(m))$ on the connection 1-forms.
 In order to explain this correspondence we observe that from each $\s{A}\in \Omega^1(X,\rr{u}(m))$  one can define (pointwise) an element $\tilde{\s{A}}\in \Omega^1(X\times \n{U}(m),\rr{u}(m))$ by means of the formula
 $$
 \tilde{\s{A}}_{(x,u)}\;:=\;{\rm Ad}(u^{-1})\;\circ\; \s{A}_x\circ \pi_*\;=\;u^{-1}\cdot(\s{A}_x\circ \pi_*)\cdot u\;.
 $$
By construction $\tilde{\s{A}}$ vanishes on vertical vectors and verifies the transformation rule $(R_u)^*\tilde{\s{A}}={\rm Ad}(u^{-1})\circ\tilde{\s{A}}$, hence 
\begin{equation}\label{eq:form_connect}
\omega_{\s{A}}\;:=\;\tilde{\s{A}}\;+\;\omega_{\rm flat}
\end{equation} 
results to be a connection 1-forms on 
$X\times \n{U}(m)$ since it verifies properties (a') and (b') of 
Proposition \ref{prop_commect}. Moreover, ${\rm pr}_2\circ\imath(x)=\n{1}$ and ${\pi}\circ\imath(x)=x$
for all $x\in X$ imply  $\imath^*\omega_{\s{A}}=\imath^* \tilde{\s{A}}=\s{A}$. In view of Lemma \ref{lemma:invol_connect} the 1-form
 $\hat{\Theta}^*\overline{\omega_{\s{A}}}$ in the \virg{Real} case or  the 1-form $\hat{\Theta}^*\sigma(\omega_{\s{A}})$ in the \virg{Quaternionic} case are still connections  and so there exists a unique $\s{A}^J\in  \Omega^1(X,\rr{u}(m))$ 
 associated to them by  $\imath^*$.
A standard  (but tedious) calculation shows that
$$
\s{A}^J\;:=\; \left\{
\begin{aligned}
&\overline{J^{-1}\cdot (\tau^*\s{A})\cdot J\;+\;J^{-1}\;{\rm d}J}&&\qquad\text{(\virg{Real} case)}\\
&\sigma\big(J^{-1}\cdot (\tau^*\s{A})\cdot J\;+\;J^{-1}\;{\rm d}J\big)&&\qquad\text{(\virg{Quaternionic} case)}\;.\\
\end{aligned}
\right.
$$
Evidently, the connection 1-form $\omega_{\langle\s{A}\rangle}$
 associated to the \emph{averaged} 1-form
$\langle\s{A}\rangle:=\frac{1}{2}(\s{A}+\s{A}^J)$ 
by means of \eqref{eq:form_connect} turns out to be equivariant in the sense of Definition \ref{def:R&Q_connec}. In particular if one start with $\s{A}=0$ (which produces  via \eqref{eq:form_connect} the flat connection $\omega_{\rm flat}$) one ends, after averaging, with the two special forms
$$
\s{A}^J_0\;:=\; \frac{1}{2}\left\{
\begin{aligned}
&\overline{J^{-1}\;{\rm d}J}&&\qquad\text{(\virg{Real} case)}\\
&\sigma\big(J^{-1}\;{\rm d}J\big)&&\qquad\text{(\virg{Quaternionic} case)}\;.\\
\end{aligned}
\right.
$$
This form
defines through the construction \eqref{eq:form_connect}
 a special  connection $\omega_{\s{A}^J_0}$
on the product bundle $X\times \n{U}(m)$ 
which is equivariant with respect to the $\hat{\Theta}$ structure induced by the map $J:X\to \n{U}(m)$.

\medskip

We point out that the construction here discussed has a quite general validity in low dimension. In effect if the dimension of the manifold $X$ does not exceed 3 and if $H_{\Z_2}^2(X,\Z(1))=0$
(see Section \ref{subsec:borel_cohom} and references therein)
 then all   \virg{Real} or  \virg{Quaternionic}  principal bundles over $(X,\tau)$ are, up to isomorphisms, of 
the form described in this example. This fact can be proved by applying the equivalence of categories stated in Corollary \ref{cor:equi_categ} to the general construction described in \cite[Section 4.2]{denittis-gomi-14-gen}.
}\hfill $\blacktriangleleft$
\end{example}

\subsection{Time-reversal equivariant connections on vector bundles}\label{sect:equiv_sect_VB}

Connections on  principal bundles define covariant derivatives on vector bundles (\cf Appendix \ref{sect:remind_connect}). 
Thus, it is not surprising that a similar correspondence extends also to the \virg{Real} or  \virg{Quaternionic} categories.

\medskip

Let us consider a vector bundle $\bb{E}$ over the involutive space $(X,\tau)$ endowed with an $\rr{R}$ or $\rr{Q}$-structure
$\Theta:\bb{E}\to\bb{E}$. In order to define connections on vector bundles one needs  the spaces
\begin{equation}\label{eq:E-val_for_TR}
\Omega^k(X,\bb{E})\;:=\;
\Gamma\big(\bb{E}\otimes\Omega^k(X)\big)\;\simeq\;
\Gamma\big(\bb{E}\big)\; \otimes_{\Omega^0(X)}\; \Omega^k(X)
\end{equation}
of $k$-forms on $X$ with values in the vector bundle $\bb{E}$.
The second  tensor product  in \eqref{eq:E-val_for_TR} has to be understood as the a product of modules over the ring $\Omega^0(X)$ of smooth  functions on $X$. This also implies that $\Omega^0(X,\bb{E})=\Gamma\big(\bb{E}\big)$.
According to Definition
\ref{def:cov_deriv}
a \emph{vector bundle connection} $\nabla$
on  $\bb{E}\to X$
is a  differential operator
\begin{equation}\label{eq:cov_deriv1_TR}
\nabla\;:\;\Gamma(\bb{E})\;\longrightarrow\;\Omega^{1}(X,\bb{E})\ \end{equation}
which verifies the
 \emph{Leibniz rule}
\eqref{eq:cov_deriv2}. We use the symbol $\rr{A}(\bb{E})$ for the space of vector bundle connections on $\bb{E}$. The $\Theta$-structure on $\bb{E}$ defines an involution $\tau_\Theta:\Gamma\big(\bb{E}\big)\to \Gamma\big(\bb{E}\big)$ on the space of sections given by $\tau_\Theta(s):=\Theta\circ s\circ\tau$ for all $s\in \Gamma\big(\bb{E}\big)$ (see \cite[Section 4.3]{denittis-gomi-14} for the \virg{Real} case and \cite[Section 2.2]{denittis-gomi-14-gen} for the \virg{Quaternionic} case).
This extends  to an involution (still denoted with the same symbol) $\tau_\Theta:\Omega^k(X,\bb{E})\to \Omega^k(X,\bb{E})$ in each degree $k$ by
\begin{equation}\label{eq:inv_forms}
\tau_\Theta(\phi)\;:=\;\Theta\circ \phi\circ\tau\;,\qquad\quad \phi\in \Omega^k(X,\bb{E})\;.
\end{equation}
In particular, on elementary tensor products one has
$$
\tau_\Theta(s\otimes \varepsilon)\;=\;(\Theta\circ s\circ\tau)\;\otimes\;\overline{(\tau^*\varepsilon)}\;,\qquad\quad s\in \Gamma\big(\bb{E}\big)\;,\ \  \ \varepsilon\in \Omega^k(X)\;.
$$

\begin{definition}[\virg{Real} and \virg{Quaternionic} equivariant vector bundle connections]\label{def:R&Q_connec_VB}
Let $(X,\tau)$ be an involutive space that verifies conditions (H.1) - (H.3) and $\bb{E}\to X$ a smooth vector bundle over $X$
endowed with a \virg{Real} or \virg{Quaternionic} structure ${\Theta}:\bb{E}\to \bb{E}$.
We say that a {vector bundle connection} $\nabla:\Gamma(\bb{E})\to\Omega^{1}(X,\bb{E})$ is \emph{equivariant} if 
$$
\nabla\;\circ\;\tau_\Theta\;=\;\tau_\Theta\;\circ\;\nabla.
$$
where $\tau_\Theta:\Omega^k(X,\bb{E})\to \Omega^k(X,\bb{E})$, $k=0,1$ is the involution defined by \eqref{eq:inv_forms}.
\end{definition}
\noindent
We will use consistently the symbols $\rr{A}_{\rr{R}}(\bb{E})$ and $\rr{A}_{\rr{Q}}(\bb{E})$ for the sets of equivariant vector bundle connections on  \virg{Real} or \virg{Quaternionic} vector bundles, respectively. We notice that the notion of equivariant connection as in Definition \ref{def:R&Q_connec_VB} is not new in the literature. For instance 
a similar (but less general) notion is already present  in \cite{freed-moore-13}.

\medskip

There is a one-to-one correspondence between connection 1-forms $\omega$ (also called linear connections) on the frame bundle
$\pi:\bb{F}(\bb{E})\to X$  and vector bundle connections $\nabla$ on the underlying vector bundle $\bb{E}\to X$ (see Appendix \ref{sect:remind_connect}).
In particular, this correspondence  is made explicit by using local trivializations and local sections as shown in Proposition \ref{prop_local_1to1}. A similar result extends also to the \virg{Real} or \virg{Quaternionic} categories.

\begin{theorem}
Let $(X,\tau)$ be an involutive space that verifies conditions (H.1) - (H.3) and $\bb{E}\to X$ a smooth vector bundle over $X$
endowed with a \virg{Real} or \virg{Quaternionic} structure ${\Theta}:\bb{E}\to \bb{E}$. Let $\pi:\bb{F}(\bb{E})\to X$
be the related frame bundle and  $\hat{\Theta}:\bb{F}(\bb{E})\to \bb{F}(\bb{E})$ the equivariant structure induced by $\Theta$ (\cf Proposition \ref{prop:equi_top_Prin_QRVB}).
There is a one-to-one correspondence between 
\emph{$\Theta$-equivariant} vector bundle connections $\nabla$
on the vector bundle $\bb{E}$ and
\emph{$\hat{\Theta}$-equivariant} connection 1-forms $\omega$ on the frame bundle
$\bb{F}(\bb{E})$, namely $\rr{A}_{\rr{T}}(\bb{E})\simeq \rr{A}_{\rr{T}}(\bb{F}(\bb{E}))$ where  the subscript $\rr{T}$  
stands for both the \virg{Real} structure $\rr{R}$ and the\virg{Quaternionic} structure $\rr{Q}$.
\end{theorem}
\proof[{proof} (sketch of)]
The simplest way to prove this theorem is  to use the \emph{local} equivalence  stated in Proposition \ref{prop_local_1to1}. Let $\{\f{U}_\alpha,f_\alpha\}$ be a system of local trivializations for the vector bundle $\bb{E}\to X$ and  $\{\varphi_{\beta,\alpha}\}$ the related transition functions (which are the same as the associated frame bundle $\bb{F}(\bb{E})$). We know that these trivializations can be chosen to be equivariant (\cf Remark \ref{rk:loc_triv}) and, in particular, one has $\tau(\f{U}_\alpha)=\f{U}_\alpha$.
For each (smooth) section $s\in\Gamma(\bb{E})$ we define
the  family $\{s_\alpha\}$ of (smooth) local sections 
$s_\alpha:\f{U}_\alpha\to\n{C}^m$ 
of the product bundle $\f{U}_\alpha\times\C^m$
defined by the relations
$f_\alpha\circ s(x)=(x,s_\alpha(x))$. These sections obey the condition $s_\beta(x)=\varphi_{\beta,\alpha}(x)\cdot s_\alpha(x)$ for all $x\in\f{U}_\alpha\cap\f{U}_\beta$.
Moreover, the equivariance of the trivializing functions
$f_\alpha\circ \Theta=\Theta_0\circ f_\alpha$ (here $\Theta_0$ is the trivial $\rr{R}$ or $\rr{Q}$ structure on the product bundle $\f{U}_\alpha\times\C^m$) implies that the transformed section $\tau_\Theta(s)$ defines a related
family of (smooth) local sections 
$\{\tau_{\Theta_0}(s_\alpha)\}$ defined, as usual, by
$\tau_{\Theta_0}(s_\alpha):=\Theta_0\circ s_\alpha\circ\tau$. 

Let $\omega\in \rr{A}(\bb{F}(\bb{E}))$ be a connection 1-form for the frame bundle and $\nabla\in \rr{A}(\bb{E})$
the related vector bundle connection on $\bb{E}$. If $\s{A}_\alpha\in\Omega^{1}(\f{U}_\alpha,\rr{u}(m))$ are the local 1-forms which define $\omega$
(according to Theorem \ref{theo:loc_connect}) one has from  
Proposition \ref{prop_local_1to1} that 
the covariant derivative $\nabla(s)\in\Omega^{1}(X,\bb{E})$ is uniquely determined by the family of 1-forms $\nabla(s_\alpha)\in\Omega^{1}(\f{U}_\alpha,\n{C}^m)$ given by
$$
\nabla(s_\alpha)\;:=\;{\rm d}s_\alpha\;+\;\s{A}_\alpha\;\cdot\;s_\alpha\;.
$$
Let us assume that the connection 1-form $\omega$ is 
equivariant in the sense of Definition \ref{def:R&Q_connec}.
This implies, by Theorem \ref{theo:loc_connect_equiv}, that
$\s{A}_\alpha\in \Omega^1_{\rr{T}}(\f{U}_\alpha,\rr{u}(m))$ and a straightforward calculation shows that $\s{A}_\alpha\cdot \tau_{\Theta_0}(s_\alpha)=\tau_{\Theta_0}(\s{A}_\alpha \cdot s_\alpha)$. We point out that in the last equality we have tacitly extended  the map $\tau_{\Theta_0}$ from $\Omega^{0}(\f{U}_\alpha,\n{C}^m)$ to $\Omega^{k}(\f{U}_\alpha,\n{C}^m)$ as in the case of equation \eqref{eq:inv_forms}. Since the relation
${\rm d}(\tau_{\Theta_0}(s_\alpha))=\tau_{\Theta_0}({\rm d}s_\alpha)$ can be easily verified, one obtains that 
$\nabla(\tau_{\Theta_0}(s_\alpha))=\tau_{\Theta_0}(\nabla(s_\alpha))$ which implies the equivariance of $\nabla$ in the sense of Definition \ref{def:R&Q_connec_VB}.

Conversely, if $\nabla$ is an equivariant vector bundle connection in the sense of Definition \ref{def:R&Q_connec_VB} then similar considerations imply that $\s{A}_\alpha\in \Omega^1_{\rr{T}}(\f{U}_\alpha,\rr{u}(m))$, namely the equivariance of the connection 1-form $\omega$  in the sense of Definition \ref{def:R&Q_connec}.
\qed

\subsection{Grassmann-Berry connection and equivariance condition}\label{sect:bott-chern-berry}

In this section  we focus on the special situation of vector bundles (or principal bundles)
which arise from a topological quantum system with an isolated family of $m$ energy bands as described in Definition \ref{def:tsq}. 
In this case the  spectral calculus leads to a (smooth) projection valued map $X\ni x\mapsto P(x)\subset {\rm Mat}_N(\C)$ such that ${\rm Tr}(P(x))=m$ for all $x\in\ X$ and $N\geqslant m$. 
We already mentioned in the introduction that the
Serre-Swan Theorem associates to the map $x\mapsto P(x)$  a (smooth) rank $m$ complex vector bundle $\bb{E}\to X$ endowed with a Hermitian structure (that is usually called Bloch-bundle). The relevant fact here is that the map $x\mapsto P(x)$ also provides a nice recipe to build a special connection, the \emph{Grassmann-(Bott-Chern)-Berry connection} which is very useful in applications (\cf \cite[Section 2.1.3]{chruscinski-jamiolkowski-04}).

\medskip

The Bloch-bundle $\bb{E}\to X$ can be seen (by construction) as a {subbundle} of the trivial \virg{ambient} bundle $X\times\C^N\to X$ by projecting in each fiber on the range of $P(x)$. In particular this leads to the natural inclusion $\imath:\Gamma(\bb{E})\hookrightarrow \Gamma(X\times\C^N)$ and to the restriction map $P :\Gamma(X\times\C^N)\to \Gamma(\bb{E})$ defined pointwise by
 $s(x)\mapsto (P\circ s)(x):=P(x)\cdot s(x)$ for all $s\in \Gamma(X\times\C^N)$. Moreover, the restriction map $P$ 
 extends analogously 
 also to the spaces \eqref{eq:E-val_for_TR} giving maps $P : \Omega^k(X,X\times\C^N)\to \Omega^k(X,\bb{E})$. The
Grassmann-Berry connection $\nabla_{\rm GB}$ is, by definition, the compression of the  flat connection (exterior derivative) ${\rm d}$ on the trivial vector bundle $X\times\C^N$ with the maps $\imath$ and $P$. More precisely $\nabla_{\rm GB}$  is defined by the following sequence
$$
\Gamma(\bb{E})\;\stackrel{\imath}{\hookrightarrow}\; \Gamma(X\times\C^N)\;\stackrel{{\rm d}}{\rightarrow}\;\Omega^1(X,X\times\C^N)\;\stackrel{P}{\rightarrow}\;\Omega^1(X,\bb{E})\;
$$
which justifies the use of the notation $\nabla_{\rm GB}\equiv P\circ {\rm d}$. 
The vector bundle connection $\nabla_{\rm GB}$ turns out to be compatible with the Hermitian structure of $\bb{E}$, meaning that
$$
\langle \nabla_{\rm GB}(s_1);s_2\rangle_x\;+\; \langle s_1;\nabla_{\rm GB}(s_2)\rangle_x\;=\;{\rm d}\ \langle s_1;s_2\rangle_x\;,\qquad\forall\  s_1,s_2\in \Gamma(\bb{E})\;,\ \ \ \ \forall\ x\in X\;.
$$
This follows since
the map $x\mapsto P(x)$ is pointwise self-adjoint $P(x)=P(x)^*$ 
and $P(s)\cdot s(x)=s(x)$ if $s\in \Gamma(\bb{E})$. 

\medskip

Let  $\{\f{U}_\alpha\}$ be  an open cover of $X$ and for each $\f{U}_\alpha$ let $\{\psi_1^{(\alpha)},\ldots,\psi_m^{(\alpha)}\}$ be a (smooth) \emph{orthonormal} frame  which trivializes $\bb{E}|_{\f{U}_\alpha}$. This means that the map $x\mapsto P(x)$ takes locally the form (using the Dirac notation)
$$
P|_{\f{U}_\alpha}(x)\;=\;\sum_{r=1}^m\ketbra{\psi_r^{(\alpha)}(x)}{\psi_r^{(\alpha)}(x)}\;,\qquad\quad x\in \f{U}_\alpha\;.
$$
Consider now the system of  local 1-forms 
$\s{A}^{\rm GB}_\alpha:=\{(\s{A}^{\rm GB}_\alpha)_{r,s}\}\in\Omega^{1}(\f{U}_\alpha,\rr{u}(m))$ whose components are given by
\begin{equation}\label{connect_GB3}
{({\s{A}^{\rm GB}_\alpha}|_x)_{r,s}}\;:=\;\langle\psi_r^{(\alpha)};{\rm d}\psi_s^{(\alpha)}\rangle_x\;=\;
\sum_{\ell=1}^d\langle\psi_r^{(\alpha)};\partial_{x_\ell}\psi_s^{(\alpha)}\rangle_x\; {\rm d}x^\ell
\end{equation}
once a local system of coordinates $(x_1,\ldots,x_d)$ for the open neighborhood $\f{U}_\alpha$ of the $d$-dimensional (smooth) manifold
$X$ has been chosen. Since $\langle\psi_r^{(\alpha)};\psi_s^{(\alpha)}\rangle_x=\delta_{r,s}$ for all $x\in \f{U}_\alpha$
one deduces that $\langle\psi_r^{(\alpha)};{\rm d}\psi_s^{(\alpha)}\rangle_x=-\langle{\rm d}\psi_r^{(\alpha)};\psi_s^{(\alpha)}\rangle_x$ which implies, in particular, that $\s{A}^{\rm GB}_\alpha$ is an anti-Hermitian valued (\ie a $\rr{u}(m)$-valued) 1-form.
Moreover, since two frames $\{\psi_1^{(\alpha)},\ldots,\psi_m^{(\alpha)}\}$  and $\{\psi_1^{(\beta)},\ldots,\psi_m^{(\beta)}\}$ 
are related on the overlapping $\f{U}_\alpha\cap \f{U}_\beta$ by the relation $\psi_r^{(\alpha)}=\sum_{s=1}^m(\varphi_{\beta,\alpha})_{s,r}\;\psi_s^{(\beta)}$ where $(\varphi_{\beta,\alpha})_{r,s}$ are the entries (with respect to the given local frames)
of
the system of  transition functions $\varphi_{\beta,\alpha}:\f{U}_\alpha\cap \f{U}_\beta\to \n{U}(m)$, one can easily show that the 
 local 1-forms 
$\s{A}^{\rm GB}_\alpha$  transform according to the relation \eqref{eq:conn_trans_rul_bis}. This means that the collection $\{ \s{A}^{\rm GB}_\alpha\}$ defines a unique connection  on the Bloch-bundle $\bb{E}$ (\cf Proposition \ref{prop_local_1to1}).
We will  show that this connection is exactly the Grassmann-Berry connection $\nabla_{\rm GB}$.

\medskip

Each $s\in \Gamma(\bb{E})$ can be locally expanded as 
$$
s_\alpha(x)\;:=\;s|_{\f{U}_\alpha}(x)\;=\;\sum_{r=1}^ms_r^{(\alpha)}(x)\; \ket{\psi_r^{(\alpha)}(x)}
\;,\qquad\quad x\in \f{U}_\alpha\;.
$$
where the (smooth) functions  $s_r^{(\alpha)}(x):=\langle\psi_r^{(\alpha)}; s\rangle_x$ are defined by the fiberwise Hermitian product. The Leibniz's rule provides
$$
{\rm d}s_\alpha(x)\;=\; \sum_{r=1}^m\left({\rm d}s_r^{(\alpha)}(x)\right)\; \ket{\psi_r^{(\alpha)}(x)}
\;+\; \sum_{r=1}^ms_r^{(\alpha)}(x)\; \ket{{\rm d}\psi_r^{(\alpha)}(x)}
$$
where $\ket{{\rm d}\psi_r^{(\alpha)}(x)}:=\sum_{\ell=1}^d\ket{\partial_{x_\ell}\psi_r^{(\alpha)}(x)}\; {\rm d}x^\ell$.
The definition of $\nabla_{\rm GB}$ and a simple calculation provide
$$
\nabla_{\rm GB}(s_\alpha)(x)\;=\;P(x)\cdot {\rm d}s_\alpha(x)\;=\; \sum_{r=1}^m\left({\rm d}s_r^{(\alpha)}(x)\;+\; \sum_{k=1}^m\langle\psi^{(\alpha)}_r;{\rm d}\psi^{(\alpha)}_k\rangle_x\; s_k^{(\alpha)}(x)\right)\ket{\psi_r^{(\alpha)}(x)}
$$
Finally a comparison with the \eqref{connect_GB3} shows that  $\nabla_{\rm GB}$ acts on the local components 
$(s_1^{(\alpha)},\ldots,s_m^{(\alpha)})$
of the section
$s_\alpha$ by means of the  1-forms $\s{A}_\alpha^{\rm GB}$ as described in Proposition \ref{prop_local_1to1}. This proves that 
the Grassmann-Berry connection $\nabla_{\rm GB}$ is uniquely defined by the family $\{\s{A}_\alpha^{\rm GB}\}\subset \Omega^{1}(\f{U}_\alpha,\rr{u}(m))$.

\begin{remark}{\upshape
The vector bundle connection $\nabla_{\rm GB}$ specifies, and is specified by,  a unique connection 1-form $\omega_{\rm GB}$ on the principal frame bundle  $\bb{F}(\bb{E})\to X$ associated with $\bb{E}$ (\cf Proposition \ref{prop_local_1to1}). In particular
$\omega_{\rm GB}$ can be reconstructed from the family of  local  1-forms $\{\s{A}_\alpha^{\rm GB}\}$ (\cf Theorem \ref{theo:loc_connect}). In order to understand the geometric content of the connection $\omega_{\rm GB}$ it is important to compute the horizontal distribution $\bb{F}(\bb{E})\ni p\mapsto H_p\subset \bb{F}(\bb{E})|_p$ associated with $\omega_{\rm GB}$ by the relation $H_p={\rm Ker}({\omega_{\rm GB}}_p)$ (\cf Proposition \ref{prop_commect}).
We observe that the choice of a local frame $\{\psi_1^{(\alpha)},\ldots,\psi_m^{(\alpha)}\}$ over 
$\f{U}_\alpha$ defines a canonical local section $\sigma_\alpha:X\to \bb{F}(\bb{E})|_{\f{U}_\alpha}\simeq \f{U}_\alpha\times\n{U}(m)$ 
and the  local relation ${\s{A}^{\rm GB}_\alpha}=\sigma_\alpha^*\omega_{\rm GB}$ holds  over $\f{U}_\alpha$.
This means that the horizontal space $H_{\sigma_\alpha(x)}$ is determined by the pushforward ${\sigma_\alpha}_*{\rm v}$ of tangent vectors ${\rm v}\in T_xX$ such that ${\s{A}^{\rm GB}_\alpha}|_x({\rm v})=0$. The remaining  horizontal  spaces $H_{p}$ for $p\in\pi^{-1}(x)$ can be computed from $H_{\sigma_\alpha(x)}$ via the right action of the structure group $\n{U}(m)$ (\cf Appendix \ref{sect:remind_connect}).
Said differently, a path $\tilde{\gamma}_\alpha:[0,1]\to\f{U}_\alpha$ such that ${\s{A}^{\rm GB}_\alpha}(\dot{\tilde{\gamma}}_\alpha)=0$ has a horizontal lift which is simply given by $\gamma_\alpha:=\sigma_\alpha\circ{\tilde{\gamma}}_\alpha$.
}\hfill $\blacktriangleleft$
\end{remark}

Now we want to consider the case in which the projection $P(x)$ is subjected to an even or odd time-reversal symmetry of type
\eqref{eq:tqsA3}. As  discussed in Example \ref{ex:PB_prod} we can assume that this symmetry acts at level of the 
 trivial \emph{ambient} bundle $X\times\C^N$ by $\Theta:(x,{\rm v})=(\tau(x),J(x)\cdot\overline{\rm v})$
and restrict to the subbundle $\bb{E}$ since the constraint $P(\tau(x))\cdot J(x)=J(x)\cdot\overline{P(x)}$ implies $P\circ\Theta=\Theta\circ P$. This also implies that $P\circ\tau_\Theta=\tau_\Theta\circ P$ where the map $\tau_\Theta$ is described by  
 \eqref{eq:inv_forms}. Each section $s\in\Gamma(\bb{E})$ is a map $s:x\mapsto (x,{\rm v}(x))$ subjected to the constraint $P(x){\rm v}(x)={\rm v}(x)$. The transformed section is  $\tau_\Theta(s)(x)=(x,J(\tau(x))\;\overline{{\rm v}(\tau(x))})$ while  $\nabla_{\rm GB}$ acts as $\nabla_{\rm GB}(s)(x)=(x,P(x)\;{\rm dv}(x))$. By using these definitions a straightforward calculation shows that
$$
 \nabla_{\rm GB}\;\circ\;\tau_\Theta\;-\;\tau_\Theta\;\circ\;\nabla_{\rm GB}\;=\;P\cdot {\rm d}(J\circ \tau)\;=\;\pm\;P\cdot\overline{{\rm d}J^*}
 $$
 where the last equality is a consequence of the constraint $\overline{J\circ \tau}=\; \pm J^*$ (\cf equation \eqref{eq:tqsA3}) and the sign $\pm$ depends on the parity of the structure $\Theta$. Then, we have proved that:
 \begin{proposition}\label{prop:equivGB_conn}
 Let $\bb{E}\to X$ be a rank $m$ \virg{Real} or \virg{Quaternonic} vector bundle over $(X,\tau)$ associated to a 
 projection valued map $x\mapsto P(x)$ subjected to the constraints $(P\circ\tau)\cdot J=J\cdot\overline{P}$ and $\overline{J\circ \tau}=\; \pm J^*$. The associated Grassmann-Berry connection $\nabla_{\rm GB}$ is invariant (in the sense of Definition \ref{def:R&Q_connec_VB}) if and only if $P\cdot\overline{{\rm d}J^*}=0$.
 \end{proposition} 
 
\noindent
In many physical situations of interest the unitary-valued map $x\mapsto J(x)=J_0$ takes a constant value and consequently the 
Grassmann-Berry connection turns out to be invariant.

\subsection{Time-reversal equivariant Holonomy}\label{sect:equiv_holonm}

The aim of this subsection is to summarize a few basic properties of the parallel
transport with respect to a time-reversal equivariant connections.
We refer to Appendix \ref{sect:remind_connect} for a review of the most important notions concerning parallel transport and holonomy. 

\medskip

Let $\pi:\bb{P}\to X$ a principal $\n{U}(m)$-bundle endowed with the connection $\omega\in\Omega^1(\bb{P},\rr{u}(m))$
and for each (piecewise smooth) curve $\tilde{\gamma}:[0,1]\to X$
let ${\rm PT}_{\tilde{\gamma}}:\bb{P}_{\tilde{\gamma}(0)}\to \bb{P}_{\tilde{\gamma}(1)}$ be the parallel transport
along $\tilde{\gamma}$ with respect to $\omega$. We recall the notation $\bb{P}_{\tilde{\gamma}(t)}:=\pi^{-1}(\tilde{\gamma}(t))$ for the fibers of $\bb{P}$ along the path $\tilde{\gamma}$.
We recall also the transformation property of the parallel transport
 ${\rm PT}_{\tilde{\gamma}}\circ R_u=R_u\circ {\rm PT}_{\tilde{\gamma}}$ which is valid for every $u\in\n{U}(m)$. Moreover, if $\tilde{\gamma}_1\cdot \tilde{\gamma}_2$ is the \emph{concatenation} of paths  and $\tilde{\gamma}^{-1}$  is the \emph{reversed} path one has ${\rm PT}_{\tilde{\gamma}_1\cdot \tilde{\gamma}_2}={\rm PT}_{\tilde{\gamma}_2}\circ {\rm PT}_{\tilde{\gamma}_1}$ and ${\rm PT}_{\tilde{\gamma}^{-1}}={\rm PT}_{\tilde{\gamma}}^{-1}$.

\medskip

Now,  we assume that $\bb{P}$ has a time-reversal structure $\hat{\Theta}$ and $\omega$ is a  time-reversal equivariant connection. For a given path $\tilde{\gamma}:[0,1]\to X$ on the involutive space $(X,\tau)$ let $\tau(\tilde{\gamma}):[0,1]\to X$ be the new path defined pointwise by $\tau(\tilde{\gamma})(t):=\tau(\tilde{\gamma}(t))$.

\begin{proposition}\label{prop:par_trans}
Let $(\bb{P},\hat{\Theta})$ be a principal bundle with time-reversal structure  on the involutive manifold $(X,\tau)$.
The parallel transport associated with an equivariant connection $\omega$ on $\bb{P}$ verifies
$$
{\rm PT}_{\tau(\tilde{\gamma})}\;\circ\;\hat{\Theta}\;=\;\hat{\Theta}\;\circ\;{\rm PT}_{\tilde{\gamma}}
$$
for each path $\tilde{\gamma}:[0,1]\to X$.
\end{proposition}
\proof
Let $p_0\in {\bb{P}}_{\tilde{\gamma}(0)}$ be a fixed point
and $\gamma:[0,1]\to\bb{P}$ the unique horizontal lift (with respect to $\omega$) of $\tilde{\gamma}$ such that $\gamma(0)=p_0$ (\cf Theorem \ref{theo:par_transp}). By exploiting the equivariance of $\omega$ one shows that $\gamma_\Theta:=\hat{\Theta}\circ\gamma$ is the unique horizontal lift of $\tau(\tilde{\gamma})$ such that $\gamma_\Theta(0)=\hat{\Theta}(p_0)$. In other words
one has ${\rm PT}_{\tau(\tilde{\gamma})}(\hat{\Theta}(p_0))=\gamma_\Theta(1)=\hat{\Theta}(\gamma(1))=\hat{\Theta}\big({\rm PT}_{\tilde{\gamma}}(p_0)\big)$. 
The result follows by observing that the last relation holds for all $p_0\in {\bb{P}}_{\tilde{\gamma}(0)}$.
\qed

\medskip

Now, let $\tilde{\gamma}:[0,1]\to X$ be a loop, \ie $\tilde{\gamma}(0)=x_0=\tilde{\gamma}(1)$. Clearly the parallel transport defines a map 
${\rm PT}_{\tilde{\gamma}}:\bb{P}_{x_0}\to \bb{P}_{x_0}$ of the fiber at $x_0$ into itself. Therefore, since the structure group $\n{U}(m)$
acts transitively on the fibers, we can define a map $\rr{hol}_{\tilde{\gamma}}:\bb{P}_{x_0}\to\n{U}(m)$ such that $\rr{hol}_{\tilde{\gamma}}(p)=u$ is the element in the group which gives ${\rm PT}_{\tilde{\gamma}}(p)=R_u(p)=p\cdot u$. We refer to $\rr{hol}_{\tilde{\gamma}}$ as the \emph{holonomy map} associated to the loop $\tilde{\gamma}$. It follows that 
$$
\rr{hol}_{\tilde{\gamma}}\;\circ R_u\;=\;u^{-1}\cdot\rr{hol}_{\tilde{\gamma}}\cdot u\;,\qquad\quad \forall\ u\in\n{U}(m)\;.
$$
In the case of a principal bundle with a  time-reversal structure we have the following immediate consequence of Proposition \ref{prop:par_trans}:
\begin{corollary}\label{cor:par_trans}
Let $(\bb{P},\hat{\Theta})$ be a principal bundle with time-reversal structure on the involutive manifold $(X,\tau)$. Assume that $\bb{P}$ in endowed with an equivariant connection and consider the related  parallel transport and holonomy maps.
For each loop $\tilde{\gamma}:[0,1]\to X$ such that $\tilde{\gamma}(0)=x_0=\tilde{\gamma}(1)$ and each point $p\in \bb{P}_{x_0}$
it holds that:
$$
\begin{aligned}
\rr{hol}_{\tau(\tilde{\gamma})}\big(\hat{\Theta}(p)\big)&\;=\;\overline{\rr{hol}_{\tilde{\gamma}}(p)} &&\qquad\text{(\virg{Real} case)}\\
\rr{hol}_{\tau(\tilde{\gamma})}\big(\hat{\Theta}(p)\big)&\;=-Q\cdot \overline{\rr{hol}_{\tilde{\gamma}}(p)}\cdot Q\;&&\qquad\text{(\virg{Quaternionic} case)}\;.
\end{aligned}
$$
\end{corollary}
\proof
By definition ${\rm PT}_{\tau(\tilde{\gamma})}(\hat{\Theta}(p))=\hat{\Theta}(p)\cdot \rr{hol}_{\tau(\tilde{\gamma})}\big(\hat{\Theta}(p)\big)$. As a consequence of  Proposition \ref{prop:par_trans} one has that 
$$
{\rm PT}_{\tau(\tilde{\gamma})}(\hat{\Theta}(p))\;=\;\hat{\Theta}\big({\rm PT}_{\tilde{\gamma}}(p)\big)\;=\;
\hat{\Theta}\big(p\cdot \rr{hol}_{\tilde{\gamma}}(p)\big)\;.
$$
The result follows from the property ($\hat{\rr{R}}$) for the \virg{Real} case or from the property ($\hat{\rr{Q}}$) for the \virg{Quaternionic} case as given in Definition \ref{def_princ_R&QB}.
\qed

\medskip

\noindent
In the particular case that: (i) $\tilde{\gamma}:[0,1]\to X^\tau$ in a loop in the fixed point set  of the
involutive space $(X,\tau)$; (ii) $p\in\bb{P}_{x_0}$ is a fixed point of $\hat{\Theta}$, namely $\hat{\Theta}(p)=p$ one deduces from Corollary   \ref{cor:par_trans} that
$$
\begin{aligned}
\rr{hol}_{\tilde{\gamma}}(p)&\;\in\;\n{O}(m)\;\subset\;\n{U}(m) &&\qquad\text{(\virg{Real} case)}\\
\rr{hol}_{\tilde{\gamma}}(p)&\;\in\;\n{U}(2m)^\sigma\;\subset\;\n{U}(2m)\;&&\qquad\text{(\virg{Quaternionic} case)}\;
\end{aligned}
$$
where $\n{O}(m)$ is the orthogonal group and  $\n{U}(2m)^\sigma$ is the invariant subgroup with respect to the action $u\mapsto\sigma(u)=-Q\cdot\overline{u}\cdot Q$. We recall that 
$\n{U}(2m)^\sigma\subset S\n{U}(2m)$ 
and $\n{U}(2m)^\sigma\simeq\n{S}p(m)$ 
(\cf \cite[Remark 2.1]{denittis-gomi-14-gen}).

\subsection{Curvature of a time-reversal equivariant connections}\label{sect:equiv_curv}
This section is devoted to the study of the transformation properties of the curvature $F_\omega$ associated to a 
time-reversal equivariant connection $\omega$ by the
structural equation 
$$
F_\omega\;:=\;{\rm d}\omega\;+\;\frac{1}{2}\; [\omega\wedge \omega]\;.
$$
\begin{proposition}[Equivariant curvature]\label{prop:equiv_curv}
Let $(X,\tau)$ be an involutive space that verifies conditions (H.1) - (H.3) and $\pi:\bb{P}\to X$ a smooth principal $\n{U}(m)$-bundle over $X$
endowed with a \virg{Real} (resp. \virg{Quaternionic}) structure $\hat{\Theta}:\bb{P}\to \bb{P}$. Let $\omega\in\rr{A}_{\rr{R}}(\bb{P})$ (resp. $\omega\in\rr{A}_{\rr{Q}}(\bb{P})$) be an equivariant connection 1-form and $F_\omega \in\Omega(\bb{P},\rr{u}(m))$ the related curvature. Then
\begin{equation}
\begin{aligned}
\hat{\Theta}^*\overline{F_\omega}\;&=\;{F_\omega}&&\qquad\text{(\virg{Real} case)}\\
\hat{\Theta}^*\sigma(F_\omega)\;&=\;F_\omega&&\qquad\text{(\virg{Quaternionic} case)}\;.\\
\end{aligned}
\end{equation}
\end{proposition}
\proof
Clearly $\sigma({\rm d}\omega)={\rm d}(\sigma(\omega))$ and $\sigma([\omega\wedge \omega])=[\sigma(\omega)\wedge\sigma(\omega)]$. This shows that $\sigma(F_\omega)=F_{\sigma(\omega)}$. Similarly one has $F_\omega=F_{\overline{\omega}}$. Since the pullback is compatible with the exterior derivative and the  wedge product, one has $\hat{\Theta}^*({\rm d}\omega)={\rm d}(\hat{\Theta}^*\omega)$ and $\hat{\Theta}^*[\omega\wedge \omega]= [\hat{\Theta}^*\omega\wedge \hat{\Theta}^*\omega]$
which implies $\hat{\Theta}^*\sigma(F_\omega)=F_{\hat{\Theta}^*\sigma(\omega)}$ and similarly $\hat{\Theta}^*\overline{F_\omega}=F_{\hat{\Theta}^*\overline{\omega}}$. The equivariance of $\omega$ concludes the proof.
\qed

\medskip

\noindent 
Let $\{\s{F}_\alpha\in\Omega^2(\f{U}_\alpha,\rr{g})\}$ be the collection of local 2-forms which describe the curvature $F_\omega$
in the sense of Theorem \ref{theo:loc_curvatur}. When $\omega$ is equivariant the local forms $\{\s{A}_\alpha\in\Omega^1(\f{U}_\alpha,\rr{g})\}$ verify the conditions \eqref{eq:set_equi_form_base}. In this case the same argument as in the proof of Proposition \ref{prop:equiv_curv} provides
\begin{equation}
\begin{aligned}
{\tau}^*\overline{\s{F}_\alpha}\;&=\;{\s{F}_\alpha}&&\qquad\text{(\virg{Real} case)}\\
{\tau}^*\sigma(\s{F}_\alpha)\;&=\;\s{F}_\alpha&&\qquad\text{(\virg{Quaternionic} case)}\;.\\
\end{aligned}
\end{equation}

\medskip

\begin{remark}[Equivariance of the  Grassmann-Berry curvature]{\upshape
The notion of equivariant curvature can be formulated also at level of vector bundles. Let $\nabla:\Gamma(\bb{E})\to \Omega^{1}(X,\bb{E})$ be a vector bundle connection  and $F^\nabla:=\nabla\circ\nabla:\Gamma(\bb{E})\to \Omega^{2}(X,\bb{E})$
the associated curvature (\cf equation \eqref{eq:VB_connAX}). The equivariance of $\nabla$, which in  accordance with Definition \ref{def:R&Q_connec_VB} is
expressed by  $\nabla\circ\tau_\Theta=\tau_\Theta\circ\nabla$, implies immediately a similar relation for the curvature $F^\nabla\circ\tau_\Theta=\tau_\Theta\circ F^\nabla$. Thus, equivariant vector bundle connections induce equivariant vector bundle curvatures. 

This conclusion applies also to the \emph{Grassmann-Berry curvature}. This is the curvature  associated to the Grassmann-Berry connection $\nabla_{\rm GB}=P\circ {\rm d}$
  described in Section \ref{sect:bott-chern-berry}. 
The curvature of 
$\nabla_{\rm GB}$ is the element $F_{\rm GB}\in\Omega^2(X,{\rm End}(\bb{E}))$ defined by the formula \eqref{eq:VB_connAX}, \ie
\begin{equation}\label{eq:BG_curv}
F_{\rm GB}\;=\; \nabla_{\rm GB}\circ \nabla_{\rm GB}\;=\;P\;\big({\rm d}P\wedge {\rm d}P\big)\;P\;=\;P\;{\rm d}P\wedge {\rm d}P
\end{equation}
where the last equality follows from ${\rm d}P\;P=(\n{1}-P){\rm d}P$ (for details about the derivation of \eqref{eq:BG_curv} see \cite[Section 12.5]{taubes-11}). As for the connection $\nabla_{\rm GB}$, the equivariance of the curvature  $F_{\rm GB}$ is guaranteed by the condition in Proposition \ref{prop:equivGB_conn}.
}\hfill $\blacktriangleleft$
\end{remark}

\section{Differential geometric classification of \virg{Real} vector bundles}\label{sect:R-bundles}

\subsection{A short reminder of the equivariant Borel cohomology}
\label{subsec:borel_cohom}

The proper cohomology theory for the analysis of vector bundle theories in the category of spaces with involution is the {equivariant cohomolgy} introduced by  A.~Borel in \cite{borel-60}. This cohomology provides the right tool for the topological classification of \virg{Real} vector bundles \cite{denittis-gomi-14}. A short   self-consistent summary of this cohomology theory can be found in \cite[Section 5.1]{denittis-gomi-14}.
For a more complete  introduction to the subject  we refer to
\cite[Chapter 3]{hsiang-75} and \cite[Chapter 1]{allday-puppe-93}.

\medskip

Since we need this tool we briefly recall the main steps of the
\emph{Borel construction} for the equivariant cohomology. 
The \emph{homotopy quotient} of an involutive space   $(X,\tau)$ is the orbit space
\begin{equation}\label{eq:homot_quot}
{X}_{\sim\tau}\;:=\;X\times\hat{\n{S}}^\infty /( \tau\times \vartheta)\;.
\end{equation}
Here $\vartheta$ is the \emph{antipodal map} on the infinite sphere $\n{S}^\infty$ 
(\cf \cite[Example 4.1]{denittis-gomi-14}) and $\hat{\n{S}}^\infty$ is used as short notation for the pair $(\n{S}^\infty,\vartheta)$.
The product space $X\times{\n{S}}^\infty$ (forgetting for a moment the $\Z_2$-action) has the \emph{same} homotopy type of $X$ 
since $\n{S}^\infty$ is contractible. Moreover, since $\vartheta$ is a free involution,  also the composed involution $\tau\times\vartheta$ is free, independently of $\tau$.
Let $\s{R}$ be any commutative ring (\eg, $\R,\Z,\Z_2,\ldots$). The \emph{equivariant cohomology} ring 
of $(X,\tau)$
with coefficients
in $\s{R}$ is defined by
$$
H^\bullet_{\Z_2}(X,\s{R})\;:=\; H^\bullet({X}_{\sim\tau},\s{R})\;.
$$
More precisely, each equivariant cohomology group $H^j_{\Z_2}(X,\s{R})$ is given by the
 singular cohomology group  $H^j({X}_{\sim\tau},\s{R})$ of the  homotopy quotient ${X}_{\sim\tau}$ with coefficients in $\s{R}$ and the ring structure is given, as usual, by the {cup product}.
As the coefficients of
the usual singular cohomology are generalized to \emph{local coefficients} (see \eg \cite[Section 3.H]{hatcher-02} or
\cite[Section 5]{davis-kirk-01}), the coefficients of the Borel  equivariant cohomology are also
generalized to local coefficients. Given an involutive space $(X,\tau)$ let us consider the homotopy group $\pi_1({X}_{\sim\tau})$
and the associated  \emph{group ring} $\Z[\pi_1({X}_{\sim\tau})]$. Each module $\s{Z}$ over the group $\Z[\pi_1({X}_{\sim\tau})]$ is, by definition,
a \emph{local system} on $X_{\sim\tau}$.  Using this local system one defines, as usual, the equivariant cohomology with local coefficients in $\s{Z}$:
$$
H^\bullet_{\Z_2}(X,\s{Z})\;:=\; H^\bullet({X}_{\sim\tau},\s{Z})\;.
$$
We are particularly interested in modules $\s{Z}$ whose underlying groups are identifiable with $\Z$ or $\R$. 
For each involutive space  $(X,\tau)$, there always exists a particular family of local systems $\Z(m)$ and $\R(m)$
labelled by $m\in\Z$. Here
 $\Z(m)\simeq X\times\Z$ and  $\R(m)\simeq X\times\R$ denote the $\Z_2$-equivariant local system on $(X,\tau)$  made equivariant  by the $\Z_2$-action $(x,l)\mapsto(\tau(x),(-1)^ml)$.
Because the module structure depends only on the parity of $m$, we consider only the $\Z_2$-modules ${\Z}(0)$ and ${\Z}(1)$ and similarly ${\R}(0)$ and ${\R}(1)$. Since ${\Z}(0)$ and $\R(0)$ correspond to the case of the trivial action of $\pi_1(X_{\sim\tau})$ on $\Z$ one has $H^k_{\Z_2}(X,\Z(0))\simeq H^k_{\Z_2}(X,\Z)$ and $H^k_{\Z_2}(X,\R(0))\simeq H^k_{\Z_2}(X,\R)$ \cite[Section 5.2]{davis-kirk-01}.

\medskip

We recall the two important group isomorphisms 
\begin{equation}\label{eq:iso:eq_cohom}
H^1_{\Z_2}(X,\Z(1))\;\simeq\;[X,\n{U}(1)]_{\Z_2}\;,\qquad\qquad H^2_{\Z_2}(X,\Z(1))\;\simeq\;{\rm Vec}_{\rr{R}}^1(X,\tau)\;
\end{equation}
 involving the 
first two equivariant cohomology groups. 
The first isomorphism \cite[Proposition A.2]{gomi-13} says that the first equivariant cohomology group is isomorphic to the set of $\Z_2$-homotopy classes of equivariant maps $\varphi:X\to\n{U}(1)$ where the involution on $\n{U}(1)$ is induced by the complex conjugation, \ie $\varphi(\tau(x))=\overline{\varphi(x)}$. The second isomorphism is due to B.~Kahn \cite{kahn-59} and 
expresses the equivalence between the Picard group of \virg{Real} line bundles (in the sense of \cite{atiyah-66,denittis-gomi-14}) over  $(X,\tau)$ and the second equivariant cohomology group of this space.
A more modern proof of this result can be found in \cite[Corollary A.5]{gomi-13}.

\medskip

The equivariant cohomology with coefficients in $\R(1)$
admits a nice interpretation. On an involutive space $(X,\tau)$
 the singular cohomology $H^k(X,\R)$ has a natural splitting given by two subgroups
$$
\begin{aligned}
H^k_+(X,\R)\;&:=\;\big\{a\in H^k(X,\R)\ |\ \tau^*(a)=+a\big\}\\
H^k_-(X,\R)\;&:=\;\big\{a\in H^k(X,\R)\ |\ \tau^*(a)=-a\big\}\;.
\end{aligned}
$$ 
\begin{proposition}\label{prop:Hpm}
For all $k\in\N\cup\{0\}$ the following isomorphisms hold:
$$
H^k_{\Z_2}(X,\R(0))\;\simeq\; H^k_+(X,\R)\;,\qquad\quad H^k_{\Z_2}(X,\R(1))\;\simeq\; H^k_-(X,\R)\;.
$$
These isomorphisms extend to all other groups $H^k_{\Z_2}(X,\R(m))$ by means of the   the recurrence relation  $H^k_{\Z_2}(X,\R(m))=H^k_{\Z_2}(X,\R(m+2))$.
\end{proposition}
\proof
The isomorphism
\eqref{eq:iso_imp} and Lemma \ref{lemma:vanish} imply
\begin{equation*}
H^{q}_{\Z_2}(X ,{\R}(m))\;\simeq\;H^0_{\rm group}(\Z_2,H^q(X,{\R}(m)))\;\simeq\; \big\{a\in H^q(X,{\R})\ |\ \tau^*(a)=(-1)^m a\big\}\;.
\end{equation*}
\qed

\medskip

\noindent
This result has an immediate consequence. Let $(\Omega^\bullet,{\rm d})$ be the complex of the differential forms on $X$. The associated cohomology $H_{\rm d.R}^\bullet(X):=H^\bullet(\Omega^\bullet(X),{\rm d})$ is called \emph{de Rham cohomology} and, under very broad conditions for the base space $X$ (certainly covered by (H.1) - (H.3)), 
the
celebrated \emph{de Rham Theorem}  \cite[Chapter II, Theorem 8.9]{bott-tu-82} asserts that
$$
H^{\bullet}_{\rm d.R.}(X)\;\stackrel{}{\simeq}\; H^\bullet(X,\n{R})\;.
$$
Let us consider the two subcomplex $(\Omega_\pm^\bullet(X),{\rm d})$ given by
$$
\Omega_\pm^k(X)\;:=\;\Big\{\omega\in\Omega^k(X)\;\Big|\; \tau^*\omega=\pm\omega \Big\}
$$
and the associated cohomologies
$$
H_{{\rm d.R},\pm}^\bullet(X):=H^\bullet(\Omega^\bullet_\pm(X),{\rm d})
$$
Then the de Rham isomorphism implies that 
\begin{equation}\label{eq:dR_iso_pm}
H_{{\rm d.R},\pm}^k(X)\;\stackrel{}{\simeq}\; H_{\pm}^k(X,\n{R})
\end{equation}
and as a consequence of Proposition \ref{prop:Hpm} we have that:
\begin{corollary}[Equivariant de Rham isomorphism]\label{cor:equiv_dR_sio}
Let $(X,\tau)$ a smooth manifold endowed with a smooth involution. Then
$$
H_{{\rm d.R},+}^\bullet(X)\;\stackrel{}{\simeq}\; H_{\Z_2}^k(X,\n{R}(0))\;,\qquad\quad H_{{\rm d.R},-}^\bullet(X)\;\stackrel{}{\simeq}\; H_{\Z_2}^k(X,\n{R}(1))\;.
$$
\end{corollary}

\medskip
\medskip

The next result allows us to compute $H^k_{\Z_2}(X,\R(m))$ from the knowledge of $H^k_{\Z_2}(X,\Z(m))$ .
\begin{proposition}\label{prop:cohom_real_int}
For all $k\in\N\cup\{0\}$ and $m\in\Z$ the following isomorphisms hold:
\begin{equation}\label{eq:cohom_coef1}
H^k_{\Z_2}(X,\R(m))\;\simeq\; H^k_{\Z_2}(X,\Z(m))\;\otimes_\Z\;\R\;
\end{equation}
where $\otimes_\Z$ denotes the tensor product of abelian groups.
\end{proposition}
\proof The isomorphisms $H^k_{\Z_2}(X,\s{R}(m))\simeq H^k_{\Z_2}(X,\s{R}(m+2))$, with $\s{R}=\Z,\R$ allow us to 
consider only the cases $m=0$ and $m=1$.

Let we start with the case $m=0$. In this case the claim reduces to the proof of the isomorphism
\begin{equation}\label{eq:cohom_coef2}
H^k(Y,\R)\;\simeq\; H^k(Y,\Z)\;\otimes_\Z\;\R\;
\end{equation}
for a given space $Y$. Indeed, since  the Borel  equivariant cohomology (with coefficients) is the singular cohomology of the homotopy quotient $X_{\sim\tau}$ (with coefficients) the proof of \eqref{eq:cohom_coef1} when $m=0$ follows from \eqref{eq:cohom_coef2} when $Y=X_{\sim\tau}$. 
The isomorphism \eqref{eq:cohom_coef2} follows from an application of  the \emph{universal coefficient theorem} \cite[Chapter 3,Theorem 3.2]{hatcher-02} 
which provides 
$$
H^k(Y,\R)\;\simeq\;{\rm Hom}_\Z\big(H_k(Y),\R\big)
$$
since  $n\R=\R$ for each $n\in\Z$ (\ie $\R$ is a divisible  abelian group) and so the ${\rm Ext}$ group vanishes.  
The splitting $H_k(Y)=F_k(Y)\oplus T_k(Y)$ in \emph{free} part $F_k(Y)\simeq\Z^{\beta(k)}$ (the non-negative integer $\beta(k)$ is called Betti number) and \emph{torsion} part
$T_k(Y)=\bigoplus_j\Z_{p_j}$ implies that
\begin{equation}\label{eq:cohom_coef30}
H^k(Y,\R)\;\simeq\;{\rm Hom}_\Z\big(F_k(Y),\R\big)\;\oplus\; {\rm Hom}_\Z\big(T_k(Y),\R\big)\simeq \R^{\beta(k)}
\end{equation}
where the natural isomorphisms ${\rm Hom}_\Z\big(\Z^n,\R\big)\simeq\R^n$ and ${\rm Hom}_\Z\big(\Z_p,\R\big)\simeq0$ have been used. Now, as a consequence of the isomorphism $H^k(Y,\Z)\simeq F_k(Y)\oplus T_{k-1}(Y)$ \cite[Chapter 3,Corollary 3.3]{hatcher-02} one has that 
$$
H^k(Y,\Z)\;\otimes_\Z\;\R\;\simeq\;(F_k(Y)\otimes_\Z\;\R)\;\oplus\;
(T_{k-1}(Y)\otimes_\Z\;\R)\;\simeq\;\R^{\beta(k)}
$$
and a comparison with \eqref{eq:cohom_coef30} immediately proves
\eqref{eq:cohom_coef2}.

Since the universal coefficient theorem holds also for the relative cohomology of a pair of spaces $I\subseteq Y$ we obtain with the same proof the generalization of  \eqref{eq:cohom_coef2}
\begin{equation}\label{eq:cohom_coef2_gen}
H^k(Y|I,\R)\;\simeq\; H^k(Y|I,\Z)\;\otimes_\Z\;\R\;
\end{equation}
where $H^k(Y|I,\s{R})$ is the $k$-th group of the relative cohomology of the pair $I\subseteq Y$ with coefficients in $\s{R}$.

The proof of the case $m=1$ can be reduced to the case $m=0$ 
by applying the \emph{Thom isomorphism} in the form of \cite[Proposition 2.5]{gomi-13}. Let $\tilde{I}=([-1,1], \imath)$ 
be the involutive space given by the interval $[-1,1]$ and the involution $\imath:t\to-t$. One can think of $\tilde{I}$ as the unit disk bundle of the $\Z_2$-equivariant line bundle $\pi:X\times\R\to X$ with involution $(x,t)\mapsto(\tau(x),\imath(t))$. The 
Thom isomorphism in cohomology applied to this line bundle gives
$$
H^k_{\Z_2}\big(X, \s{R}(1)\big)\;\simeq\;H^{k+1}_{\Z_2}\big((X\times\tilde{I})|(X\times \partial \tilde{I}), \s{R}(0)\big)\;,\qquad\quad\s{R}=\Z,\R\;.
$$
Therefore, one has
$$
\begin{aligned}
H^k_{\Z_2}\big(X, {\R}(1)\big)\;&\simeq\;H^{k+1}_{\Z_2}\big((X\times\tilde{I})|(X\times \partial \tilde{I}), {\R}(0)\big)&\qquad&\quad \text{(Thom isomorphism)}\\
&\simeq\;H^{k+1}_{\Z_2}\big((X\times\tilde{I})|(X\times \partial \tilde{I}), {\Z}(0)\big)\;\otimes_\Z\;\R&\qquad&\quad \text{(case}\ m=0\ \text{ and \eqref{eq:cohom_coef2_gen})}\\
&\simeq\;H^k_{\Z_2}\big(X, {\Z}(1)\big)\;\otimes_\Z\;\R&\qquad&\quad \text{(Thom isomorphism)}
\end{aligned}
$$
as expected. 
\qed

\medskip
\medskip

\begin{remark}\label{rk_torsion1}{\upshape
Proposition \ref{prop:cohom_real_int} has an  important 
consequence:  in the passage from $H^k_{\Z_2}(X,\Z(m))$ to $H^k_{\Z_2}(X,\R(m))$ all the information contained in the torsion of $H^k_{\Z_2}(X,\Z(m))$ is lost. This fact is not trivial since the groups $H^k_{\Z_2}(X,\Z(m))$ for spaces with fixed points, like  $\tilde{\n{S}}^d\equiv({\n{S}}^d,\tau)$ or $\tilde{\n{T}}^d\equiv({\n{T}}^d,\tau)$ (see Section \ref{sect:intro}), automatically have a non-trivial torsion.
}\hfill $\blacktriangleleft$
\end{remark}

\subsection{\virg{Real} Chern classes: topological point of view}
\label{sect:eq_chern_class}
The theory of characteristic classes for $\rr{R}$-vector bundles is formulated in \cite{kahn-59,krasnov-92,pitsch-scherer-13}.
According to \cite[Theorem 4.13]{denittis-gomi-14}, the involutive space $(G_m(\C^\infty),\varrho)$ 
classifies  ${\rm Vec}^m_\rr{R}(X,\tau)$. Here $\rho$ acts as the complex conjugation in each $m$-plane contained in $G_m(\C^\infty)$ (see \cite[Example 4.4]{denittis-gomi-14} for more details). To each  involutive space $(X,\tau)$ we can associate the graded cohomology ring
\beql{eq:ring_eq_cohom1}
\s{A}(X,\tau)\;:=\;H^\bullet_{\Z_2}(X,\Z(0))\;\oplus\; H^\bullet_{\Z_2}(X,\Z(1))
\eeq
where the ring structure is given by the cup product. Equation \eqref{eq:ring_eq_cohom1}, restricted to a single point $\{\ast\}$ provides a ring
\beql{eq:ring_eq_cohom1'}
\n{A}:=\s{A}(\{\ast\})\;=\;H^\bullet(\R P^\infty,\Z)\;\oplus\; H^\bullet(\R P^\infty,\Z(1))\;\simeq\; \Z[\alpha]/(2\alpha)
\eeq
where   $\alpha$ corresponds to the additive generator of $H^1(\R P^\infty,\Z(1))=\Z_2$ and it is subject to the only relation $2\alpha=0$  \cite[Proposition 2.4]{gomi-13}. Moreover, $\alpha^2$ agrees with the generator $\kappa\in H^2(\R P^\infty,\Z)$ and this justifies the more 	
suggestive
 notation $\alpha=\sqrt{\kappa}$ already used in \cite{gomi-13}.  
  The ring $\n{A}$ is crucial in the description of 
$\s{A}(G_m(\C^\infty),\varrho)$, in fact one has \cite[Th\'{e}or\`{e}me 3]{kahn-59}
\begin{equation}\label{eq:equi_univ_chern_class}
\s{A}(G_m(\C^\infty),\varrho)\;=\; \n{A}\left[{\rr{c}}^{\rr{R}}_1,\ldots,{\rr{c}}^{\rr{R}}_m\right]\;=\; \n{Z}\left[\alpha,{\rr{c}}^{\rr{R}}_1,\ldots,{\rr{c}}^{\rr{R}}_m\right]/(2\alpha)
\end{equation}
where ${\rr{c}}^{\rr{R}}_k\in H^{2k}_{\Z_2}(G_m(\C^\infty),\Z(j))$ is called $k$-th \emph{universal  \virg{Real} Chern class}.
The cohomology ring $\s{A}(G_m(\C^\infty),\varrho)$ is the polynomial algebra over the ring $\n{A}$ on $m$ generators ${\rr{c}}^{\rr{R}}_1,\ldots,{\rr{c}}^{\rr{R}}_m$ and a comparison with equation \eqref{eq:univ_chern_class} shows that 
equation \eqref{eq:equi_univ_chern_class} provides the equivariant generalization of the cohomology ring $H^\bullet\big(G_m(\C^\infty),\Z\big)$ generated by the usual Chern classes. In effect, the process of forgetting the \virg{Real} structure of the pair $(G_m(\C^\infty),\varrho)$ defines a canonical homomorphism $\jmath:H^{\bullet}_{\Z_2}(G_m(\C^\infty),\Z(j))\to H^{\bullet}(G_m(\C^\infty),\Z)$ such that $\jmath({\rr{c}}^{\rr{R}}_k)={\rr{c}}_k$ for all $k=1,\ldots,m$.

\medskip

The universal ring $\s{A}(G_m(\C^\infty),\varrho)$ and the homotopy classification \cite[Theorem 4.13]{denittis-gomi-14} allow us to define equivariant Chern classes for each element in ${\rm Vec}^m_\rr{R}(X,\tau)$. Let $(\bb{E},\Theta)$ be an $\rr{R}$-bundle over $(X,\tau)$
classified by the equivariant map $\varphi\in[X,G_m(\C^\infty)]_{\rm eq}$, then the 
\emph{$k$-th topological \virg{Real} Chern class} of $\bb{E}$ is by definition
$$
\tilde{c}^{\rr{R}}_k(\bb{E})\;:=\;\varphi^\ast({\rr{c}}^{\rr{R}}_k)\;\in\;H^{2k}_{\Z_2}(X,\Z(k))\qquad\quad k=1,2,3,\ldots\;
$$
where $\varphi^\ast:H^\bullet_{\Z_2}(G_m(\C^\infty),\Z(i))\to H^\bullet_{\Z_2}(X,\Z(j))$ is the homomorphism of cohomology groups induced by $\varphi$.
\virg{Real} Chern classes (or {$\rr{R}$-Chern classes} in short) verify all the {Hirzebruch axioms} of Chern classes (up to a change in the normalization)
and are uniquely specified by these axioms \cite[Th\'{e}or\`{e}me 2]{kahn-59} or \cite[Theorem 4.2]{pitsch-scherer-13}.
An important property  is that $\tilde{c}^{\rr{R}}_k(\bb{E})=0$ for all $k>m$ if $\bb{E}$ has rank $m$. Finally, under the map $\jmath:{\rm Vec}^m_\rr{R}(X,\tau)\to {\rm Vec}^m_\C(X)$ that forgets the \virg{Real} structure one has the identification $\jmath\tilde{c}^{\rr{R}}_k(\bb{E})=c_k(\jmath(\bb{E}))$. 

\medskip

The notion of topological \virg{Real} Chern classes extends automatically also to \virg{Real} principal bundles through the isomorphism
$$
{\rm Prin}^{\rr{R}}_{\n{U}(m)}(X,\tau)\;\simeq\ {^{\rm top}{\rm Vec}^m_{\rr{R}}}(X,\tau)
$$ 
in Proposition \ref{prop:equi_top_Prin_QRVB}. More precisely one sets
$$
\tilde{c}^{\rr{R}}_k(\bb{E})\;=\;\tilde{c}^{\rr{R}}_k(\bb{P})
$$
if $(\bb{P},\hat{\Theta})$ is the $\rr{R}$-principal bundle associated to the $\rr{R}$-bundle $(\bb{E},\Theta)$, and vice versa.

\subsection{\virg{Real} Chern-Weil theory}\label{sect:equiv_ChernÐWeil}
Let $(\bb{P},\hat{\Theta})$ be a \virg{Real} $\n{U}(m)$-bundle over the involutive space $(X,\tau)$. Given a connection 1-form $\omega\in\Omega^1(\bb{P},\rr{u}(m))$ we can associate to $\bb{P}$ \emph{differential} Chern classes $c_k(\bb{P})\in H^{2k}_{\rm d.R.}(X)$ by the prescription $c_k(\bb{P}):=cw(\bb{P},C_k)$
where  $C_k$
is the $\rm{Ad}$-invariant homogeneous polynomial of degree $k$ given by
$$
{\rm det}\left(t\n{1}\;-\;\frac{1}{\ii\; 2\pi}F_\omega\right)\;=\;\sum_{k=0}^mC_k(F_\omega\wedge\ldots\wedge F_\omega)\; t^{m-k}\;
$$
and $F_\omega\in \Omega^2(\bb{P},\rr{u}(m))$ is the curvature associated to $\omega$.
We refer to Appendix \ref{sect:ChernÐWeil} for more details.
In the case of an equivariant  connection $\omega$ in the sense of Definition \ref{def:R&Q_connec}, also the {differential} Chern classes inherit a proper symmetry under the involution $\tau$.
\begin{lemma}\label{lemma:chern_diff_equi}
Let $(\bb{P},\hat{\Theta})$ be a \virg{Real} $\n{U}(m)$-bundle over the involutive space $(X,\tau)$ and $\omega\in\rr{A}_{\rr{R}}(\bb{P})$ a \virg{Real} connection. Then
\begin{equation}\label{eq:chern_diff_equi}
c_k^{\rr{R}}(\bb{P})\;\in\;\left\{
\begin{aligned}
&H^{2k}_{{\rm d.R.},+}(X)&&\qquad\ \ \text{if}\ \  k\ \ \text{is even}\\
&H^{2k}_{{\rm d.R.},-}(X)&&\qquad\ \ \text{if}\ \  k\ \ \text{is odd}\;
\end{aligned}
\right.
\end{equation}
where $c_k^{\rr{R}}(\bb{P})$ denotes the differential $k$-th
 Chern class computed with respect to a \virg{Real} connection.
 \end{lemma}
\proof
The equivariance of the connection $\hat{\Theta}^*\overline{\omega}={\omega}$ implies the equivariance of the curvature $\hat{\Theta}^*\overline{F_\omega}={F_\omega}$ as proved in Proposition \ref{prop:equiv_curv}.
Since $F_\omega$ takes value in the Lie algebra $\rr{u}(m)$ of $m\times m$ anti-hermitian matrices one has 
$$
{\rm det}\left(t\n{1}\;-\;\frac{1}{\ii\; 2\pi}\overline{F_\omega}\right)\;=\;{\rm det}\left(t\n{1}\;+\;\frac{1}{\ii\; 2\pi}{F^t_\omega}\right)\;=\;(-1)^m\;{\rm det}\left(-t\n{1}\;-\;\frac{1}{\ii\; 2\pi}{F_\omega}\right)
$$
which implies $C_k(\overline{F_\omega}\wedge\ldots\wedge \overline{F_\omega})=(-1)^k\;C_k(F_\omega\wedge\ldots\wedge F_\omega)$. Then
$$
C_k(F_\omega\wedge\ldots\wedge F_\omega)\;=\;C_k(F_{\hat{\Theta}^*\overline{\omega}}\wedge\ldots\wedge F_{\hat{\Theta}^*\overline{\omega}})\;=\;\hat{\Theta}^*C_k(\overline{F_\omega}\wedge\ldots\wedge \overline{F_\omega})
$$
which provides
$$
\hat{\Theta}^*C_k(F_\omega\wedge\ldots\wedge F_\omega)\;=\;(-1)^k\; C_k(F_\omega\wedge\ldots\wedge F_\omega)\;.
$$
As explained in Appendix \ref{sect:ChernÐWeil}, $C_k(F_\omega\wedge\ldots\wedge F_\omega)\in\Omega^{2k}(\bb{P})$ is horizontal and invariant (\ie \emph{basic}) and it defines  uniquely a $2k$-form $\hat{C}_k(F_\omega\wedge\ldots\wedge F_\omega)\in\Omega^{2k}(X)$ such that $\pi^*\hat{C}_k(F_\omega\wedge\ldots\wedge F_\omega)=C_k(F_\omega\wedge\ldots\wedge F_\omega)$ where $\pi:\bb{P}\to X$ denotes the bundle projection. The equivariance  $\pi\circ \hat{\Theta}=\tau\circ\pi$ induces the equality 
$$
(-1)^k\Big(\pi^*\hat{C}_k(F_\omega\wedge\ldots\wedge F_\omega)\Big)\;=\;\hat{\Theta}^*\Big(\pi^*\hat{C}_k(F_\omega\wedge\ldots\wedge F_\omega)\Big)\;=\;\pi^*\Big(\tau^*\hat{C}_k(F_\omega\wedge\ldots\wedge F_\omega)\Big)
$$
which implies 
$$
\tau^*\hat{C}_k(F_\omega\wedge\ldots\wedge F_\omega)\;=\;(-1)^k\;\hat{C}_k(F_\omega\wedge\ldots\wedge F_\omega)
$$
since $\pi^*$ establishes an isomorphism between $\Omega^k(X)$ and the basic forms on $\Omega^k(\bb{P})$. This shows that $\hat{C}_k(F_\omega\wedge\ldots\wedge F_\omega)\in\Omega^{2k}_{+}(X)$ if $k$ is even and
$\hat{C}_k(F_\omega\wedge\ldots\wedge F_\omega)\in\Omega^{2k}_{-}(X)$ if $k$ is odd. The result follows observing that $c_k^{\rr{R}}(\bb{P})$ is defined as the image of $\hat{C}_k(F_\omega\wedge\ldots\wedge F_\omega)$ in the de Rham cohomology.
\qed

\medskip
\medskip

\begin{remark}{\upshape
Although in  this section we are mainly interested in the case of \virg{Real} $\n{U}(m)$-bundles we notice that the proof of Lemma \ref{lemma:chern_diff_equi} works with minimal modifications also for the \virg{Quaternionic} case. Indeed, the important point is that  the polynomials  $C_k$ are invariant under the transformation $F_\omega\mapsto Q\cdot F_\omega\cdot Q^{-1}$ since they are defined through a determinant. In conclusion equation \eqref{eq:chern_diff_equi} describe correctly also the case of \virg{Quaternionic} $\n{U}(2m)$-bundles
}\hfill $\blacktriangleleft$
\end{remark}

\medskip

\begin{corollary}
Due to the   equivariant de Rham isomorphism in Corollary \ref{cor:equiv_dR_sio} one has that
$$
c_k^{\rr{R}}(\bb{P})\;\in\;  H_{\Z_2}^{2k}(X,\n{R}(k))\;.
$$
\end{corollary}

\medskip

\noindent
Just for economy of notation, we decided to identify with the same symbol $c_k^{\rr{R}}(\bb{P})$ both the differential Chern class as element of the graded
de Rham cohomology $H^{2k}_{{\rm d.R.},\pm}(X)$ and its isomorphic image in the equivariant cohomology $H_{\Z_2}^{2k}(X,\n{R}(k))$, hoping that this will not create confusion to the reader.

\begin{theorem}[\virg{Real} Chern-Weil homomorphism]\label{theo:CWhom}
Let $(\bb{P},\hat{\Theta})$ be a \virg{Real} $\n{U}(m)$-bundle over the involutive space $(X,\tau)$  with a \virg{Real} connection  $\omega\in\rr{A}_{\rr{R}}(\bb{P})$ and related Chern classes $c_k^{\rr{R}}(\bb{P})$. Let 
$$
H^{k}_{\Z_2}(X,\Z(k))\;\stackrel{r_{\Z_2}}{\longrightarrow}\; H^{k}_{\Z_2}(X,\R(k))
$$
be the homomorphism induced by the  inclusion $\Z(k)\hookrightarrow\R(k)$. Then
\begin{equation}\label{eq:chern_class_equi_CW}
r_{\Z_2}\;\left(\tilde{c}^{\rr{R}}_k(\bb{P})\right)\;=\;c_k^{\rr{R}}(\bb{P})\;,
\end{equation}
namely the image under $r_{\Z_2}$ of the   \emph{topological} \virg{Real} Chern classes agree with the \emph{differential} Chern classes  computed with respect to a \virg{Real} connection.
\end{theorem}
\proof
Consider the commutative diagram
\beql{eq:diag00_CW}
\begin{diagram}
H^{2k}_{\Z_2}(X,\Z(k))                        &         \rTo^{\ \ \ \ \ \ \ \ r_{\Z_2}\ \ \ \ \ \ \ \ }        &H^{2k}_{\Z_2}(X,\R(k))&\simeq& H^{2k}_{\Z_2}(X,\Z(k))\;\otimes_\Z\;\R\\
       \dTo^{\jmath}                                                           &                &\dTo_{\jmath_\R} \\
H^{2k}(X,\Z)&\rTo_{r} & H^{2k}(X,\R) &\simeq&H^{2k}(X,\Z)\;\otimes_\Z\;\R\\
\end{diagram}
\eeq
where the vertical maps consist in forgetting the $\Z_2$-actions and the homomorphism $r$ is induced by the  inclusion $\Z\hookrightarrow\R$. Let us start with the \virg{universal} case $X=G_m(\C^\infty)$ endowed with the involution $\varrho$ given by the  
complex conjugation (see \cite[Example 4.4]{denittis-gomi-14} for more details).
There is no problem to think at $G_m(\C^\infty)$ as an infinite dimensional smooth manifold and the construction of the differential Chern classes applies as well (see \eg \cite{quillen-88}). We already know the equivariant cohomology of $G_m(\C^\infty)$ given by 
\eqref{eq:equi_univ_chern_class} and the ordinary cohomology of $G_m(\C^\infty)$ given by \eqref{eq:univ_chern_class}. Moreover, we have that $\jmath({\rr{c}}^{\rr{R}}_k)={\rr{c}}_k$. These data together with Proposition \ref{prop:cohom_real_int}
 allow us to conclude that the map $\jmath_\R$ is injective. Moreover, $r_{\Z_2}({\rr{c}}^{\rr{R}}_k)\neq 0$ since 
 the ${\rr{c}}^{\rr{R}}_k$'s are not subjected to any relation and 
 $$
r_{\Z_2}\left({\rr{c}}^{\rr{R}}_k\right)\;=\;\jmath^{-1}_\R\left(r\left(\jmath\left({\rr{c}}^{\rr{R}}_k\right)\right)\right)\;=\;\jmath^{-1}_\R\left(r\left({\rr{c}}_k\right)\right)=\jmath^{-1}_\R\left(\hat{{\rr{c}}}_k\right)\;.
$$
where $\hat{{\rr{c}}}_k:=r\left({\rr{c}}_k\right)\in H^{2k}(G_m(\C^\infty),\R)$ can be also identified with a de Rham class $\hat{{\rr{c}}}_k\in H^{2k}_{\rm d.R.}(G_m(\C^\infty))$ which provides the differential form of the Chern class. Since all the maps are functorial we arrive at equation \eqref{eq:chern_class_equi_CW} just by considering the pullback with respect to a $\Z_2$-equivariant map $\varphi:X\to G_m(\C^\infty)$ which classifies the \virg{Real} $\n{U}(m)$-bundle $\bb{P}$.
\qed

\medskip

\begin{remark}\label{rk_torsion2}{\upshape
In connection to Remark \ref{rk_torsion1} we notice that during the mapping \eqref{eq:chern_class_equi_CW} all the information encoded in the torsion part of $H^{2k}_{\Z_2}(X,\Z(k))$ are lost. This implies, in particular, that also in low dimension the differential  \virg{Real} Chern classes  fail to be good invariants for the classification of \virg{Real} $\n{U}(m)$-bundles. We will investigate these aspects in more details in the next section.
}\hfill $\blacktriangleleft$
\end{remark}

\subsection{Classification of \virg{Real} $\n{U}(1)$-principal bundles}\label{sect:non-triv}
Let $(X,\tau)$ be an involutive space which verifies the assumptions (H.1) - (H.3). The following classification result hols true:
\begin{equation}\label{eq:class_U1-Rbund}
{\rm Prin}^{\rr{R}}_{\n{U}(1)}(X,\tau)\;\simeq\;{{\rm Vec}^1_{\rr{R}}}(X,\tau)\;\stackrel{\tilde{c}^{\rr{R}}_1}{\simeq}\;H^{2}_{\Z_2}(X,\Z(1))
\end{equation}
The first isomorphism follows from Corollary \ref{cor:equi_categ} while the second isomorphism, given by the first (topological) \virg{Real} Chern class $\tilde{c}^{\rr{R}}_1$, has been proved in \cite{kahn-59} (see also \cite[Section 5.2]{denittis-gomi-14}).
The question that we want to address in this section is in which way we can apply the geometric differential apparatus developed in the previous sections in order to describe the classification of \virg{Real} $\n{U}(1)$-principal bundles or equivalently \virg{Real} line bundles. We start with some examples.

\medskip

\begin{example}\label{ex_1}{\upshape
Consider the circle $\n{S}^1$ endowed with the \emph{trivial} $\Z_2$-action (\ie $\tau(z)=z$ for all $z\in\n{S}^1$).
A simple calculation shows that
$$
H^{2}_{\Z_2}\left(\n{S}^1,\Z(1)\right)\;\simeq\;H^{2}\left(\n{S}^1\times\R P^\infty,\Z(1)\right)\;\simeq\;H^{1}\left(\n{S}^1,\Z(1)\right)\;\otimes_\Z\;\Z_2\;\simeq\;\Z_2
$$ 	
where we used the K\"{u}nneth formula for cohomology together with  $H^{0}(\R P^\infty,\Z(1)) =H^{2}(\R P^\infty,\Z(1)) =0$,  $H^{1}(\R P^\infty,\Z(1)) =\Z_2$ and $H^{1}(\n{S}^1,\Z(1))\simeq H^{1}(\n{S}^1,\Z)=\Z$ \cite[Lemma 2.15]{gomi-13}. This means that there are only two (up to isomorphisms)  \virg{Real} $\n{U}(1)$-principal bundles over $\n{S}^1$. We denote with $\bb{P}_0=\n{S}^1\times\n{U}(1)$ the \emph{trivial} $\rr{R}$-bundle with trivial \virg{Real} structure $\hat{\Theta}_0:(z,u)\mapsto(z,\overline{u})$ and with $\bb{P}_{\rm M}=\n{S}^1\times\n{U}(1)$ the \emph{M\"{o}bius} $\rr{R}$-bundle with \emph{non}-trivial \virg{Real} structure $\hat{\Theta}_{\rm M}:(z,u)\mapsto(z,z\overline{u})$ where we have identified $\n{S}^1\simeq\n{U}(1)\subset\C$. 
We know from Corollary \ref{corol:uniq_R-connection} that both $\bb{P}_{0}$ and $\bb{P}_{\rm M}$ admit a unique \virg{Real} connection. The construction of these connections follows as in Example \ref{ex:low_dim}.
For the trivial bundle $\bb{P}_{0}$ one has  the  \emph{trivial} (flat)  connection $\omega_0:=({\rm pr}_2)^*\theta$ 
where $\theta\in \Omega^1_{\rm left}(\n{U}(1),\rr{u}(1))$ is the  {Maurer-Cartan}  1-form   $\theta_u=u^{-1}\cdot{\rm d}u$ 
and 
 ${\rm pr}_2:X\times\n{U}(m)\to X$. In the case of  the M\"{o}bius bundle $\bb{P}_{\rm M}$ one has $\omega_{\rm M}:=\frac{1}{2}z{\dd}z^{-1}+\omega_0=-\frac{1}{2}z^{-1}{\dd}z+\omega_0$. The curvatures associated to this two connections vanish identically: $F_{\omega_0}=0$ is a consequence of the fact that $\omega_0$ is a flat connection (\cf Proposition \ref{prop:curv_PB} (ii));
$F_{\omega_{\rm M}}=0$ follows observing that $\dd\big(z^{-1}{\dd}z\big)={\dd}z^{-1}\wedge{\dd}z=0$ (just identify $z=\expo{\ii\vartheta}$ with $\vartheta\in\R/2\pi\Z$), $z^{-1}{\dd}z\wedge z^{-1}{\dd}z=0$ and $z^{-1}{\dd}z\wedge\omega_0+\omega_0\wedge z^{-1}{\dd}z=0$. This means that the differential \virg{Real} Chern classes $c_1^{\rr{R}}(\bb{P}_0)$ and
$c_1^{\rr{R}}(\bb{P}_{\rm M})$, built from  $F_{\omega_0}$ and $F_{\omega_{\rm M}}$ respectively, vanish. Since  $c_1^{\rr{R}}(\bb{P}_0)=c_1^{\rr{R}}(\bb{P}_{\rm M})=0$ as elements of $H^2_{\Z_2}(\n{S},\R(1))\simeq0$
 we can not use these objects to distinguish $\bb{P}_0$ from $\bb{P}_{\rm M}$. In effect, this is a consequence of the fact that $H^{2}_{\Z_2}(\n{S}^1,\Z(1))$ consists of torsion only as already discussed in Remark \ref{rk_torsion2}.
  However, we can still distinguish $\bb{P}_0$ from $\bb{P}_{\rm M}$ by
computing the holonomy. Let $\tilde{\gamma}:[0,1]\to\n{S}^1$ be the loop given by $\tilde{\gamma}(t):=\expo{\ii 2\pi t}$.
The horizontal lift with respect to $\omega_0$ and starting point $p_0=(1,u_0)$  is simply given by $\gamma(t):=(\tilde{\gamma}(t),u_0)$ which implies that $\rr{hol}_{\tilde{\gamma}}^{\omega_0}(p_0)=1\in\n{U}(1)$ independently of the choice of $p_0$.
In particular this also prove the triviality of the  holonomy group ${\rm Hol}_{p_0}(\bb{P}_0,\omega_0)=\{1\}$  (independently of $p_0$).
On the other side the horizontal lift of $\tilde{\gamma}$ with respect to $\omega_{\rm M}$ (and starting point $p_0$) has the form $\gamma(t):=(\tilde{\gamma}(t),g(t)u_0)$ with $t\mapsto g(t)\in\n{U}(1)$ such that $g(0)=1$ and $g(1)=\expo{-\frac{\ii}{2}\int_0^{2\pi}\dd\theta}=\expo{-\ii\pi}=-1$ (\cf equation \eqref{eq:holo1}). This proves that $\rr{hol}_{\tilde{\gamma}}^{\omega_{\rm M}}(p_0)=-1\in\n{U}(1)$ and ${\rm Hol}_{p_0}(\bb{P}_{\rm M},\omega_{\rm M})=\Z_2$ (independently of $p_0$).
As a final comment, let us recall \cite[Proposition 4.5]{denittis-gomi-14} which provides the identification between the set of equivalence classes of \virg{Real}  
$\n{U}(1)$-principal bundles over the trivial involutive space $\n{S}^1$ and $\n{O}(1)$-principal bundles
over $\n{S}^1$. The latter are classified by the (first) Stiefel-Whitney class $w_1\in H^1(\n{S}^1,\Z_2)\simeq\Z_2$ and under the map which forget the \virg{Real} structure one has the identification $\jmath:\tilde{c}_1^{\rr{R}}\to w_1$ between the topological \virg{Real} Chern class and the corresponding Stiefel-Whitney class \cite[Th\'{e}or\`{e}me]{kahn-59} or \cite[Theorem 3.2.1]{krasnov-92}. 
}\hfill $\blacktriangleleft$
\end{example}

\medskip

\begin{example}\label{ex_2}{\upshape
Let $\n{C}P^1$ be the complex \emph{projective line} endowed with the involution $\kappa:\n{C}P^1\to \n{C}P^1$ given on the homogeneous coordinates by $\kappa:[z:w]\mapsto[\overline{z}:\overline{w}]$. Under the usual identification $\n{C}P^1\simeq\n{S}^2\subset\R^3$
given by the  stereographic projection
$$
\n{S}^2\;\ni\;(x_0,x_1,x_2)\;\longmapsto\;\left[\frac{x_1+\ii x_2}{1-x_0}:1\right]\;=\;\left[1:\frac{x_1-\ii x_2}{1+x_0}\right]\;\in\;\n{C}P^1
$$
(the equality of the  equivalence classes is well defined for $x_0\neq\pm1 $) one obtains an involutive sphere $\hat{\n{S}}^2\equiv(\n{S}^2,\kappa)$ with involution given by $\kappa:(x_0,x_1,x_2)\mapsto(x_0,x_1,-x_2)$. The fixed-point set of this action in the equatorial circle
$$
\big(\hat{\n{S}}^2\big)^\kappa\;:=\;\left\{(x_0,x_1,0)\in\n{S}\right\}\;\simeq\;\n{S}^1\;.
$$
Let us denote with $\imath:\n{S}^1\hookrightarrow \big(\hat{\n{S}}^2\big)^\kappa\subset\hat{\n{S}}^2$ the inclusion of $\n{S}^1$ in the 
fixed-point set of $\hat{\n{S}}^2$. The sphere $\hat{\n{S}}^2$ is equivariantly isomorphic to the reduced suspension of the involutive circle $\tilde{\n{S}}^1\equiv(\n{S}^1,\tau)$ along the invariant direction $x_0$. This fact allows us to use the suspension
formula (for the reduced cohomology) in the following calculation: 
\begin{equation}\label{eq:cohomShat} 
H^{2}_{\Z_2}\left(\hat{\n{S}}^2,\Z(1)\right)\;\simeq\;\tilde{H}^{2}_{\Z_2}\left(\hat{\n{S}}^2,\Z(1)\right)\;\simeq\;\tilde{H}^{1}_{\Z_2}\left(\tilde{\n{S}}^1,\Z(1)\right)\;\simeq\;\Z\;.
\end{equation}
Here the first isomorphism follows from $H^{2}_{\Z_2}(\{\ast\},\Z(1))=0$ the second is due to the suspension isomorphism and the third is a consequence of ${H}^{1}_{\Z_2}(\tilde{\n{S}}^1,\Z(1))=\Z_2\oplus\Z$ and $H^{1}_{\Z_2}(\{\ast\},\Z(1))=\Z_2$ (\cf \cite{gomi-13}). We notice that the induced homomorphism $\imath^*:H^{2}_{\Z_2}(\hat{\n{S}}^2,\Z(1))\simeq\Z\to H^{2}_{\Z_2}({\n{S}}^1,\Z(1))\simeq\Z_2$ (see the computation in Example \ref{ex_1}) agree with the unique non-trivial homomorphism between $\Z$ and $\Z_2$.
The computation \eqref{eq:cohomShat} and the isomorphism \eqref{eq:class_U1-Rbund} imply that there are 
(up to isomorphisms) exactly $\Z$ \virg{Real} $\n{U}(1)$-principal bundles over $\hat{\n{S}}^2$. These bundles are completely classified by the 
topological \virg{Real} Chern classes $\tilde{c}^{\rr{R}}_1$. Moreover, as a consequence of Theorem \ref{theo:CWhom}, the absence of torsion in $H^{2}_{\Z_2} (\hat{\n{S}}^2,\Z(1) )$ also 
implies that these classes are  completely described by the differential \virg{Real} Chern classes. Let $\bb{P}^{(k)}\to \hat{\n{S}}^2$ be a representative of a  \virg{Real} $\n{U}(1)$-principal bundle over $\hat{\n{S}}^2$ endowed with a \virg{Real} connection $\omega^{(k)}$ and 
(differential) \virg{Real} Chern classes $c_1^{\rr{R}}\big(\bb{P}^{(k)}\big)\simeq k\in\Z$ with isomorphism given by the integration
$$
\int_{\n{S}^2}c_1^{\rr{R}}\big(\bb{P}^{(k)}\big)\;=\;k\;.
$$
The inclusion map $\imath$ provides the   restricted \virg{Real} $\n{U}(1)$-principal bundle $\bb{P}^{(k)} |_{\imath(\n{S}^1)}\to\n{S}^1$ which has been classified in Example \ref{ex_1} with the help of the holonomy. More precisely, given the loop  $\tilde{\gamma}:[0,1]\to\n{S}^1$ defined by $\tilde{\gamma}(t):=\imath\big(\expo{\ii2\pi t}\big)$ one has
$$
\rr{hol}_{\tilde{\gamma}}^{\omega^{(k)}}(p_0)\;=\;\expo{-\int_{\n{S}^1}\omega^{(k)}}\;=\;\expo{-\ii2\pi\int_{\n{S}^2}c_1^{\rr{R}}\big(\bb{P}^{(k)}\big)}\;=\;(-1)^{k}\;,\qquad\quad\text{(independently of $p_0$)}\;.
$$
The above calculation  is no more than an application of the Stokes' theorem to equation \eqref{eq:holo1}.
}\hfill $\blacktriangleleft$
\end{example}

\medskip

\begin{example}\label{ex_3}{\upshape
Let $\hat{\T}^2\equiv(\n{T}^2,\eta)$ be the two dimensional torus endowed with the involution $\eta:(z_1,z_2)\mapsto(z_1,\overline{z_2})$.
By using the standard notation used in this paper we can write $\hat{\T}^2=\n{S}^1\times\tilde{\n{S}}^1$ and the fixed-point set is given by
$\big(\hat{\T}^2\big)^\eta=\n{S}^1_+\sqcup \n{S}^1_-$ with $\n{S}^1_\pm:=\n{S}^1\times\{\pm 1\}$. We observe that $\hat{\T}^2$ can be identified with the total space of the \virg{Real} $\n{U}(1)$-principal bundle $\bb{P}_0\to\n{S}^1$ described in Example \ref{ex_1}. The Gysin exact sequence \cite[Corollary 2.11]{gomi-13} leads to 
\begin{equation}\label{eq:gysin_split}
H^{2}_{\Z_2}\left(\hat{\n{T}}^2,\Z(1)\right)\;\simeq\;H^{2}_{\Z_2}\left({\n{S}}^1,\Z(1)\right)\;\oplus\; H^{1}\left(\n{S}^1\times\R P^\infty,\Z\right)\;\simeq\;\Z_2\;\oplus\;\Z
\end{equation}
where the last isomorphism comes from the computation in Example \ref{ex_1} and the  K\"{u}nneth formula (for cohomology) which gives
$H^{1}(\n{S}^1\times\R P^\infty,\Z)\simeq H^{1}(\n{S}^1,\Z)\otimes_\Z H^{0}(\R P^\infty,\Z)\simeq\Z\otimes_\Z\Z\simeq\Z$. This shows that there are non-trivial \virg{Real} $\n{U}(1)$-principal bundles over $\hat{\n{T}}^2$. From the exact sequence  \cite[Proposition 2.3]{gomi-13} 
$$
\ldots\;H^{2}_{\Z_2}\left(\hat{\n{T}}^2,\Z(1)\right)\;\stackrel{\jmath}{\longrightarrow}\;H^{2}\left({\n{T}}^2,\Z\right)\;\stackrel{\nu_1}{\longrightarrow}\;H^{2}_{\Z_2}(\hat{\n{T}}^2,\Z(0))\stackrel{\nu_2}{\longrightarrow}H^{3}_{\Z_2}\left(\hat{\n{T}}^2,\Z(1)\right)\;\stackrel{}{\longrightarrow}\;H^{3}\left({\n{T}}^2,\Z\right)\;\ldots
$$
we can extrapolate that the map that  forgets the $\Z_2$-action $\jmath:H^{2}_{\Z_2}(\hat{\n{T}}^2,\Z(1))\to H^{2}({\n{T}}^2,\Z)$ is surjective. In fact from 
 $H^{k}({\n{T}}^2,\Z)\simeq\Z^{\binom{2}{k}}$ for $k=0,1,2$ and $H^{k}({\n{T}}^2,\Z)\simeq0$ for $k>2$ one has that the group homomorphism $\nu_2$ is  
 surjective.
The Gysin exact sequence \cite[Corollary 2.11]{gomi-13} and the  K\"{u}nneth formula (for cohomology)  provide
$$
\begin{aligned}
H^{2}_{\Z_2}(\hat{\n{T}}^2,\Z(0))\;&\simeq\;H^{2}\left({\n{S}}^1\times\R P^\infty,\Z\right)\;\oplus\;H^{1}_{\Z_2}\left({\n{S}}^1,\Z(1)\right)\\
&\simeq\;\left(H^{0}\left({\n{S}}^1,\Z\right)\;\otimes_\Z\; H^{2}\left(\R P^\infty,\Z\right)\right)\;\oplus\;\left(H^{1}\left({\n{S}}^1,\Z\right)\;\otimes_\Z\; H^{1}\left(\R P^\infty,\Z(1)\right)\right)\\
&\simeq\;\left(\Z\;\otimes_\Z\;\Z_2\right)\;\oplus\;\left(\Z\;\otimes_\Z\;\Z_2\right)\;\simeq\; \Z_2\;\oplus\;\Z_2\;.
\end{aligned}
$$
and similarly 
$$
\begin{aligned}
H^{3}_{\Z_2}(\hat{\n{T}}^2,\Z(1))\;&\simeq\;H^{3}\left({\n{S}}^1\times\R P^\infty,\Z(1)\right)\;\oplus\;H^{2}_{\Z_2}\left({\n{S}}^1,\Z(0)\right)\\
&\simeq\;\left(H^{0}\left({\n{S}}^1,\Z\right)\;\otimes_\Z\; H^{3}\left(\R P^\infty,\Z(1)\right)\right)\;\oplus\;\left(H^{0}\left({\n{S}}^1,\Z\right)\;\otimes_\Z\; H^{2}\left(\R P^\infty,\Z\right)\right)\\
&\simeq\;\left(\Z\;\otimes_\Z\;\Z_2\right)\;\oplus\;\left(\Z\;\otimes_\Z\;\Z_2\right)\;\simeq\; \Z_2\;\oplus\;\Z_2\;.
\end{aligned}
$$
Hence, the surjectivity of $\nu_2:\Z_2 \oplus \Z_2\to\Z_2 \oplus \Z_2$ implies also the injectivity of $\nu_2$ and consequently $\nu_1=0$. This proves that $\jmath$ is surjective. Since the group $H^{2}({\n{T}}^2,\Z)\simeq\Z$ classifies $\n{U}(1)$-principal bundles over ${\n{T}}^2$ via the first Chern class and the homomorphism  $\jmath$ reduces (topological) \virg{Real} Chern classes to (topological)  Chern classes (see \cite[Section 5.6]{denittis-gomi-14}) one has that the map $\jmath\circ \tilde{c}_1^{\rr{R}}:{\rm Prin}^{\rr{R}}_{\n{U}(1)}(\hat{\n{T}}^2)\to \Z$ provides the integer part of the classification. Moreover, this integer can be also detected by looking at the differential \virg{Real} Chern class
${c}_1^{\rr{R}}$ which, in particular, provides a representative for the differential Chern class which classifies $\n{U}(1)$-principal bundles over ${\n{T}}^2$. In order to give an interpretation of the torsion part we can look at the first isomorphism in \eqref{eq:gysin_split} which provides the decomposition into the free 
and the torsion part of $H^{2}_{\Z_2}(\hat{\n{T}}^2,\Z(1))$. This isomorphism is a consequence of the 
splitting of the Gysin
exact sequence and shows that the torsion part is the contribution from the fixed point set $\n{S}^1_+\sqcup \n{S}^1_-$. As a consequence  the torsion invariant can be detected just by looking at the  restriction of the principal bundle on the fixed point set and this procedure produces two disjoint copies of \virg{Real}  $\n{U}(1)$-principal bundles of the type studied in Example \ref{ex_1}. This allows us to conclude that the torsion part of  $H^{2}_{\Z_2}(\hat{\n{T}}^2,\Z(1))$ can be identified with the Stiefel-Whitney class of the $\n{O}(1)$-principal bundle over $\n{S}^1_+\sqcup \n{S}^1_-$ obtained  by restricting on the fixed point set and by forgetting the \virg{Real} structure. Moreover, this invariant can be detected by computing the holonomy around  $\n{S}^1_+$ and $\n{S}^1_-$ associated to every \virg{Real} connection.
}\hfill $\blacktriangleleft$
\end{example}

\begin{example}\label{ex_4}{\upshape
Let $\check{\T}^2\equiv(\n{T}^2,\xi)$ be the two dimensional torus endowed with the involution $\xi:(z_1,z_2)\mapsto(z_1,z_1\overline{z_2})$.
This space agrees with the total space of the {M\"{o}bius} $\rr{R}$-bundle $\pi:\bb{P}_{\rm M}\to\n{S}^1$ studied in Example \ref{ex_1}. The fixed-point set is given by $\big(\check{\T}^2\big)^\xi:=\{(z_1,z_1^2)\ |\ z_1\in\n{S}^1\}\simeq\n{S}^1$. The classification of \virg{Real} $\n{U}(1)$-principal bundle over $\check{\T}^2$ is provided by $H^2_{\Z_2}(\check{\T}^2,\Z(1))=H^2_{\Z_2}(\bb{P}_{\rm M},\Z(1))$ and this can be computed by means of the Gysin sequence \cite[Corollary 2.11]{gomi-13}
$$
H^0_{\Z_2}\left(\n{S}^1,\Z(0)\right)\stackrel{\chi_{\rm M}}{\longrightarrow}
H^2_{\Z_2}\left(\n{S}^1,\Z(1)\right)\stackrel{\pi^*}{\longrightarrow}
H^2_{\Z_2}\left(\bb{P}_{\rm M},\Z(1)\right)\stackrel{\pi_*}{\longrightarrow}
H^1_{\Z_2}\left(\n{S}^1,\Z(0)\right)\stackrel{\chi_{\rm M}}{\longrightarrow}
H^3_{\Z_2}\left(\n{S}^1,\Z(1)\right)
$$
where $\chi_{\rm M}$ is implemented by the cup product by the   (topological) \virg{Real} Chern class $\tilde{c}_1^{\rr{R}}(\bb{P}_{\rm M})\in H^2_{\Z_2}\left(\n{S}^1,\Z(1)\right)\simeq\Z_2$. Moreover, since $\bb{P}_{\rm M}$ is the non-trivial element one has $\tilde{c}_1^{\rr{R}}(\bb{P}_{\rm M})\simeq -1$. Since
$
H^{k}_{\Z_2}(\n{S}^1,\Z(m))\simeq H^{k}(\n{S}^1\times\R P^\infty,\Z(m))$ 
one obtains with the help of the K\"{u}nneth formula (for cohomology) that $
H^{k}_{\Z_2}(\n{S}^1,\Z(0))\simeq\Z\otimes_\Z\Z\simeq\Z$ if $k=0,1$. Moreover, $H^{3}_{\Z_2}(\n{S}^1,\Z(1))\simeq \Z\otimes_\Z\Z_2\simeq\Z_2$. Then, the Gysin sequence reads
$$
\ldots\;\Z\;\stackrel{\chi_{\rm M}}{\longrightarrow}\;\Z_2\;\stackrel{\pi^*}{\longrightarrow}\;
H^2_{\Z_2}\left(\bb{P}_{\rm M},\Z(1)\right)\;\stackrel{\pi_*}{\longrightarrow}\;\Z\;\stackrel{\chi_{\rm M}}{\longrightarrow}\;
\Z_2\;\ldots
$$
The non-triviality of $\bb{P}_{\rm M}$ (and its associated \virg{Real} Chern class) assures that the homomorphism $\chi_{\rm M}$ is surjective. Hence $0={\rm Ker}(\pi^*)={\rm Im}(\pi^*)$ and $\pi_*$ is injective. This also implies that $H^2_{\Z_2}\left(\bb{P}_{\rm M},\Z(1)\right)\simeq{\rm Ker}(\chi_{\rm M})\simeq\Z$. In conclusion one has that the \virg{Real} $\n{U}(1)$-principal bundles over $\check{\T}^2$
are classified by $H^2_{\Z_2}(\check{\T}^2,\Z(1))\simeq\Z$ and the absence of torsion assures that the differential \virg{Real} Chern class associated to a \virg{Real} connection provides a  faithful invariant (\cf Theorem \ref{theo:CWhom}).
}\hfill $\blacktriangleleft$
\end{example}

\subsection{ \virg{Real} $\n{U}(1)$-principal bundles and flat connections}\label{sect:real_flat_connections}
Consider a \virg{Real} $\n{U}(1)$-principal bundle $(\bb{P},\hat{\Theta})$ and a \virg{Real} connection $\omega\in\rr{A}_\rr{R}(\bb{P})$. We recall that $\omega$ is called \emph{flat} if the associated curvature vanishes, $F_\omega=0$. The study of \virg{Real} flat connections provides important information for the classification of 
\virg{Real} principal bundles for reasons that will be clarified in the next section.

\medskip

The first question that we want to address  is how many {inequivalent} \virg{Real} flat connections there are. We recall that two connections are called \emph{gauge equivalent} if they are connected by a 
{gauge transformation} of the \virg{Real} principal bundle, \ie by {local} maps from the base space $X$ to the structure group $\n{U}(m)$ which are suitably $\Z_2$-equivariant in a way compatible with the \virg{Real} structure. We denote with ${\rm Flat}_\rr{R}(\bb{P})$ the set of gauge equivalence classes of 
flat \virg{Real} connections. The following result shows that in general ${\rm Flat}_\rr{R}(\bb{P})$ contains more than one element.

\begin{lemma}\label{lemma:flat_real_conn}
Let $X\times\n{U}(1)$ be the product $\n{U}(1)$-principal bundle endowed with the trivial  \virg{Real} structure $\hat{\Theta}_0:(x,u)\to(\tau(x),\overline{u})$. Then
$$
{\rm Flat}_\rr{R}\big(X\times\n{U}(1)\big)\;\simeq\; H^1_{\Z_2}\big(X,\R(1)\big)\;/\;H^1_{\Z_2}\big(X,\Z(1)\big)
$$
where the equivalence  relation is given by the  natural homomorphism $H^1_{\Z_2}\big(X,\Z(1)\big)\to H^1_{\Z_2}\big(X,\R(1)\big)$ induced by $\Z(1)\hookrightarrow\R(1)$.
\end{lemma}
\proof
Since we are concerned with a product $\n{U}(1)$-bundle the space of the \virg{Real} connections is described by
$
\rr{A}_\rr{R}\big(X\times\n{U}(1)\big)\simeq\Omega_-^1(X)
$ and a flat connection $\omega\in \Omega_-^1(X)$ is specified by the closedness condition $\dd\omega=0$.
Let $\tilde{\n{R}}$ be the real line endowed with the involution (reflection) $r\mapsto-r$ and ${\rm Map}_{\Z_2}(X,\tilde{\n{R}})\simeq\Omega_-^0(X)$ the associated group of $\Z_2$-equivariant maps. An element $g\in{\rm Map}_{\Z_2}(X,\tilde{\n{R}})$ acts on a (\virg{Real}) connection by $\omega\mapsto\omega+\ii\dd g$ and preserves the flatness. By Corollary \ref{cor:equiv_dR_sio} one has
$$
\frac{\text{\virg{Real} flat connections}}{{\rm Map}_{\Z_2}(X,\tilde{\n{R}})}\;\simeq\;\frac{{\rm Ker}\big(\dd:\Omega_-^1(X)\to \Omega_-^2(X)\big)}{{\rm Im}\big(\dd:\Omega_-^0(X)\to \Omega_-^1(X)\big)}\;=\;H_{{\rm d.R},-}^1(X)\;\simeq\;H_{\Z_2}^1\big(X,\n{R}(1)\big)\;.
$$
The 
gauge transformation group on the product bundle is identifiable with  set ${\rm Map}_{\Z_2}(X,\tilde{\n{S}}^1)$ of  $\Z_2$-equivariant maps between the involutive space $(X,\tau)$ and the structure group $\n{U}(1)$ endowed with the complex conjugation, the latter being isomorphic to the TR sphere $\tilde{\n{S}}^1$.
A map $f\in {\rm Map}_{\Z_2}(X,\tilde{\n{S}}^1)$ acts on a (\virg{Real} ) connection by $\omega\mapsto\omega+\frac{1}{ 2\pi}\dd \log(f)$.
The exact sequence \cite[Appendix A.4]{gomi-13}
$$
{\rm Map}_{\Z_2}\big(X,\tilde{\n{R}}\big)\;\stackrel{{\rm exp \ii2\pi}}{\longrightarrow}\;{\rm Map}_{\Z_2}\big(X,\tilde{\n{S}}^1\big)\;\stackrel{}{\longrightarrow}\;H^1_{\Z_2}\big(X,\Z(1)\big)\;\stackrel{}{\longrightarrow}\;0
$$
gives the isomorphism $H^1_{\Z_2}\big(X,\Z(1)\big)\simeq {\rm Map}_{\Z_2}\big(X,\tilde{\n{S}}^1\big)/ {\rm Map}_{\Z_2}\big(X,\tilde{\n{R}}\big)$ and this concludes the proof.
\qed

\medskip

\begin{example}[Flat \virg{Real} connections on the 1 dimensional TR sphere]\label{ex_5}{\upshape
As an application of the previous result let us consider the case of the 1 dimensional TR sphere $\tilde{\n{S}}^1$ as a base space. In this case every \virg{Real} $\n{U}(m)$-principal bundle is isomorphic to the trivial product bundle (see Section \ref{sect:non-triv_low-dim}). In Example \ref{ex_2} we computed ${H}^{1}_{\Z_2}(\tilde{\n{S}}^1,\Z(1))=\Z_2\oplus\Z$. Here the free part $\Z$ is generated by the identity map $\tilde{\n{S}}^1\to \tilde{\n{S}}^1$ and the torsion part $\Z_2$ is given by the constant maps $\tilde{\n{S}}^1\to \pm 1$.
By  Proposition \ref{prop:cohom_real_int} we can compute ${H}^{1}_{\Z_2}(\tilde{\n{S}}^1,\R(1))=\R$ and the generator can be identified with the standard 1-form on $\n{S}^1$. Hence the \emph{moduli space} of flat \virg{Real} connections on $\n{S}^1\times \n{U}(1)$ is identifiable with $\R/\Z$. In particular this shows that 
there are several possible choices of inequivalent flat \virg{Real} connections. 
}\hfill $\blacktriangleleft$
\end{example}

\medskip

In the following we need  the Borel cohomology with coefficient system $\R/\Z(1)$. We can describe (or define) this cohomology theory as a relative Borel cohomology \cite{gomi-13}, \ie 
\begin{equation}\label{new_cohom_coeffi}
H^k_{\Z_2}\big(X,\R/\Z(1)\big)\;:=\; H^k_{\Z_2}\big(X\times \tilde{I},X\times\partial \tilde{I},\R/\Z\big)
\end{equation}
where $\tilde{I}$ is the interval $[-1,1]$ with the involution $t\to-t$.
Some relevant properties of this cohomology will be discussed in Appendix \ref{sec:new_cohom}. The relevance of the definition \eqref{new_cohom_coeffi} is given by the following result which generalizes Lemma \ref{lemma:flat_real_conn}.

\begin{proposition}\label{prop:flat_real_conn}
The group of \virg{Real} $\n{U}(1)$-principal bundles with flat connection over
the involutive space $(X, \tau)$ is isomorphic to $H^1_{\Z_2}\big(X,\R/\Z(1)\big)$.
Moreover, given a 
 \virg{Real} $\n{U}(1)$-principal bundle $(\bb{P},\hat{\Theta})$ over $(X,\tau)$ one has
$$
{\rm Flat}_\rr{R}\big(\bb{P}\big)\;\simeq\; {\rm Flat}_\rr{R}\big(X\times\n{U}(1)\big)\;.
$$
\end{proposition}
\proof[{proof} (sketch of)]
Let $\bb{P}\to X$ be a $\n{U}(1)$-principal bundle equipped with a 
flat connection
$\omega$. This implies that the transition functions which define $\bb{P}$ are locally constant (see Appendix \ref{sect:remind_connect}) and so we can identify the set of the
isomorphism classes of 
flat line bundles with $H^1(X;\C^\times)=H^1(X;\R/\Z)$ \cite[Section 2.2] {putman-12}. 
One  can generalize this result by incorporating the \virg{Real} structure as in \cite[Appendix A]{gomi-13}. More precisely one can identify the set of the isomorphism classes of 
flat \virg{Real} line bunldes with
the sheaf cohomology $H^1(\Z_2^\bullet
\times X; \tilde{\n{S}}^1_\bullet)$ on the simplicial space $\Z_2^\bullet
\times X$. The identification    $H^1(\Z_2^\bullet
\times X; \tilde{\n{S}}^1_\bullet)\simeq H^1_{\Z_2}(X,\R/\Z(1))$ completes the {first part of the proof. The second part follows by observing that ${\rm Flat}_\rr{R}\big(\bb{P}\big)$ is a torsor over ${\rm Flat}_\rr{R}\big(X\times\n{U}(1)\big)$ and the identification is provided by the election of a 
 flat connection on $\bb{P}$.} 
\qed

\medskip

\noindent
The connection between Lemma \ref{lemma:flat_real_conn} and Proposition \ref{prop:flat_real_conn} can be deduced by the exact sequence
\begin{equation}\label{eq:flat_exact_seq}
H^1_{\Z_2}\big(X,\Z(1)\big)\;\longrightarrow\;H^1_{\Z_2}\big(X,\R(1)\big)\;\longrightarrow\;H^1_{\Z_2}\big(X,\R/\Z(1)\big)\;\longrightarrow\;H^2_{\Z_2}\big(X,\Z(1)\big)
\end{equation}
associated to the short exact sequence $0\to\Z\to\R\to\R/\Z\to0$ of coefficients.

\subsection{Classification of \virg{Real} principal bundles in low dimensions}\label{sect:non-triv_low-dim}
In this section we focus on involutive base spaces $(X,\tau)$ which satisfy assumptions (H.1) - (H.3) together with:   
\begin{itemize}
\vspace{1.3 mm}
\item[(H.5)] The fixed-point set $X^\tau$ has dimension 0. \vspace{1.3 mm}
\end{itemize}
Under this extra assumption we know from \cite[Theorem 1.3]{denittis-gomi-14} that \virg{Real} $\n{U}(m)$-principal bundles
are completely classified by $H^2_{\Z_2}(X,\Z(1))$ if ${\rm dim}(X)\leqslant 3$. In this section we want to describe such a  classification by applying the  differential geometric apparatus developed in the previous sections. First, let us start with the cases ${\rm dim}(X)\leqslant 2$ which allow for a complete description.

\medskip
\medskip

\noindent
{\bf 1-dimesional case.} In this case conditions (H.1) and (H.2) imply that $X$ is isomorphic to the interval $I:=[0,1]$ or to the sphere $\n{S}^1$. Since $I$ is contractible (and equivariantly contractible) it leads only to trivial \virg{Real} $\n{U}(m)$-principal bundles. On the other side on the circle $\n{S}^1$ there are only three different involutions \cite{kwun-tollefson-75}: the \emph{trivial involution} which lead to the trivial involutive space $\n{S}^1\equiv (\n{S}^{1}, \rm{Id}_{\n{S}^{1}})$; the \emph{TR involution} $\tau:(x_0,x_1)\to(x_0,-x_1)$ which leads to the TR involutive space $\tilde{\n{S}}^1\equiv (\n{S}^{1}, \tau)$; the \emph{antipodal involution} $\theta:(x_0,x_1)\to(-x_0,-x_1)$ which leads to the free involutive space $\check{\n{S}}^1\equiv (\n{S}^{1}, \theta)$. The case of the  {TR involution} is trivial since $H^{2}_{\Z_2}(\tilde{\n{S}}^1,\Z(1))=0$
as proved in \cite[eq. (5.26)]{denittis-gomi-14}. Also the case of the antipodal involution is trivial. Indeed, since $\theta$ acts freely on $\check{\n{S}}^1$ one has $H^k_{\Z_2}(\check{\n{S}}^1,\Z)=H^k(\n{S}^1,\Z)$ where we used the isomorphism $\n{S}^1\simeq\check{\n{S}}^1/\theta$.
Since $H^k(\n{S}^1,\Z)=0$ if $k\neq0,1$ one deduces from the exact sequence in \cite[Proposition 2.3]{gomi-13} that also $H^2_{\Z_2}(\check{\n{S}}^1,\Z(1))=0$. Therefore, the only non-trivial case is provided by the  trivial involution which has been described in detail in Example \ref{ex_1}.

\medskip
\medskip

\noindent
{\bf 2-dimesional case.}
Consider the case of an involutive space $(X,\tau)$ which verifies conditions (H.1) - (H.3), which is closed (\ie compact with no boundary) and which has a finite number of fixed point (hence in particular  (H.5) is verified). In this setting one can prove that $H^k_{\Z_2}(X,\Z(1))=0$ if ${\rm dim}(X)=2$ \cite[Proposition 4.9]{denittis-gomi-14-gen}. This means that  only trivial \virg{Real} $\n{U}(m)$-principal bundles can be constructed over such a $(X,\tau)$.

\medskip
\medskip

\noindent
More in general we have the following classification principle:

\begin{theorem}\label{theo:class_princ}
Let $(X,\tau)$ be an involutive space that verifies assumption (H.1) - (H.5) and ${\rm dim}(X)\leqslant 3$. 
Then:
\begin{enumerate}
\item[(i)] If $H^2_{\Z_2}(X,\Z(1))$ has no torsion then 
${\rm Prin}^{\rr{R}}_{\n{U}(m)}(X,\tau)$ is classified by the \emph{differential} \virg{Real} Chern class associated to any  \virg{Real} connection.
 ;\vspace{1.2mm}
\item[(ii)]  If $H^2_{\Z_2}(X,\Z(1))$ is pure torsion then 
the \virg{Real} principal bundles are distinguished by the holonomy $\rm{Hol}(\bb{P},\omega)$ with respect to a flat \virg{Real} connection $\omega$ (modulo \emph{gauge} transformations).
\end{enumerate}
\end{theorem}
\proof
From \cite[Theorem 1.3]{denittis-gomi-14} we know that
$$
{\rm Prin}^{\rr{R}}_{\n{U}(m)}(X,\tau)\;\simeq\;
{\rm Prin}^{\rr{R}}_{\n{U}(1)}(X,\tau)\;\simeq\; H^2_{\Z_2}(X,\Z(1))
$$
with isomorphism  given by the topological (first) \virg{Real} Chern class $\tilde{c}_1^{\rr{R}}$. Then
(i) is  a direct consequence of Theorem \ref{theo:CWhom}.  

The proof of (ii) is more complicated. We start by considering the exact sequence of maps
\beql{eq:diag00_flat_holon}
\begin{diagram}
H^{1}_{\Z_2}\big(X,\R/\Z(1)\big)                        &         \rTo^{}        &H^{2}_{\Z_2}\big(X,\Z(1)\big)& \rTo^{}& H^{2}_{\Z_2}\big(X,\R(1)\big)\\
       \dTo^{\jmath}                                                           &                &\dTo^{\jmath}&                &\dTo^{\jmath} \\
H^{1}\big(X,\R/\Z\big)&\rTo & H^{2}\big(X,\Z\big) & \rTo^{}&H^{2}\big(X,\R\big)\\
\end{diagram}
\eeq
associated to the short exact sequence $0\to\Z\to\R\to\R/\Z\to0$ of coefficients. The vertical maps $\jmath$
forget the \virg{Real} structure and the injectivity of $\jmath:H^{1}_{\Z_2}(X,\R/\Z(1))\to H^{1}_{\Z_2}(X,\R/\Z)$ has been proved in Proposition \ref{prop:inj_foget_R/Z}. We know that $H^2_{\Z_2}(X,\Z(1))$ classifies \virg{Real}
principal $\n{U}(1)$-bundle and $H^{1}_{\Z_2}(X,\R/\Z(1))$ classifies \virg{Real}
principal $\n{U}(1)$-bundle endowed with a flat connection modulo gauge transformations (\cf Proposition \ref{lemma:flat_real_conn}). The exactness at $H^{2}_{\Z_2}\big(X,\Z(1)\big)$ and the isomorphism $H^{2}_{\Z_2}(X,\R(1))\simeq H^{2}_{\Z_2}(X,\Z(1))\otimes_\Z\R$ imply that
the torsion part of $H^{2}_{\Z_2}(X,\Z(1))$ classifies \virg{Real} principal $\n{U}(1)$-bundle that can be endowed with a flat connection modulo gauge transformations\footnote{A similar result for the complex case is proved in \cite[Lemma 2.6]{putman-12}.}.
In case where a \virg{Real} principal $\n{U}(1)$-bundle  $(\bb{P},\hat{\Theta})$ is classified by a pure torsion class, we can thus choose a 
flat \virg{Real} connection $\omega$ to get an element in $H^{1}_{\Z_2}\big(X,\R/\Z(1)\big) $. This element injects into $H^{1}(X,\R/\Z)$, which is the
set of the isomorphism classes of flat
principal $\n{U}(1)$-bundle. The
universal coefficient theorem for cohomology \cite[Chapter 3,Theorem 3.2]{hatcher-02}  and the Hurewicz theorem \cite[Chapter 2,Theorem 2A.1]{hatcher-02} lead to 
$$
H^{1}\big(X,\R/\Z\big)\;\simeq\;{\rm Hom}_\Z\big(H_1(X),\R/\Z\big)\;\simeq\;{\rm Hom}_\Z\big(\pi_1(X),\n{U}(1)\big)\;.
$$
Under this  isomorphism the element in $H^{1}\big(X,\R/\Z\big)$ which classifies the flat
principal $\n{U}(1)$-bundle $(\bb{P},\omega)$ is given by the 
$\rm{Hol}(\bb{P},\omega)$ with respect the flat connection $\omega$ (\cf \cite[Chapter I, Section 2]{kobayashi-87}). Indeed since $X$ is path-connected one has that the holonomy is independent on the choice of a particular point in the fibers (Proposition \ref{prop:holonI} (iii))
and the fact that $\rm{Hol}(\bb{P},\omega)\in{\rm Hom}_\Z\big(\pi_1(X),\n{U}(1)\big)$ is a consequence of the  
flatness of  $\omega$ which assures that homotopic loops provide the same holonomy (Proposition \ref{prop:holonI} (v)). 
\qed

\medskip

\begin{remark}[The mixed case]\label{rk_mixed}{\upshape
The case in which  the cohomology $H^2_{\Z_2}(X,\Z(1))=F\oplus T$ has both non-trivial \emph{free} part $F$ and \emph{torsion} part $T$ is  more  involved and no general results are available at the moment. A possible strategy can be developed along the following arguments: The exact sequence \eqref{eq:flat_exact_seq} implies that
$$
T\;\simeq\;{\rm Im}\Big(H^{1}_{\Z_2}\big(X,\R/\Z(1)\big)\;\to\; H^{2}_{\Z_2}\big(X,\Z(1)\big)\Big)
$$
and this contribution can be studied upon a choice of a 
 \virg{Real} flat connection. On
the other side the free part $F$  injects into $H^{2}_{\Z_2}(X,\R(1))$. The topological \virg{Real} Chern class
of a principal $\n{U}(1)$-bundle can be decomposed (but \emph{non} canonically) into its free part and the torsion part. The free part which injects
into $H^{2}_{\Z_2}(X,\R(1))$ can be detected by a Chern class realized by a
differential form. The torsion part can be studied through the holonomy associated to a 
\virg{Real} flat connection.
}\hfill $\blacktriangleleft$
\end{remark}

\appendix

\section{\virg{Real} structures associated to  the generalized harmonic oscillator}\label{sect:model_HO}
In this section we investigate the topology of the generalized \emph{harmonic oscillator}
by developing the idea sketched in \cite[Example 2.2.3]{chruscinski-jamiolkowski-04}.
Let $X$ be a manifold and for each point $x\in X$ consider the  \emph{generalized harmonic oscillator}
\begin{equation}\label{eq:HO_gen}
H_{\phi,\nu}(x)\;:=\;\frac{1}{2}\left( \rr{p}^2_\phi\;+\;\nu(x)^2\;\rr{q}^2 \right)\;=\;\frac{1}{2}\left[\rr{p}^2+\phi(x)\;(\rr{p}\rr{q}+\rr{q}\rr{p})\;+\; \left(\nu(x)^2+\phi(x)^2\right)\;\rr{q}^2\right]
\end{equation}
acting on the Hilbert space $L^2(\R,\dd r)$. In \eqref{eq:HO_gen} $\phi,\nu:X\to \R$ are two real-valued smooth maps, 
 $(\rr{q}\psi)(r)=r\psi(r)$ is the \emph{position} operator, $(\rr{p}\psi)(r)=-\ii\frac{\dd}{\dd r}\psi(r)$ is the 
\emph{momentum} operator and  by definition $\rr{p}_\phi:=\rr{p}+\phi(x)\rr{q}$. The function $\nu$ is called \emph{frequency}
and we assume that $\nu(x)>0$ for all $x\in X$.
We refer to $\phi$ as the \emph{anomaly}.
 The operator \eqref{eq:HO_gen} differs from the usual harmonic oscillator exactly when $\phi\neq0$. However, since $\rr{p}_\phi=U_\phi\rr{p}{U_\phi}^{*}$ and $[U_\phi,\rr{q}]=0$
with $U_\phi=\expo{- \frac{\ii}{2}\phi(x)\;\rr{q}^2}$ a unitary operator, one has that $H_{\phi,\nu}(x)$ is unitarily equivalent to the
harmonic oscillator of {frequency} $\nu$ (and mass $m=1$). This immediately implies that its spectrum is given by
\begin{equation}\label{eq:HO_spec_bis}
\sigma\Big(H_{\phi,\nu}(x)\Big)\;=\;\left\{\lambda_n(x):=\nu(x)\left(n+\frac{1}{2}\right)\ \Big|\ n\in\N\cup\{0\}\right\}
\end{equation}
with a corresponding complete set  of eigenvectors $H_{\phi,\nu}(x)\psi^{(x)}_n=\lambda_n(x)\psi^{(x)}_n$, called \emph{Hermite functions}, and given by
\begin{equation}\label{eq:herm_fun_bis}
\psi^{(x)}_n(r)\;:=\;C_n\;\nu(x)^{\frac{1}{4}}\;  H_n\left(r\; \nu(x)^{\frac{1}{2}}\right)\;
\expo{-\frac{r^2}{2}\big(\nu(x)+\ii\;\phi(x)\big)}
\end{equation}
where $C_n:=(n!2^n\sqrt{\pi})^{-\frac{1}{2}}$ is a normalization constant and
$$
H_n(r)\;:=\;(-1)^n\; \expo{r^2}\;\frac{\dd^n}{\dd r^n}\; \expo{-r^2}
$$ 
is the $n$-th Hermite polynomial \cite[Section 8.95]{gradshteyn-ryzhik-07}.

\medskip

\noindent
{\bf Time reversal symmetry.} 
The \emph{complex conjugation} $C$ acts on $L^2(\R,\dd r) $ as $(C\psi)(r)=\overline{\psi(r)}$. Then one has
$C\rr{p}C=-\rr{p}$ and $C\rr{q}C=\rr{q}$ which implies $C\rr{p}_\phi C=-\rr{p}_{-\phi}$ and so 
$H_{-\phi,\nu}(x)=C H_{\phi,\nu}(x) C$. Let us assume that $X$ can be endowed with an involution $\tau$ such that $\phi\circ\tau=-\phi$ and $\nu\circ\tau=\nu$. In this case one has that
\begin{equation}\label{eq:herm_invol}
H_{\phi,\nu}\big(\tau(x)\big)\;=\;C\;H_{\phi,\nu}(x)\;C\;,\qquad\quad x\in X
\end{equation}
and a comparison with \eqref{eq:tqsA3} shows that the Hamiltonian \eqref{eq:HO_gen} describes a 
topological quantum system
with a \virg{Real} time-reversal symmetry over the involutive space $(X,\tau)$.

\medskip

\noindent
{\bf \virg{Real} line-bundles.} For each energy level $n$ the  projection $P_n(x):=\ketbra{\psi^{(x)}_n}{\psi^{(x)}_n}$ defines a continuous map $X\ni x\mapsto P_n(x)$ and, as a consequence of the  Serre-Swan Theorem,
an associated line bundle $\bb{L}_n\to X$ (\cf Section \ref{sect:bott-chern-berry}). The line bundles $\bb{L}_n$ are trivial as complex line bundles since they have a global section $x\mapsto \psi^{(x)}_n$. However, the complex conjugation endows each $\bb{L}_n$ with a \virg{Real} structure which can lead to non-trivial effects. The topology of $\bb{L}_n$ can be investigated by the associated Grassmann-Berry connection that is  automatically \virg{Real} as a consequence of  Proposition \ref{prop:equivGB_conn}.
In Section \ref{sect:bott-chern-berry} we described  how the Grassmann-Berry connection is determined by the (local) 1-forms
$$
\s{A}^{(n)}(x)\;:=\;\sum_{j=1}^3\;\bra{\psi^{(x)}_n}\partial_{x_j}\ket{\psi^{(x)}_n}\; \dd x_j\;
$$
where $x=(x_1,\ldots,x_d)$ are local coordinates for $X$.
For the calculation of the scalar products $\bra{\psi^{(x)}_n}\partial_{x_j}\ket{\psi^{(x)}_n}$ we need the following relations:
$$
\int_{\R}\dd y\; y^2\;H_n\left(y\right)^2\; \expo{-y^2}\;=\;\left(n+\frac{1}{2}\right)\;C_n^{-2}
\;,\qquad\quad
\int_{\R}\dd y\; y\;H_n\left(y\right)\;H_n'\left(y\right)\; \expo{-y^2}\;=\;n\;C_n^{-2}\;.
$$
Both are consequences of the orthogonal condition $\int_{\R}\dd y H_n\left(y\right) H_m\left(y\right) \expo{-y^2}\delta_{n,m}\;C_n^{-2}$ \cite[eq. 8.959, 2.]{gradshteyn-ryzhik-07}.
The first  follows from the  identity
$y^2 \;H_n\left(y\right)=\frac{1}{4}H_{n+2}\left(y\right)+n(n-1)H_{n-2}\left(y\right)+(n+\frac{1}{2})H_{n}\left(y\right)$ \cite[eq. 8.952, 2.]{gradshteyn-ryzhik-07} while the second comes from $y\;H'_n\left(y\right)=nH_n(y)+2n^2H_{n-2}(y)$ \cite[eq. 8.952, 1. \& 3.]{gradshteyn-ryzhik-07}. A direct computation provides
$
\bra{\psi^{(x)}_n}\partial_{x_j}\ket{\psi^{(x)}_n}\;=\; -\ii\frac{2n+1}{4}\;\frac{1}{\nu(x)}\; \frac{\partial\phi(x)}{\partial x_j}
$, namely
$$
\s{A}^{(n)}(x)\;=\;-\ii\frac{2n+1}{4}\;\frac{1}{\nu(x)}\; \dd\phi(x)\;.
$$
The associated curvature is then given by
$$
\begin{aligned}
\s{F}^{(n)}(x)\;=\;\dd\s{A}^{(n)}(x)\;=\;\ii\frac{2n+1}{4}\;\frac{1}{\nu(x)^2}\;\dd \nu(x)\wedge\dd \phi(x)\;. 
\end{aligned}
$$

\medskip

\noindent
{\bf A  realization of Example \ref{ex_3}.} Just to give an example let us set the parameter space $X$ to be the torus $\n{T}^2\simeq[-\pi,\pi]^2$ parametrized by two angles $(\theta_1,\theta_2)$. We can also endow  $\n{T}^2$ with the involution $\eta:(\theta_1,\theta_2)\mapsto(-\theta_1,\theta_2)$. In this way we obtain the involutive space $\hat{\T}^2\equiv(\n{T}^2,\eta)$ described in Example
\ref{ex_3}. Let $f,g:\n{S}^1\to \R$ be two smooth functions and set $\nu(\theta_1,\theta_2)=\delta+f(\theta_2)^2$, with $\delta>0$ an  arbitrary constant and 
$\phi(\theta_1,\theta_2)=\sin(\theta_1)g(\theta_2)$. 
With this choice $H_{\phi,\nu}(\theta_1,\theta_2)$ turns out to have a \virg{Real} time-reversal symmetry over the involutive space $\hat{\T}^2$ and the
Grassmann-Berry connection becomes  
$$
\s{A}^{(n)}(\theta_1,\theta_2)\;=\;-\ii\frac{2n+1}{4}\;\frac{1}{\delta+f(\theta_2)^2}\; \left(\cos(\theta_1)g(\theta_2)\dd\theta_1\;+\; \sin(\theta_1)g'(\theta_2)\dd\theta_2\right)\;.
$$
However, one can check that the holonomy around the fixed point set vanishes.

\section{A reminder of the theory of connections}\label{sect:remind_connect}
Usually, the theory of connections is developed first for \emph{principal} bundles and then  is adapted to vector bundles. Although this procedure is not strictly necessary (connection can be defined directly for vector bundles) it is of course the most general and the most suitable for generalizations.  
For sake of completeness we review in this section   the basis of the theory of 
connections. A standard reference on this subject is \cite[vol. I \& II]{kobayashi-nomizu}
while
a more friendly introduction, with many links to the physics of quantum systems, can be found in 
\cite{bohm-mostafazadeh-koizumi-niu-zwanziger-03}.

\medskip
\medskip

\noindent
{\bf Principal bundles and vector bundles.} 
We start by fixing some basic notations: Let $\n{G}$ be a finite-dimensional Lie group with identity $e$. The associated Lie algebra $\rr{g}=T_{e}\n{G}$ is, by definition, the tangent space of $\n{G}$ at the identity. Since $\rr{g}$  carries the structure of a finite dimensional vector space we can consider the group of the linear transformations $\n{GL}(\rr{g})$ and the \emph{adjoint} representation ${\rm Ad}:\n{G}\to \n{GL}(\rr{g})$ which associates to each $g\in\n{G}$  
the differential at the identity ${\rm Ad}(g)$ of the mapping $\n{G}\ni h\mapsto ghg^{-1}\in \n{G}$. We are mainly interested in \emph{matrix} Lie groups $\n{G}$ like $\n{GL}_{\n{F}}(m)$ (the general linear group of invertible matrices of size $m$ over $\n{F}=\R,\C,\n{H}$) with Lie algebras $\rr{gl}_{\n{F}}(m)={\rm Mat}_{\n{F}}(m)$ (the full matrix algebra), or $\n{O}(m)$ (the real orthogonal group) with Lie algebra $\rr{o}(m)=\{\xi\in {\rm Mat}_{\n{R}}(m)\ |\ \xi+\xi^t=0\}$ (antisymmetric matrices), or 
 $\n{U}(m)$ (the complex unitary group) with Lie algebra $\rr{u}(m)=\{\xi\in {\rm Mat}_{\n{C}}(m)\ |\ \xi+\xi^\ast=0\}$ (anti-hermitian matrices) or finally $\n{U}_{\n{H}}(m)\simeq \n{S}{\rm p}(m)$ (the quaternionic unitary group equiv. the complex symplectic group) with Lie algebra $\rr{u}_{\n{H}}(m)=\{\xi\in {\rm Mat}_{\n{H}}(m)\ |\ \xi+\xi^\star=0\}$ (quaternionic skew-Hermitian matrices with respect to the quaternionic conjugate). In all these cases one has that ${\rm Ad}(g)[\xi]=g\cdot \xi\cdot g^{-1}$ is given by the matrix multiplication for all $g\in\n{G}$ and $\xi\in\rr{g}$.

\medskip

Let $\n{G}$ be an arbitrary finite-dimensional Lie group. A (smooth) \emph{principal} $\n{G}$-bundle is a fiber bundle $\pi:\bb{P}\to X$ in which $\pi$ is a smooth  map between smooth manifolds. Furthermore the total space $\bb{P}$ is endowed with a smooth right $\n{G}$-action $\bb{P}\times\n{G}\ni(p,g)\mapsto R_g(p)\equiv pg\in \bb{P}$ such that: (i) for each $x\in X$ the fiber  $\bb{P}_x:=\pi^{-1}(x)$ is an orbit for the $\n{G}$-action; (ii) every point $x\in X$ has a neighborhood $\f{U}$  and a smooth isomorphism $h:\pi^{-1}(\f{U})\to\f{U}\times \n{G}$
which is \emph{fiber-preserving} and \emph{equivariant} meaning that $h$ maps $\bb{P}_x$ to $\{x\}\times \n{G}$ for each $x\in\f{U}$  and $h(R_g(p))=R_g(h(p) )$
for all $p\in \pi^{-1}(\f{U})$ and $g\in\n{G}$ where $\n{G}$ acts on the right on $\f{U}\times \n{G}$ by $R_g((p,g'))\equiv(p,g') g:=(p,g'\cdot g)$. The group $\n{G}$ is usually called \emph{structure group} of the principal bundle $\bb{P}$.
Let us remark that from the above definition also follows that: $\pi$ is surjective and open; the right action of $\n{G}$ on $\bb{P}$ is \emph{free}; the orbit space $\bb{P}/\n{G}$ is isomorphic to the \emph{base} space $X$ (as smooth manifolds). Property (ii) is called
\emph{local triviality} and
says that a principal bundle is locally (isomorphic to) a product space. A principal bundle $\pi:\bb{P}\to X$ is called \emph{trivial} if it is globally isomorphic to the product principal bundle $X\times\n{G}\to X$. We recall that an isomorphism between principal bundles over the same base space is a smooth isomorphism of the total spaces which is fiber-preserving and equivariant with respect to the right action of $\n{G}$.  We denote with ${\rm Prin}_{\n{G}}(X)$ the family of isomorphism classes of {principal} $\n{G}$-bundles over $X$. A classical result says that a {principal} $\n{G}$-bundle over $X$ is trivial if and only if it has a global (smooth) section $\rr{s}:X\to\bb{P}$ \cite[8.3]{steenrod-51}.

\medskip

Vector bundles and principal bundles are strongly related objects. First of all, we see how  to each vector bundle is associated a canonical principal bundle called \emph{frame bundle} (see \eg \cite[vol. I, Chapt. I, Example 5.2]{kobayashi-nomizu}).   Let $\bb{E}\to X$ be a rank $m$ (smooth) vector bundle over $\n{F}$. This means that  each fiber $\bb{E}_x$ is isomorphic to the vector space $\n{F}^m$.
A \emph{frame} at a point $x\in X$ is an ordered basis $\{{\rm v}_1(x),\ldots,{\rm v}_m(x)\}$ for the vector space $\bb{E}_x$ and it is represented by a linear isomorphism $F:\n{F}^m\to \bb{E}_x$ such that $F({\rm e}_j)={\rm v}_j(x)$ where $\{{\rm e}_1,\ldots,{\rm e}_m\}$ is the canonical basis of $\n{F}^m$.
The set of all frames at $x$, denoted $\bb{F}_x$, has a natural right action by the general linear group $\n{GL}_{\n{F}}(m)$: a group element $g\in \n{GL}_{\n{F}}(m)$ acts on the frame $F$ via composition to give a new frame $F\circ g:\n{F}^m\to \bb{E}_x$.
This action of $\n{GL}_{\n{F}}(m)$ on $\bb{F}_x$ is both free and transitive. 
The frame bundle of $\bb{E}$, denoted by $\bb{F}(\bb{E})$, is given by the  disjoint union
$\bb{F}(\bb{E}) = \coprod_{x\in X}\bb{F}_x$.
Each point in $\bb{F}(\bb{E})$ is a pair $(x, F)$ where $x\in X$ and $F$ is a frame at $x$. There is a natural projection $\pi :\bb{F}(\bb{E})\to X$ which sends $(x, F)$ to $x$. The group $\n{GL}_{\n{F}}(m)$ acts on $\bb{F}(\bb{E})$ on the right as above and the action is clearly free with orbits given  just by the fibers $\bb{F}_x=\pi^{-1}(x)$.
The frame bundle $\bb{F}(\bb{E})$ can be given a natural topology and bundle structure determined by that of $\bb{E}$. Let $\{\f{U}_\alpha, f_\alpha\}$ be a local trivialization of $\bb{E}$. These data determine bijections $h_\alpha:\pi^{-1}(\f{U}_\alpha)\to \f{U}_\alpha \times\n{GL}_{\n{F}}(m)$ given by $h_\alpha(x, F):=(x, f_\alpha(x)\circ F)$ and
with these bijections we can endow  each $\pi^{-1}(\f{U}_\alpha)$ with the topology of $\f{U}_\alpha \times\n{GL}_{\n{F}}(m)$. The topology on $\bb{F}(\bb{E})$ is the \emph{final} topology induced by the inclusion maps  $\pi^{-1}(\f{U}_\alpha)\to \bb{F}(\bb{E})$.
With all of the above data the frame bundle $\bb{F}(\bb{E})$ becomes a (smooth) principal fiber bundle over $X$ with structure group $\n{GL}_{\n{F}}(m)$.  From the construction is also evident that isomorphic vector bundles provide isomorphic  frame bundles. 
On the other side is also true that to each  principal $\n{GL}_{\n{F}}(m)$-bundle 
is associated a canonical rank $m$ vector bundle called \emph{associated} vector bundle \cite[vol. I, Chapt. I, Proposition 5.4]{kobayashi-nomizu}. Let $\pi:\bb{P}\to X$ be a principal $G$-bundle and  $N$ a (smooth)  manifold on which $\n{G}$ acts on the left: $N\times\n{G}\ni(n,g)\mapsto L_g(n)\equiv gn\in N$. Let $\bb{E}_N:=(\bb{P}\times N)/\n{G}$ be the orbit space for the $\n{G}$-action given by $g:(p,n)\mapsto(R_g(p),L_{g^{-1}}(n))$
for all $p\in\bb{P}$, $n\in N$ and $g\in\n{G}$. By combining the canonical projection $\bb{P}\times N\to \bb{P}$ with the bundle projection $\pi$
one obtains a map $\bb{P}\times N\to X$ which factors through the equivalence relation and defines a projection $\pi_N:\bb{E}_N\to X$. 
Since each point $x\in X$ has a neighborhood $\f{U}$ such that $\pi^{-1}(\f{U})\simeq\f{U}\times\n{G}$ we can identify the $\n{G}$-action on  
$\pi^{-1}(\f{U})\times N$ with the mapping $g:((x,g'), n)\mapsto ((x,g'\cdot g), L_{g^{-1}}(n))$ for all $x\in\f{U}$, $g,g'\in\n{G}$ and $n\in N$.
This leads to an isomorphism $\pi_N^{-1}(\f{U})\simeq \f{U}\times N$. We can therefore introduce a topological (smooth) structure on  $\bb{E}_N$
by requiring that $\pi_N^{-1}(\f{U})$ is an open submanifold of   $\bb{E}_N$ which is (smoothly) isomorphic to $\f{U}\times N$.
 With all these data $\pi_N:\bb{E}_N\to X$ turns out to be a (smooth) fiber bundle over $X$ with typical fiber $N$ and structure group $\n{G}$.
When $\n{G}=\n{GL}_{\n{F}}(m)$ and $N=\n{F}^{m}$ (with the natural left-action) then the {associated} bundle 
has the structure of a rank $m$ vector bundle with typical fiber $\n{F}^{m}$. From the construction, it also follows that isomorphic principal bundles define isomorphic {associated} bundles. 

\medskip

The  frame bundle construction  and  the  {associated} bundle construction are, up to isomorphisms, mutually inverse mappings between principal bundles and vector bundles. As a result, one obtains that
\begin{equation}\label{eq:isoPV1}
{\rm Prin}_{\n{GL}_{\n{F}}(m)}(X)\;\simeq\;{\rm Vec}_{\n{F}}^m(X)\;,\qquad\quad\n{F}\;=\;\R\;,\;\C\;,\n{H}\;,
\end{equation}
where, by virtue of Theorem \ref{theo:top-smooth}, we  can consider \eqref{eq:isoPV1} as an isomorphism in the topological or in the smooth category. One of the implication of \eqref{eq:isoPV1} is that a vector bundle is trivial if and only if the associated frame bundle has a (global) section.

\medskip

The notion of {associated} bundle plays an important role when the structure group of a principal $\n{G}$-bundle $\pi:\bb{P}\to X$ has a closed subgroup $\n{K}\subset\n{G}$. 
In this case  $\n{G}$ acts in a natural way on the left on the  (left) coset space $\n{G}/\n{K}$ and one can construct  the {associated} bundle
$\pi_{\n{G}/\n{K}}:\bb{E}_{\n{G}/\n{K}}\to X$ with typical fiber $\n{G}/\n{K}$. On the other hand, being a subgroup of $\n{G}$, $\n{K}$ 
acts on the right  on the total space $\bb{P}$ and one can define the quotient space $\bb{P}/\n{K}$ associated with this action.
One can prove that  there exists a (smooth) isomorphism $\bb{P}/\n{K}\simeq \bb{E}_{\n{G}/\n{K}}$ and the structure group of $\pi:\bb{P}\to X$ is reducible to $\n{H}$ if and only if the associated bundle $\bb{P}/\n{K}\to X$ admits a section 
\cite[vol. I, Chapt. I, Proposition 5.5 and Proposition 5.6]{kobayashi-nomizu}. When $X$ is a paracompact space the  reductions 
\begin{equation*}
{\rm Prin}_{\n{GL}_{\n{R}}(m)}(X)\simeq{\rm Prin}_{\n{O}(m)}(X)\;,\quad{\rm Prin}_{\n{GL}_{\n{C}}(m)}(X)\simeq{\rm Prin}_{\n{U}(m)}(X)\;,\quad{\rm Prin}_{\n{GL}_{\n{H}}(m)}(X)\;\simeq\;{\rm Prin}_{\n{U}_{\n{H}}(m)}(X)\;,\qquad
\end{equation*}
hold true and, combined with  \eqref{eq:isoPV1}, they correspond to the existence of a unique (up to isomorphisms) metric for real, complex or quaternionic vector bundles \cite[Chapt. 3, Theorem 9.5]{husemoller-94}.

\medskip

As for vector bundles, also principal bundles are completely characterized by the set of the \emph{transition functions}. Let $\{\f{U}_\alpha,h_\alpha\}$ be a trivialization for the principal $\n{G}$-bundle $\pi:\bb{P}\to X$. For $\f{U}_\alpha\cap\f{U}_\beta\neq \emptyset$  the composition $h_\beta \circ h_\alpha^{-1}: (\f{U}_\alpha\cap\f{U}_\beta)\times\n{G} \to (\f{U}_\alpha\cap\f{U}_\beta)\times\n{G}$ is a (smooth) isomorphism 
which acts as $h_\beta \circ h_\alpha^{-1}(x,g)=(x,\varphi_{\beta,\alpha}(x)\cdot g)$ where $g\in\n{G}$, $x\in \f{U}_\alpha\cap\f{U}_\beta$
and $\varphi_{\beta,\alpha}:\f{U}_\alpha\cap\f{U}_\beta\to\n{G}$ are (smooth) maps. The family of transition functions $\{\varphi_{\beta,\alpha}\}$ verifies the \emph{cocycle} conditions
\begin{equation}\label{eq:isoPV3}
\begin{aligned}
&\varphi_{\beta,\alpha}(x)\;=\;\varphi_{\beta,\gamma}(x)\cdot\varphi_{\gamma,\alpha}(x)&\qquad&\text{for all}\ x\in \f{U}_\alpha\cap\f{U}_\beta\cap\f{U}_\gamma\\
&\varphi_{\alpha,\alpha}(x)\;=\;e&\qquad&\text{for all}\ x\in \f{U}_\alpha\;.
\end{aligned}
\end{equation}
Two families $\{\varphi'_{\beta,\alpha}\}$ and $\{\varphi_{\beta,\alpha}\}$ of transition functions are called isomorphic if there exists a family of (smooth) functions $\lambda_\alpha:U_\alpha\to\n{G}$ such that the following  \emph{coboundary} conditions hold
\begin{equation}\label{eq:isoPV4}
\varphi'_{\beta,\alpha}(x)\;=\;\lambda_{\beta}(x)^{-1}\cdot\varphi_{\beta,\alpha}(x)\cdot\lambda_{\alpha}(x) \quad\qquad \text{for all}\ x\in \f{U}_\alpha\cap\f{U}_\beta\;. 
\end{equation}
The \emph{fiber bundle construction theorem} \cite[vol. I, Chapt. I, Proposition 5.2]{kobayashi-nomizu} asserts a complete equivalence (up to isomorphisms) between principal $\n{G}$-bundles $\pi:\bb{P}\to X$ and families of functions
$\varphi_{\beta,\alpha}:\f{U}_\alpha\cap\f{U}_\beta\to\n{G}$ which verify the \emph{cocycle} conditions \eqref{eq:isoPV3}.
Moreover, as a consequence of the isomorphism \eqref{eq:isoPV1}, one has that each principal $\n{GL}_{\n{F}}(m)$-bundle
shares with its  associated vector bundle the same family of transition functions.

\medskip
\medskip

\noindent
{\bf Connection on principal bundles.} 
Let $\pi:\bb{P}\to X$ be a principal $\n{G}$-bundle. For each fixed $p\in\bb{P}$ the (smooth) map $G\ni g\mapsto R_g(p)\in\bb{P}$ between manifolds
induces an injective map $V_p:\rr{g}\to T_p\bb{P}$ between tangent spaces and the quotient space by the image of $V_p$ is mapped isomorphically onto $T_{\pi(p)}X$ by the differential $\pi_*$ of $\pi$. In other words, there is an exact sequence of vector spaces 
\begin{equation}\label{eq:exact1}
0\;\longrightarrow\;\rr{g}\;\stackrel{V_p}{\longrightarrow}\;T_p\bb{P}\;\stackrel{\pi_*}{\longrightarrow}\;T_{\pi(p)}X\;\longrightarrow\;0\;.
\end{equation}
The space $V_p(\rr{g})\subset T_p\bb{P}$ is called the \emph{vertical} subspace of $T_p\bb{P}$ and contains vectors tangent to the fiber $\bb{P}_{\pi(p)}$ through $p$. A \emph{horizontal distribution} (also called \emph{Ehresmann connection}) is an assignment $\bb{P}\in p\mapsto H_p\subset T_p\bb{P}$ of subspaces of $T_p\bb{P}$ such that:
\begin{enumerate}
\item[(a)] $T_p\bb{P}=H_p\;\oplus_p\;V_p(\rr{g})$\ \  (direct sum);
\item[(b)] $H_{R_g(p)}=(R_g)_*H_p$ for every $p\in\bb{P}$ and $g\in\n{G}$;
\item[(c)] $H_p$ depend smoothly on $p$.
\end{enumerate}
The spaces $H_p$ are called  \emph{horizontal} subspaces of $T_p\bb{P}$. The symbol $\oplus_p$ means that the decomposition depicted in (a) is valid pointwise. Moreover, in view of the exact sequence \eqref{eq:exact1}, (a) is equivalent to $H_p\simeq T_{\pi(p)}X$ for all $p\in\bb{P}$.
Condition (b) means that the {horizontal} subspaces are permuted by the right action of $\n{G}$ on $T\bb{P}$, namely the distribution $p\mapsto H_p$ $\n{G}$-invariant. Let us point out that  the vertical subspaces $V_p(\rr{g})$ are uniquely defined by the bundle projection 
$\pi:\bb{P}\to X$, while the choice of a {horizontal distribution} $H_p$ is usually not unique.
\medskip

To each {horizontal distribution} $p\mapsto H_p$ one can associate a 1-form  on $\bb{P}$ with value in the Lie algebra $\rr{g}$ of $\n{G}$, \ie an element  $\omega\in\Omega^1(\bb{P},\rr{g})$.
For each $\xi\in \rr{g}$ the mapping $\R\ni t\mapsto {\rm e}^{t\xi}\in \n{G}$ defines a 1-parameter subgroup of $\n{G}$ and
\begin{equation}\label{eq:defi_vert_vec}
\xi^*_p\;:=\;\left.\frac{{\rm d}}{{\rm d}t} R_{{\rm e}^{t\xi}}(p)\right|_{t=0}\;=\;V_p(\xi)\;,\qquad\quad p\in\bb{P}
\end{equation}
is a vertical vector. The map $p\mapsto \xi^*_p$  defines a (smooth) vector field $\xi^*\in\Gamma(T\bb{P})$ and the assignment $\xi\mapsto \xi^*$ is a Lie algebra homomorphism between $\rr{g}$ and $\Gamma(T\bb{P})$   \cite[vol. I, Chapt. I, Proposition 4.1]{kobayashi-nomizu}.  $\xi^*$ is called the \emph{fundamental} vector field associated to $\xi$
and for each $g\in\n{G}$, $(R_g)_*\xi^*$ is the fundamental vector field associated to ${\rm Ad}(g^{-1})[\xi]$ \cite[vol. I, Chapt. I, Proposition 5.1]{kobayashi-nomizu}. Since $\n{G}$ acts freely on $\bb{P}$, one has that $\xi^*_p\neq 0$ for all $p\in\bb{P}$ (if $\xi\neq 0$) and the 
assignment $ \xi\mapsto \xi^*_p$  defines a linear isomorphism between $\rr{g}$ and $V_p(\rr{g})$ for each $p\in\bb{P}$. For each  vector field
${\rm w}\in \Gamma(T\bb{P})$ let ${\rm w}_p={\rm w}^H_p+{\rm w}^V_p$ be the splitting induced by (a).
We can define pointwise a 1-form $\omega\in\Omega^1(\bb{P},\rr{g})$ by saying that $\omega_p({\rm w}_p)\in\rr{g}$ is the unique element such that
\begin{equation}\label{eq:defi_connect}
\omega_p({\rm w}_p)^*_p\;:=\;\left.\frac{{\rm d}}{{\rm d}t} R_{{\rm e}^{t\omega_p({\rm w}_p)}}(p)\right|_{t=0}\;=\;{\rm w}^V_p\;.
\end{equation}
The 1-form $\omega\in\Omega^1(\bb{P},\rr{g})$ defined by \eqref{eq:defi_connect}
 is called the \emph{connection} (1-form) associated to the {horizontal distribution} $p\mapsto H_p$.
It is clear from the definition that  $\omega_p({\rm w}_p)=0$ if and only if ${\rm w}_p\in H_p$. The following classical result holds true \cite[vol. I, Chapt. II, Proposition 1.1]{kobayashi-nomizu}:
\begin{proposition}\label{prop_commect}
Let $\omega\in\Omega^1(\bb{P},\rr{g})$ be a  \emph{connection} 1-form associated to a {horizontal distribution} $p\mapsto H_p$. Then
\begin{enumerate}
\item[\upshape{(a')}] $\omega_p\circ V_p\;=\;{\rm Id}_{\rr{g}}$\ \  for all $p\in\bb{P}$;
\item[\upshape{(b')}] $(R_g)^*\omega\;=\;{\rm Ad}(g^{-1})\circ \omega$ for all $g\in\n{G}$.
\end{enumerate}
Moreover, each 1-form $\omega\in\Omega^1(\bb{P},\rr{g})$ which verifies \upshape{(a')}-\upshape{(b')} defines a {horizontal distribution} $p\to H_p$ such that $H_p={\rm Ker}(\omega_p)$.
\end{proposition}
\noindent
A second crucial result concerns the existence of connections \cite[vol. I, Chapt. II, Theorem 2.1]{kobayashi-nomizu}:
\begin{theorem}\label{teo:exist_connect}
If $X$ is paracompact then each principal $\n{G}$-bundle $\pi:\bb{P}\to X$ admits at least one connection  $\omega\in\Omega^1(\bb{P},\rr{g})$.
\end{theorem}
\begin{remark}\label{rk:convex_connect}{\upshape
We denote with $\rr{A}(\bb{P})\subset \Omega^1(\bb{P},\rr{g})$ the space of connections on $\bb{P}$.
Theorem \ref{teo:exist_connect} assures that $\rr{A}(\bb{P})\neq\emptyset$. Moreover, from Proposition \ref{prop_commect} it follows that
for every collection $\{\omega_1,\ldots,\omega_N\}\subset \rr{A}(\bb{P})$ of connections and numbers $\{a_1,\ldots,a_N\}\subset\C$ such that 
$\sum_{j=1}^Na_j=1$ then the convex combination
$\omega:=\sum_{j=1}^n\;a_j\;\omega_j$ is still a connection. The space  $\rr{A}(\bb{P})$ is closed under \emph{convex} combinations.
}\hfill $\blacktriangleleft$
\end{remark}

\medskip

Once a connection $\omega$ on $\pi:\bb{P}\to X$ has been given we can lift \virg{horizontally} vector fields on $X$ to {horizontal vector fields}
on $\bb{P}$. More precisely, let $\tilde{\rm w}\in\Gamma(TX)$ be a vector field on $X$. The \emph{horizontal  lift} of $\tilde{\rm w}$ is the unique vector filed ${\rm w}\in\Gamma(T\bb{P})$ such that ${\rm w}_p\in H_p$ and $\pi_*: {\rm w}_p=\tilde{\rm w}_p$ for all $p\in\bb{P}$. The horizontal  lift is invariant under the right action of $\n{G}$ and each horizontal vector field is the horizontal  lift of an element in $\Gamma(TX)$ \cite[vol. I, Chapt. II, Proposition 1.2 and Proposition 1.3]{kobayashi-nomizu}.

\medskip

A connection 1-form $\omega$ on $\bb{P}$ can be described by family of local 1-forms on the base manifold $X$. Let $\{\f{U}_\alpha,h_\alpha\}$
be a local trivialization for $\pi:\bb{P}\to X$ and $\{\varphi_{\beta,\alpha}\}$ the related family of transition functions. For each $\alpha$ let $\Omega^1(\f{U}_\alpha,\rr{g})$ be the set of local (smooth) 1-forms defined on the open set  $\f{U}_\alpha\subset X$ and with value in the Lie algebra $\rr{g}$ and $\sigma_\alpha\in\Gamma(\f{U}_\alpha)$ the local (smooth) \emph{canonical section} defined by $\sigma_\alpha(x):=h_\alpha^{-1}(x,e)$, with $e\in\n{G}$ the identity element. The pullback $\s{A}_\alpha:=(\sigma_\alpha)^*\omega$ associates to a connection $\omega\in\Omega^1(\bb{P},\rr{g})$  a  1-forms  $\s{A}_\alpha\in\Omega^1(\f{U}_\alpha,\rr{g})$.
Let $\Omega^1_{\rm left}(\n{G},\rr{g})$ be the space of \emph{left-invariant} (smooth)
 1-forms  on $\n{G}$ with value in the Lie algebra $\rr{g}$. With left-invariant one means that
 $\varpi\in \Omega^1_{\rm left}(\n{G},\rr{g})$ if and only if $(L_g)^*\varpi=\varpi$ for every $g\in\n{G}$.
The \emph{Maurer-Cartan}  1-form $\theta\in \Omega^1_{\rm left}(\n{G},\rr{g})$ is defined by $\theta_g:=(L_{g^{-1}})_*$.
More precisely, let $\xi\in\rr{g}=T_e\n{G}$ and $\tilde{\xi}\in\Gamma(T\n{G})$ the associated vector field  $\tilde{\xi}_g=(L_{g})_*\xi$, then the \emph{Maurer-Cartan} 1-form is defined by  $\theta_g(\tilde{\xi}_g)=\xi$ for all $g\in\n{G}$ and all $\xi\in\rr{g}$.
For each non empty $\f{U}_\alpha\cap\f{U}_\beta\neq\emptyset$ the expression $\theta_{\beta,\alpha}:=(\varphi_{\beta,\alpha})^*\theta$
is an element of $\Omega^1(\f{U}_\alpha\cap\f{U}_\beta,\rr{g})$. The following characterization holds true \cite[vol. I, Chapt. II, Proposition 1.4]{kobayashi-nomizu}:
\begin{theorem}\label{theo:loc_connect}
Let $\pi:\bb{P}\to X$ be a principal $\n{G}$-bundle with  local trivializations $\{\f{U}_\alpha,h_\alpha\}$ and transition functions $\{\varphi_{\beta,\alpha}\}$.
There is a one-to-one correspondence between connections $\omega\in \Omega^1(\bb{P},\rr{g})$ and collections of 1-forms $\{\s{A}_\alpha\in\Omega^1(\f{U}_\alpha,\rr{g})\}$ which verify the transformation rule
\begin{equation}\label{eq:conn_trans_rul}
\s{A}_\alpha\;=\;{\rm Ad}\big(\varphi_{\beta,\alpha}^{-1}\big)\;\circ\; \s{A}_\beta\;+\;\theta_{\beta,\alpha}\;,\qquad\text{on}\ \ \f{U}_\alpha\cap\f{U}_\beta\;.
\end{equation}
\end{theorem}
\noindent
As a consequence of this theorem we can identify connections   with collections $\s{A}:=\{\s{A}_\alpha\}$ of local 1-forms on $X$ subjected to the 
transformation rule \eqref{eq:conn_trans_rul}. Let us also recall that in the case $\n{G}=\n{GL}_{\n{F}}(m)$ then the 
{Maurer-Cartan} 1-form has the form $\theta_g=g^{-1}\cdot{\rm d}g$ for all $g\in\n{G}$ and the transformation rule \eqref{eq:conn_trans_rul} reads
\begin{equation}\label{eq:conn_trans_rul_bis}
\s{A}_\alpha\;=\; \varphi_{\beta,\alpha}^{-1}\;\cdot\; \s{A}_\beta\;\cdot\; \varphi_{\beta,\alpha}\;+\;\varphi_{\beta,\alpha}^{-1}\;\cdot\;{\rm d }\varphi_{\beta,\alpha}\;,\qquad\text{on}\ \ \f{U}_\alpha\cap\f{U}_\beta\;. 
\end{equation}

\medskip
\medskip

\noindent
{\bf Basic forms and the space of connections.} 
Let $\pi:\bb{P}\to X$ be a principal $\n{G}$-bundle and $r:\n{G}\to\n{GL}(\s{V})$  a representation of $\n{G}$ on the (finite dimensional) vector space $\s{V}$. Let $\Phi\in\Omega^{k}(\bb{P},\s{V})$ be a differential $k$-form on $\bb{P}$ with value in $\s{V}$.
We say that: (i) $\Phi$ is \emph{horizontal} if for all $p\in\bb{P}$, $\Phi_p({\rm w}_1,\ldots,{\rm w}_k)=0$ for all $k$-tuples of tangent vectors
$\{{\rm w}_1,\ldots,{\rm w}_k\}\subset T_p\bb{P}$ for which at least one is vertical (\ie tangent to the fiber through $p$); (ii)  $\Phi$ is $r$-\emph{equivariant} if $R^*_g\Phi=r(g^{-1})\circ \Phi$ for all $g\in\n{G}$; (iii) $\Phi$ is \emph{invariant} if it is equivariant with respect to the trivial representation $r\equiv{\rm Id}_{\s{V}}$; (iv)
$\Phi$ is \emph{basic} if it is both invariant and horizontal.
We use the notation $\Omega^k_{\rm hor}(\bb{P},\s{V},r)\subset \Omega^k(\bb{P},\s{V})$ for the set of the {horizontal} $r$-{equivariant} $k$-forms on $\bb{P}$. The following result is similar in spirit to Theorem \ref{theo:loc_connect}:
\begin{proposition}\label{prep:basic_conn}
Let $r:\n{G}\to\n{GL}(\s{V})$ be a representation of $\n{G}$ on the (finite dimensional) vector space $\s{V}$ and
$\pi:\bb{P}\to X$ be a principal $\n{G}$-bundle with  local trivializations $\{\f{U}_\alpha,h_\alpha\}$ and transition functions $\{\varphi_{\beta,\alpha}\}$. 
\begin{enumerate}
\item[(i)] There is a one-to-one correspondence between 
{horizontal} $r$-{equivariant} $k$-form 
$\Phi\in \Omega^k_{\rm hor}(\bb{P},\s{V},r)$ and collections of k-forms $\{\s{F}_\alpha\in\Omega^k(\f{U}_\alpha,\s{V})\}$ which verify the transformation rule
\begin{equation}\label{eq:conn_trans_rul_F}
\s{F}_\alpha\;=\;
r\big(\varphi_{\beta,\alpha}^{-1}\big)\;\circ\; \s{F}_\beta\;,\qquad\text{on}\ \ \f{U}_\alpha\cap\f{U}_\beta\;.
\end{equation}
Here $\s{F}_\alpha$ is the pullback of $\Phi$ by the local section
$\sigma_\alpha\in\Gamma(\f{U}_\alpha)$ defined by $\sigma_\alpha(x):=h_\alpha^{-1}(x,e)$, with $e\in\n{G}$ the identity element. \vspace{1.3 mm}

\item[(ii)] The pullback $\pi^*:\Omega^k(X,\s{V})\to \Omega^k(\bb{P},\s{V})$ gives an isomorphism onto the basic form on $\bb{P}$.\vspace{1.3 mm}
\end{enumerate}
\end{proposition}

Under the assumption of the above  proposition 
we can consider the vector bundle $\pi_{\s{V}}:\bb{E}_{\s{V}}\to X$
associated to $\pi:\bb{P}\to X$ by the left action of $\n{G}$ on $\s{V}$ given by $L_g({\rm v})=r(g)\cdot {\rm v}$ for all $g\in\n{G}$ and ${\rm v}\in\s{V}$. Item (i) in Proposition \ref{prep:basic_conn} in the case $k=0$ states that there is a one-to-one correspondence between the set of $r$-equivariant functions $f:\bb{P}\to\s{V}$ and the set of sections $s:X\to \bb{E}_{\s{V}}$, namely $\Omega^0_{\rm hor}(\bb{P},\s{V},r)\simeq \Gamma( \bb{E}_{\s{V}})$ (see \eg \cite[vol. I, Chapt. II, Example 5.2]{kobayashi-nomizu}). As a  particularly interesting case let us consider the situation in which the vector space 
$\s{V}$ coincides with the Lie algebra $\rr{g}$ of $\n{G}$ and the representation $r$ is the adjoint representation ${\rm Ad}:\n{G}\to\n{GL}(\rr{g})$. From (b') in Proposition \ref{prop_commect} if follows that any connection $\omega$ is {\rm Ad}-invariant.  Combining Theorem \ref{theo:loc_connect} with Proposition \ref{prep:basic_conn} one obtains that:
\begin{corollary}\label{cor:aaf_prin}
The space of connections $\rr{A}(\bb{P})$ is an \emph{affine} space for the vector space $\Omega^1_{\rm hor}(\bb{P},\rr{g},{\rm Ad})$. That is, given any pair of connections $\omega_1,\omega_2\in \rr{A}(\bb{P})$ one has that $\omega_1-\omega_2\in \Omega^1_{\rm hor}(\bb{P},\rr{g},{\rm Ad})$.
\end{corollary}

\medskip
\medskip

\noindent
{\bf Flat connections.} 
Let $\bb{P}=X\times\n{G}$ be the trivial principal $\n{G}$-bundle over $X$. For each $g\in\n{G}$ the set $X\times\{g\}$ is a submanifold of $\bb{P}$. This fact allows us to define the \emph{flat}
horizontal distribution $ \bb{P}\in(x,g)\mapsto H_{(x,g)}$ by the identification $H_{(x,g)}:= T_{(x,g)}(X\times\{g\})$ of the 
horizontal subspaces with the tangent spaces to the 
submanifolds $X\times\{g\}$. Let $\theta\in \Omega^1_{\rm left}(\n{G},\rr{g})$ be the
{Maurer-Cartan}  1-form on $\n{G}$ and ${\rm pr}_2:X\times\n{G}\to \n{G}$ the 
natural projection and set $\omega_{\rm flat}:=({\rm pr}_2)^*\theta$. It is straightforward to verify that $\omega_{\rm flat}\in\Omega^1(\bb{P},\rr{g})$ is the connection 1-form associated to the {flat}
horizontal distribution and it is therefore called \emph{canonical flat} connection. 

\medskip

The notion of \emph{flat connection} can be extended to non-trivial principal bundle using the local trivializations. Let $\pi:\bb{P}\to X$ be a principal $\n{G}$-bundle with  local trivializations $\{\f{U}_\alpha,h_\alpha\}$ and transition functions $\{\varphi_{\beta,\alpha}\}$.
We say that $\bb{P}$ has a \emph{flat}
horizontal distribution $\bb{P}\in p\mapsto H_{p}$ if and only if on every  $\f{U}_\alpha$  the induced 
horizontal distribution on $\bb{P}|_{\f{U}_\alpha}=\pi^{-1}(\f{U}_\alpha)$ is isomorphic (via $h_\alpha$) with the standard {flat}
horizontal distribution on $\f{U}_\alpha\times\n{G}$.
Similarly, a connection 1-form $\omega\in\Omega^1(\bb{P},\rr{g})$ is said \emph{flat} if and only if $\omega|_{\bb{P}|_{\f{U}_\alpha}}=h^*_\alpha \omega_{\rm flat}$, namely if and only if the restriction of $\omega$ to 
$\bb{P}|_{\f{U}_\alpha}$ is induced (via $h_\alpha$)  by the 
{Maurer-Cartan}  1-form $\omega_{\rm flat}$ on $\f{U}_\alpha\times\n{G}$.
Finally, let us recall that a principal $\n{G}$-bundle has a flat connection if and only if it admits a system of \emph{constant} transition functions $\varphi_{\beta,\alpha}:\f{U}_\alpha\cap \f{U}_\beta\to\n{G}$.

\medskip
\medskip

\noindent
{\bf Parallel transport and holonomy group.} 
Given a connection $\omega$ in a principal $\n{G}$-bundle $\pi:\bb{P}\to X$ 
we can define the notion of {parallel transport} of fibers of $\bb{P}$ along any given ($C^1$-piecewise) curve $\tilde{\gamma}:[0,1]\to X$
on the base manifold $X$. A \emph{(horizontal)  lift} of $\tilde{\gamma}$ is a horizontal curve ${\gamma}:[0,1]\to \bb{P}$ such that $\pi\circ \gamma=\tilde{\gamma}$. Here a horizontal curve in $ \bb{P}$ means a ($C^1$-piecewise) curve whose tangent vectors are  in the horizontal spaces 
defined by $\omega$. The  lift of a given curve is usually not unique. Nevertheless, the following result holds \cite[vol. I, Chapt. II, Proposition 3.1]{kobayashi-nomizu}: 

\begin{theorem}\label{theo:par_transp}
Let $\pi:\bb{P}\to X$ be a principal $\n{G}$-bundle,  $\omega\in \Omega^1(\bb{P},\rr{g})$ a connection 1-form and $\tilde{\gamma}:[0,1]\to X$
a ($C^1$-piecewise) curve in $X$. For any $p_0\in\pi^{-1}(\tilde{\gamma}(0))$ there exists a unique (horizontal)  lift ${\gamma}$ of $\tilde{\gamma}$ such that ${\gamma}(0)=p_0$.
\end{theorem}
\noindent
In order to prove this result one starts with a ($C^1$-piecewise) curve $\rho:[0,1]\to \bb{P}$ such that $\rho(0)=p_0$ and $\pi\circ \rho=\tilde{\gamma}$. Such a curve always exists since  $\bb{P}$ is locally trivial and one can use local sections to lift locally $\tilde{\gamma}$.
The (horizontal)  lift $\gamma$, if it exists, must be of the form $\gamma(t):=R_{g(t)}(\rho(t))=\rho(t)g(t)$ for a suitable curve $[0,1]\ni t\mapsto g(t)\in\n{G}$ such that $g(0)=e$. Let $\dot{\gamma}_t$ be the tangent vector to the curve ${\gamma}$ at the point $\gamma(t)$. The condition of horizontality for $\gamma$ is given by $\omega(\dot{\gamma}_t)=0$ for all $t\in[0,1]$.
Differentiating the expression for $\gamma(t)$, one gets (with the help of the Leibniz's rule) the differential equation $\dot{g}(t)\cdot g(t)^{-1}=-\omega(\dot{\rho}_t)$ (see \cite[vol. I, Chapt. II, Proposition 3.1]{kobayashi-nomizu}). Here   $\dot{g}(t)\cdot g(t)^{-1}$  has to be understood as a curve in the Lie algebra $\rr{g}$ and $\dot{\rho}_t$ is the tangent vector to the curve ${\rho}$ at the point $\rho(t)$. This differential equation has a unique solution in $\rr{g}$ which verifies $g(0)=e$ and that can be formally written as
\begin{equation}\label{eq:holo1}
g(t)\;=\;\f{P}\;\expo{-\int_{\tilde{\gamma}(0)}^{\tilde{\gamma}(t)}\; \omega(\dot{\rho}_t)}
\end{equation}
where  the symbol $\f{P}$ stands for the \emph{path-ordered} product\footnote{If the Lie group $\n{G}$ has a trace ${\rm Tr}$ then the quantity ${\rm Tr}(g(t))$ with $g(t)$ as in equation \eqref{eq:holo1} is known as \emph{Wilson line variable}. More interesting is the case in which the path $t\mapsto\tilde{\gamma}(t)$ is a closed loop. In this case the quantity ${\rm Tr}(g(1))$ is called \emph{Wilson loop variable} (\cf \cite{giles-81}).}.

\medskip

\begin{remark}[Local expression for $g(t)$]{\upshape
It is useful to express the equation  \eqref{eq:holo1} locally with respect to a trivializing open cover $\{\f{U}_\alpha\}$ of the base space $X$. Let $\sigma_\alpha\in\Gamma(\f{U}_\alpha)$ be the  
{canonical section} defined by $\sigma_\alpha(x):=h_\alpha^{-1}(x,e)$ with $e\in\n{G}$ the identity. Given a path $\tilde{\gamma}_\alpha:[0,1]\to\f{U}_\alpha$ we can realize a lift   $\gamma_\alpha:[0,1]\to\pi^{-1}(\f{U}_\alpha)$ by the prescription $\gamma_\alpha(t):=\sigma_\alpha(\tilde{\gamma}_\alpha(t))g(t)$ for a suitable (smooth) map $g:[0,1]\to \n{G}$.
The condition of horizontality is fixed by the equation \eqref{eq:holo1} where $\dot{\rho}_t$ has to be replaced by 
$\dot{\sigma}_\alpha(\tilde{\gamma}_\alpha(t))$, \ie by the tangent vector to the curve ${\sigma}_\alpha(\tilde{\gamma}_\alpha(t))$. Since $\omega(\dot{\sigma}_\alpha(\tilde{\gamma}_\alpha(t)))=(\sigma_\alpha)^*\omega(\dot{\tilde{\gamma}}_\alpha(t))$ one has that
\begin{equation}\label{eq:holo1_bis}
g(t)\;=\;\f{P}\;\expo{-\int_{\tilde{\gamma}(0)}^{\tilde{\gamma}(t)}\;\s{A}_\alpha\big(\dot{\tilde{\gamma}}_\alpha(t)\big)}
\end{equation}
where we used the definition $\s{A}_\alpha:=(\sigma_\alpha)^*\omega$. Therefore locally the horizontal condition of a lift can be described in terms of the local forms $\s{A}_\alpha\in\Omega^1(\f{U}_\alpha,\rr{g})$ associated to the connection $\omega$. Moreover, one can prove that on the overlapping $\f{U}_\alpha\cap\f{U}_\beta$ the transformation rule \eqref{eq:conn_trans_rul} 
 combines in a properly  with the integral structure of \eqref{eq:holo1_bis} in such a way that it is possible to glue together (and smoothly) expressions of the form 
\eqref{eq:holo1_bis} for paths intersecting several open sets of the cover. Formula \eqref{eq:holo1_bis} has many applications in physics due to its relation with the notion of \emph{Berry phase} (see \eg \cite{bohm-mostafazadeh-koizumi-niu-zwanziger-03}).
}\hfill $\blacktriangleleft$
\end{remark}

Using the result of Theorem \ref{theo:par_transp} we can define the \emph{parallel transport} along the path $\tilde{\gamma}$ as the map
$$
{\rm PT}_{\tilde{\gamma}}\;:\; \bb{P}_{\tilde{\gamma}(0)}\;\longrightarrow\;\bb{P}_{\tilde{\gamma}(1)}\;,\qquad\quad {\rm PT}_{\tilde{\gamma}}(p)\;:=\;\gamma(1)
$$
 where $\gamma$ is the unique (horizontal)  lift of $\tilde{\gamma}$ such that $\gamma(0)=p$. The parallel transport ${\rm PT}_{\tilde{\gamma}}$ is  an isomorphism between the fibers $\bb{P}_{\tilde{\gamma}(0)}$ and $\bb{P}_{\tilde{\gamma}(1)}$. In fact this is a consequence of the 
 the relation ${\rm PT}_{\tilde{\gamma}}\circ R_g=R_g\circ {\rm PT}_{\tilde{\gamma}}$ which is valid for every $g\in\n{G}$ \cite[vol. I, Chapt. II, Proposition 3.2]{kobayashi-nomizu}. Let $\tilde{\gamma}_1\cdot \tilde{\gamma}_2$ be the \emph{concatenation} of the curves  $\tilde{\gamma}_1$ and $\tilde{\gamma}_2$ and $\tilde{\gamma}^{-1}$  the \emph{reversed} curve with respect to  $ \tilde{\gamma}$.
Then the relations ${\rm PT}_{\tilde{\gamma}_1\cdot \tilde{\gamma}_2}={\rm PT}_{\tilde{\gamma}_2}\circ {\rm PT}_{\tilde{\gamma}_1}$ and ${\rm PT}_{\tilde{\gamma}^{-1}}={\rm PT}_{\tilde{\gamma}}^{-1}$ hold 
\cite[vol. I, Chapt. II, Proposition 3.3]{kobayashi-nomizu}.

\medskip

For each $x\in X$ let ${\rm Loop}(x)$ be the \emph{loop space} at $x$, that is, the set of all ($C^1$-piecewise) curves in $X$ starting and ending at $x$. For each $\tilde{\gamma}\in {\rm Loop}(x)$ the corresponding parallel transport ${\rm PT}_{\tilde{\gamma}}$ is an isomorphism of the fiber $\bb{P}_x=\pi^{-1}(x)$. 
Let $p\in \bb{P}_x$ be a given point. Each $\tilde{\gamma}\in {\rm Loop}(x)$ determines an element $g\in\n{G}$ such ${\rm PT}_{\tilde{\gamma}}(p)=R_g(p)$. This mapping, which can be written as
 $$
 {\rm Loop}(x)\;\ni\; \tilde{\gamma}\;\longmapsto\;  g\;:=\;\f{P}\;\expo{-\oint_{\tilde{\gamma}}\; \omega(\dot{\rho}_t)}\,\in\n{G}
 $$
with the notation in \eqref{eq:holo1},
respects the group composition and defines a subgroup of $\n{G}$ 
$$
{\rm Hol}_p(\bb{P},\omega)\;:=\;\big\{g\in\n{G} \ |\ {\rm PT}_{\tilde{\gamma}}(p)=R_g(p),\ \ \tilde{\gamma}\in {\rm Loop}(x)\big\}
$$
which is called the \emph{holonomy group} of the connection $\omega$ with reference point $p\in\bb{P}$. The \emph{restricted} 
{holonomy group} is given by
$$
{\rm Hol}^0_p(\bb{P},\omega)\;:=\;\big\{g\in\n{G} \ |\ {\rm PT}_{\tilde{\gamma}}(p)=R_g(p),\ \ \tilde{\gamma}\in {\rm Loop}_0(x)\big\}
$$
where ${\rm Loop}_0(x)\subset {\rm Loop}(x)$ is the subset of closed paths which are homotopic to the constant path at $x$.
The following facts are true \cite[vol. I, Chapt. II, Proposition 4.1 and Theorem 4.2]{kobayashi-nomizu}:
\begin{proposition}\label{prop:holonI}
Let $\pi:\bb{P}\to X$ be a principal $\n{G}$-bundle and  $\omega\in \Omega^1(\bb{P},\rr{g})$ a connection 1-form. Then:
\begin{enumerate}
\item[(i)] Let $p_1,p_2\in \bb{P}$ such that $\pi(p_1)=\pi(p_2)$ and $p_1=R_g(p_2)$ for a $g\in \n{G}$. Then 
$$
{\rm Hol}_{p_1}(\bb{P},\omega)\;=\; g^{-1}\; {\rm Hol}_{p_2}(\bb{P},\omega)\; g\;=\;{\rm Ad}(g^{-1})\left[{\rm Hol}_{p_2}(\bb{P},\omega)\right]\;,
$$
namely the holonomy groups ${\rm Hol}_{p_1}(\bb{P},\omega)$ and ${\rm Hol}_{p_2}(\bb{P},\omega)$ are conjugate in $\n{G}$. The same holds true for the {restricted} 
{holonomy groups}.\vspace{1.3 mm}

\item[(ii)] Let $p_1,p_2\in \bb{P}$ such that $\pi(p_1)\neq\pi(p_2)$. If $p_1$ and $p_2$ can be joined by a horizontal curve then ${\rm Hol}_{p_1}(\bb{P},\omega)\simeq {\rm Hol}_{p_2}(\bb{P},\omega)$ and similarly for the {restricted} 
{holonomy groups}.\vspace{1.3 mm}
\end{enumerate}
Now, let assume that $X$ is path-connected and paracompact, then:
\begin{enumerate}
\item[(iii)] ${\rm Hol}_{p_1}(\bb{P},\omega)\simeq {\rm Hol}_{p_2}(\bb{P},\omega)$ for all pairs $p_1,p_2\in\bb{P}$ and the same is true for the {restricted} 
{holonomy groups}.\vspace{1.3 mm}
\item[(iv)] ${\rm Hol}_{p}(\bb{P},\omega)$ and ${\rm Hol}_{p}^0(\bb{P},\omega)$ are both Lie subgroups of $\n{G}$.
Moreover
${\rm Hol}_{p}^0(\bb{P},\omega)$ is a normal subgroups of ${\rm Hol}_{p}(\bb{P},\omega)$ and the coset group ${\rm Hol}_{p}(\bb{P},\omega)/{\rm Hol}_{p}^0(\bb{P},\omega)$ is countable.\vspace{1.3 mm}
\item[(v)] If $\omega$ is a flat connection then ${\rm Hol}_{p}^0(\bb{P},\omega)=\{e\}$ (homotopy invariance).\vspace{1.3 mm}
\end{enumerate}
\end{proposition}

\noindent
In the construction of the holonomy group one can replace the $C^1$-condition for the closed paths $\tilde{\gamma}$ with a stronger $C^k$-condition (with $k=2,3,\ldots,\infty$). As a consequence of \cite[vol. I, Chapt. II,  Theorem 7.2]{kobayashi-nomizu} all these holonomy groups coincide.

\medskip
\medskip

\noindent
{\bf Connection on vector bundles.}
We now consider the relevant case  $\n{G}=\n{GL}_{\n{F}}(m)$ 
with associated Lie algebra $\rr{g}={\rm Mat}_{\n{F}}(m)$.
Let $\bb{E}\to X$ be a  vector bundle with typical fiber $\n{F}^m$
and $\pi:\bb{F}(\bb{E})\to X$ the associated frame bundle.
A connection 1-form $\omega$ on $\bb{F}(\bb{E})$
is usually called a \emph{linear connection}. This
 is an element of $\Omega^1(\bb{F}(\bb{E}), {\rm Mat}_{\n{F}}(m))$, \ie a matrix-valued 1-form of type
$$
\omega\;=\;
\left(\begin{array}{ccc}
\omega_{11} & \cdots & \omega_{1m} \\
\vdots &  & \vdots \\
\omega_{m1} & \cdots & \omega_{mm}\end{array}\right)\;,\qquad\quad \omega_{ij}\;\in\; \Omega^1\big(\bb{F}(\bb{E})\big)\;.
$$
Let $\imath$ be the identical representation of $\n{GL}_{\n{F}}(m)$ on the vector space $\n{F}^m$ and $\Omega^k_{\rm hor}(\bb{F}(\bb{E}),\n{F}^m,\imath)$ the space of the $\imath$-equivariant horizontal $\n{F}^m$-valued $k$-forms on the frame bundle $\bb{F}(\bb{E})$. 
As shown in \cite[vol. I, Chapt. II, Example 5.2]{kobayashi-nomizu}, one has the following identification
\begin{equation}\label{eq_E-val_for}
\Omega^k_{\rm hor}(\bb{F}(\bb{E}),\n{F}^m,\imath)\;\simeq\;
\Gamma\big(\bb{E}\otimes\Omega^k(X)\big)\;\simeq\;
\Gamma\big(\bb{E}\big)\; \otimes_{\Omega^0(X)}\; \Omega^k(X)
\end{equation}
where the latter  tensor product  is the tensor product of modules over the ring $\Omega^0(X)$ of smooth  functions on $X$. The identification \eqref{eq_E-val_for} is defined by associating to each $\hat{\phi}\in \Omega^k_{\rm hor}(\bb{F}(\bb{E}),\n{F}^m,\imath)$ and each $x\in X$ a map $\phi_x:(T_xX)^{\times k}\to \bb{E}_x$ defined by 
\begin{equation}\label{eq:conn-equiv-deriv_bis}
\phi_x\big(\tilde{\rm w}_1,\ldots,\tilde{\rm w}_k\big)\;:=\;p\circ \hat{\phi}_p \big({\rm w}_1,\ldots,{\rm w}_k\big)
\end{equation}
where for each $k$-tuple $\{\tilde{\rm w}_1,\ldots,\tilde{\rm w}_k\}\subset T_xX$ one chooses a $k$-tuple $\{{\rm w}_1,\ldots,{\rm w}_k\}\subset  T_p\bb{F}(\bb{E})$, $p\in\pi^{-1}(x)$, such that $\pi_*{\rm w}_j=\tilde{\rm w}_j$. Due to  the $\imath$-equivariance of 
$\hat{\phi}$, this definition turns out to be  independent on the choice of the particular $p\in \pi^{-1}(x)={\rm Iso}(\n{F}^m,\bb{E}_x)$.
The isomorphism \eqref{eq_E-val_for}  can be written in the short form
$\Omega^k_{\rm hor}(\bb{F}(\bb{E}),\n{F}^m,\imath)\simeq \Omega^k(X,\bb{E})$ which emphasizes the fact that  elements in $\Omega^k_{\rm hor}(\bb{F}(\bb{E}),\n{F}^m,\imath)$ are differential forms on $X$ with values in the vector bundle $\bb{E}$.

\begin{definition}[Vector bundle connection]\label{def:cov_deriv}
A \emph{connection} on the vector bundle $\bb{E}\to X$
is a  differential operator
\begin{equation}\label{eq:cov_deriv1}
\nabla\;:\;\Gamma(\bb{E})\;\longrightarrow\;\Omega^{1}(X,\bb{E})\ \end{equation}
such that the \emph{Leibniz rule}
\begin{equation}\label{eq:cov_deriv2}
\nabla(s f)\;=\; \nabla(s) f\;+\;s\;\otimes\;{\rm d}f
\end{equation}
holds for all (smooth) functions $f:X\to\n{F}$  and all (smooth) sections $s:X\to\bb{E}$. The symbol ${\rm d}:\Omega^k(X)\to \Omega^{k+1}(X)$ denotes the usual  exterior derivative for forms on $X$. The image $\nabla(s)$ is called the \emph{covariant derivative} of the section $s$.
\end{definition}

This definition is justified by the fact there is a one-to-one correspondence between
 linear connection 1-forms $\omega\in\Omega^1(\bb{F}(\bb{E}), {\rm Mat}_{\n{F}}(m))$ of the frame bundle $\bb{F}(\bb{E})$ and vector bundle connections $\nabla$ in the sense of Definition \ref{def:cov_deriv}. For this consider $s\in\Gamma(\bb{E})$ and let $\hat{s}:\bb{F}(\bb{E})\to\n{F}^m$ be the associated $\imath$-equivariant map under the isomorphism \eqref{eq_E-val_for}. Then we define 
 $$
 \hat{{\rm d}}_\omega(\hat{s})\;:=\;{\rm d}\hat{s}\;+\;\omega\cdot \hat{s}
 $$
  where  $\cdot$ is the usual multiplication between the matrix $\omega$ and the column vector $\hat{s}$. Clearly  $\hat{{\rm d}}_\omega(\hat{s})\in \Omega^1(\bb{F}(\bb{E}),\n{F}^m)$. Moreover, one can check that  $\hat{{\rm d}}_\omega(\hat{s})$ is also $\imath$-equivariant and horizontal, namely  $\hat{{\rm d}}_\omega(\hat{s})\in\Omega^1_{\rm hor}(\bb{F}(\bb{E}),\n{F}^m,\imath)$. At this point the  covariant derivative
 $\nabla(s)$ of the section $s$ is simply defined as the image of $\hat{{\rm d}}_\omega(\hat{s})$ under the isomorphism \eqref{eq_E-val_for}. In other words, the differential operator
 $\hat{{\rm d}}_\omega$ associated to the connection 1-form $\omega$ defines uniquely a vector bundle connection $\nabla$
 by means of the commutative diagram
\beql{eq:diag1_bis}
\begin{diagram}
\Omega^0_{\rm hor}(\bb{F}(\bb{E}),\n{F}^m,\imath)                         &         \rTo^{\ \ \ \ \ \ \ \ \hat{{\rm d}}_\omega\ \ \ \ \ \ \ \ }        &\Omega^1_{\rm hor}(\bb{F}(\bb{E}),\n{F}^m,\imath)\\
       \dTo^{\simeq}                                                           &                &\dTo_{\simeq} \\
\Omega^0(X,\bb{E})&\rTo_{\nabla} & \Omega^1(X,\bb{E}) &\\
\end{diagram}
\eeq
where the prescription \eqref{eq:conn-equiv-deriv_bis} fixes the vertical isomorphisms.
 Finally, let us point out that
as a consequence of Theorem \ref{teo:exist_connect} every (smooth)
vector bundle $\bb{E}\to X$ over a paracompact base space $X$ posses a vector bundle connection.

 \medskip

A choice of a vector bundle connection for $\bb{E}\to X$ is generally not unique. Let us denote with $\rr{A}(\bb{E})$ the space of vector bundle connections. If $\nabla_1,\nabla_2\in \rr{A}(\bb{E})$
 then their difference is a (smooth) linear operator, namely
$(\nabla_1 - \nabla_2)(sf)=(\nabla_1 - \nabla_2)(s)f$
for all  functions $f$ on $X$ and all sections $s$ of $\bb{E}$. It follows from the \emph{fundamental lemma of differential geometry} (see \eg \cite{besse-87}) that the difference $\nabla_1 - \nabla_2$ is induced by a 1-form on $X$ with values in the endomorphism bundle ${\rm End}(\bb{E})\to X$,
namely $\nabla_1 - \nabla_2\in\Omega^1(X, {\rm End}(\bb{E}))\equiv\Gamma({\rm End}(\bb{E})\otimes \Omega^1(x))$.
Conversely, if $\nabla\in \rr{A}(\bb{E})$   and $A\in \Omega^1(X, {\rm End}(\bb{E}))$ then $\nabla+A$ is still a vector bundle connection for $\bb{E}$.
Said differently, the space $\rr{A}(\bb{E})$ is an affine space for the vector space $\Omega^1(X, {\rm End}(\bb{E}))$. This is the vector bundle version of Corollary \ref{cor:aaf_prin}.

\medskip

Let $\nabla$ be a vector bundle connection for $\bb{E}\to X$. For each $s\in\Gamma(\bb{E})$ and $x\in X$ the evaluation of $\nabla(s)$ on the point $x$ defines  a linear map
$\nabla(s)_x:T_xX\to\bb{E}_x$. For $\tilde{\rm w}\in T_xX$ we shall write
$$
\nabla_{\tilde{\rm w}}(s)\;:=\;\nabla(s)(\tilde{\rm w})\;\in\;\bb{E}_x
$$
and this is called the \emph{directional derivative} of $s$ at the point $x$ in the direction $\tilde{\rm w}$.
The following result holds true \cite[vol. I, Chapt. III, Proposition 2.8 and Proposition 7.5]{kobayashi-nomizu}:
\begin{proposition}
Let $\bb{E}\to X$ be a vector bundle with a vector bundle connection $\nabla$.
The directional derivative associated with $\nabla$ verifies the following properties:
\begin{enumerate}
\item[(i)] $\nabla_{\tilde{\rm w}+\tilde{\rm w}'}(s)=\nabla_{\tilde{\rm w}_1}(s)\;+\;\nabla_{\tilde{\rm w}'}(s)$
\item[(ii)] $\nabla_{\tilde{\rm w}}(s+s')=\nabla_{\tilde{\rm w}}(s)\;+\;\nabla_{\tilde{\rm w}}(s')$
\item[(iii)] $\nabla_{\lambda\tilde{\rm w}}(s)=\nabla_{\tilde{\rm w}}(\lambda s)=\lambda\nabla_{\tilde{\rm w}}(s)$
\item[(iv)] $\nabla_{\tilde{\rm w}}(sf)= \nabla_{\tilde{\rm w}}(s)\;f\;+\;s\; \tilde{\rm w}(f)$
\end{enumerate}
for all $\tilde{\rm w},\tilde{\rm w}'\in T_xX$, $s,s'\in \Gamma(\bb{E})$, $\lambda\in\n{F}$ and $f\in C^\infty(X)$.
Moreover, each map $\Gamma(TM)\times\Gamma(\bb{E})\to\Gamma(\bb{E})$ which verifies (i)-(iv)
defines a unique vector bundle connection on $\bb{E}$.
\end{proposition}
\noindent
We recall that a section $s\in\Gamma(\bb{E})$ is called \emph{parallel} with respect to $\nabla$ if and only if $\nabla_{\tilde{\rm w}}(s)=0$ for all tangent vectors $\tilde{\rm w}\in T_xX$ and all $x\in X$.

\medskip

The vector bundle connection $\nabla$ of 
$\bb{E}\to X$ can be expressed in function of the local trivialization $\{\f{U}_\alpha,f_\alpha\}$ and the related transition functions $\{\varphi_{\beta,\alpha}\}$ (which are the same as the associated frame bundle $\bb{F}(\bb{E})$). There is a one-to-one correspondence between (smooth) sections $s\in\Gamma(\bb{E})$
and families $\{s_\alpha\}$ of (smooth) sections $s_\alpha:\f{U}_\alpha\to\n{F}^m$ satisfying $s_\beta(x)=\varphi_{\beta,\alpha}(x)\cdot s_\alpha(x)$ for all $x\in\f{U}_\alpha\cap\f{U}_\beta$.
In fact the section $s$ and the functions $s_\alpha$ are related by $f_\alpha\circ s(x)=(x,s_\alpha(x))$ for each $x\in X$.
\begin{proposition}\label{prop_local_1to1}
Let $\nabla$ be a vector bundle connection for the vector bundle 
$\bb{E}\to X$. Let $\omega\in\Omega^1(\bb{F}(\bb{E}), {\rm Mat}_{\n{F}}(m))$ be the associated connection 1-form for the frame bundle of $\bb{E}$. For each $s\in\Gamma(\bb{E})$ let $\{s_\alpha\}$ be the family of local functions associated with the 
local trivialization $\{\f{U}_\alpha,f_\alpha\}$. The covariant derivative $\nabla(s)\in\Omega^{1}(X,\bb{E})$ corresponds uniquely to the family of 1-forms $\nabla(s_\alpha)\in\Omega^{1}(\f{U}_\alpha,\n{F}^m)$ \footnote{Notice that $\Omega^{k}(\f{U}_\alpha,\n{F}^m)$ is  used as short notation for $\Omega^{k}(\f{U}_\alpha,\f{U}_\alpha\times\n{F}^m)\simeq \Omega^{k}(\f{U}_\alpha)\otimes\n{F}^m$ for product bundles.} given by
$$
\nabla(s_\alpha)\;:=\;{\rm d}s_\alpha\;+\;\s{A}_\alpha\;\cdot\;s_\alpha
$$
where the $\s{A}_\alpha\in\Omega^{1}(\f{U}_\alpha,{\rm Mat}_{\n{F}}(m))$ are the local 1-forms which define $\omega$
according to Theorem \ref{theo:loc_connect}.
\end{proposition}

\medskip

Sometimes it is convenient to extend the definition of $\nabla$
given by equations \eqref{eq:cov_deriv1} and \eqref{eq:cov_deriv2}
 to arbitrary $\bb{E}$-valued forms, thus regarding it as a differential operator on  the full 
 exterior algebra $\Omega^\bullet(X,\bb{E}):=\Gamma\big(\bb{E}\otimes\Omega^\bullet(X)\big)$. In fact, given a vector bundle connection $\nabla$ satisfying \eqref{eq:cov_deriv1} and \eqref{eq:cov_deriv2}, there exists a unique extension 
 $$
 {\bf  d}^\nabla\;:\;\Omega^k(X,\bb{E})\;\longrightarrow\; \Omega^{k+1}(X,\bb{E})
 $$
 such that
$$
 {\bf  d}^\nabla(s\wedge\theta)\;=\;{\bf  d}^\nabla(s)\;\wedge\;\theta\;+\;(-1)^{k}\; s\;\wedge\;{\rm d}\theta
$$
where $s\in \Omega^k(X,\bb{E})$
 is a $\bb{E}$-valued $k$-form and $\theta\in\Omega^j(X)$ is an ordinary $j$-form (see \eg \cite[Appendix C]{milnor-stasheff-74} or \cite[Section 12.4]{taubes-11}). 
 In other words, ${\bf  d}^\nabla$ is a derivation on the sheaf of graded modules $\Omega^\bullet(X,\bb{E})$. Although ${\rm d}^2=0$, this is generally not the case for ${\bf  d}^\nabla$. The composition
\begin{equation}\label{eq:VB_connAX}
F^\nabla\;:=\;{\bf d}^\nabla\;\circ\;{\bf d}^\nabla:\Gamma(\bb{E})\:\longrightarrow\Omega^2(X, \bb{E})
\end{equation}
verifies $F^\nabla(s\wedge\theta)=F^\nabla(s)\wedge\theta$ and in particular it 
turns out to be  $C^\infty(\bb{E})$-linear \cite[Appendix C, Lemma 5]{milnor-stasheff-74}, \ie
$$
F^\nabla(f\;s)\;=\;f\;F^\nabla(s)\;,\qquad\quad f\in C^\infty(\bb{E})\;,\ \ s\in\Gamma(\bb{E})\;.
$$
This implies that  $F^\nabla$ is a smooth section of the vector bundle ${\rm End}(\bb{E},\Omega^2(X, \bb{E}))\simeq {\rm End}(\bb{E})\otimes_{\Omega^0(X)} \Omega^2(X)$. Hence one has 
$$
 F^\nabla\;\in\;\Gamma\big({\rm End}(\bb{E},\Omega^2(X, \bb{E}))\big)\;\simeq\; \Gamma\big({\rm End}(\bb{E})\big)\; \otimes_{\Omega^0(X)}\; \Omega^2(X)=:\Gamma({\rm End}(\bb{E})\otimes\Omega^2(X))\;.
 $$
 The operator $F^\nabla$ is known as (vector bundle) \emph{curvature}.  One has that
  ${\bf  d}^\nabla\circ {\bf  d}^\nabla=0$ if and only if the 
vector bundle connection $\nabla$ is flat.

\section{Chern-Weil theory}\label{sect:ChernÐWeil}

\noindent
{\bf The curvature form.}
Let $\omega\in\rr{A}(\bb{P})$ be a connection 1-form for the principal $\n{G}$-bundle $\pi: \bb{P}\to X$. The wedge product
$\omega\wedge \omega$ defines an element in $\Omega^2(\bb{P},\rr{g}\otimes\rr{g})$ and in order to obtain 
a two form with value in $\rr{g}$ we can \virg{contract} the tensor product with the help of the Lie bracket
$$
[\cdot\;,\;\cdot]\;:\;\rr{g}\otimes\rr{g}\;\longrightarrow\; \rr{g}
$$
defined by $\xi_1\otimes\xi_2\mapsto[\xi_1,\xi_2]$ for all 
$\xi_1,\xi_2\in\rr{g}$. The \emph{cuevature form} $F_\omega\in \Omega^2(\bb{P},\rr{g})$ associated to $\omega$ is defined by the \emph{structural equation} 
\begin{equation}\label{eq:struct_eq}
F_\omega\;:=\;{\rm d}\omega\;+\;\frac{1}{2}\; [\omega\wedge \omega]\;.
\end{equation}
The main properties of the curvature are listed below (see \cite[vol. I, Chapt. II, Section 5]{kobayashi-nomizu}):

\begin{proposition}\label{prop:curv_PB}
Let $\pi:\bb{P}\to X$ be a principal $\n{G}$-bundle and  $\omega\in \rr{A}(\bb{P})$ a connection 1-form and $F_\omega\in \Omega^2(\bb{P},\rr{g})$ the related curvature. Then:
\vspace{1mm}
\begin{enumerate}
\item[(i)] $F_\omega\in \Omega^2_{\rm hor}(\bb{P},\rr{g},{\rm Ad})$, namely $F_\omega$ vanishes on horizontal vector fields and 
$R_g^*F_\omega={\rm Ad}(g^{-1})\circ F_\omega=g^{-1}\cdot F_\omega \cdot g$;\vspace{1mm}
\item[(ii)] The connection $\omega$ is flat if and only if $F_\omega=0$;
 \vspace{1mm}
\item[(iii)] $F_\omega$ verifies the \emph{Bianchi identity}
$$
{\rm d}F_\omega\;=\;[F_\omega\wedge \omega]
$$
and ${\rm d}F_\omega$ vanishes on triplets of horizontal vectors.
\end{enumerate}
\end{proposition}

\medskip

\noindent
As for the connection, also the curvature admits a local expression.

\begin{theorem}\label{theo:loc_curvatur}
Let $\pi:\bb{P}\to X$ be a principal $\n{G}$-bundle with  local trivializations $\{\f{U}_\alpha,h_\alpha\}$ and transition functions $\{\varphi_{\beta,\alpha}\}$. Let $\omega\in \rr{A}(\bb{P})$ be a connection 1-form and $F_\omega\in \Omega^2(\bb{P},\rr{g})$ the related curvature. Let $\{\s{A}_\alpha\in\Omega^1(\f{U}_\alpha,\rr{g})\}$ be the family of local 1-forms which provide a local description of $\omega$ in the sense of Theorem \ref{theo:loc_connect}. Then the curvature $F_\omega$ is described by the collection of local 2-forms
$\{\s{F}_\alpha\in\Omega^2(\f{U}_\alpha,\rr{g})\}$ given by
$$
\s{F}_\alpha\;:=\;{\rm d}\s{A}_\alpha\;+\;\frac{1}{2}[\s{A}_\alpha\wedge\s{A}_\alpha]\;.
$$
\end{theorem}

\medskip
\medskip

In the case of a vector bundle $\bb{E}\to X$,  a connection 1-form $\omega$ on the associated frame bundle
$\bb{F}(\bb{E})\to X$ induces a vector bundle connection 
$
\nabla: \Gamma(\bb{E})\to\Omega^{1}(X,\bb{E})
$
according to Definition \ref{def:cov_deriv}. Similarly, the curvature $F_\omega$ is identified to the  unique 2-form
$
F^\nabla\in\Omega^{2}(X,{\rm End}(\bb{E}))
$ given by formula
\eqref{eq:VB_connAX}. Let 
$\tilde{\rm w},\tilde{\rm w}'\in TX$ be vector fields and $s\in \Gamma(\bb{E})$ a section, then
\begin{equation}\label{eq:curv_VB}
F^\nabla(\tilde{\rm w},\tilde{\rm w}')(s)\;=\;\Big(\nabla_{\tilde{\rm w}}\; \nabla_{\tilde{\rm w}'}\;-\;
\nabla_{\tilde{\rm w}'}\; \nabla_{\tilde{\rm w}}\;-\;\nabla_{[{\tilde{\rm w}}, {\tilde{\rm w}'}]}\Big)(s)\;
\end{equation}
where $[\cdot,\cdot]$ is the Lie bracket of vector fields \cite[Section 12.6]{taubes-11}.

\medskip
\medskip

\noindent
{\bf The Chern-Weil homomorphism.}
Given a principal $\n{G}$-bundle $\pi:\bb{P}\to X$ with a connection 1-form $\omega$ the \emph{Chern-Weil homomorphism} provides a way to associate certain closed differential forms on $X$. The corresponding classes in the de Rham cohomology $H^\bullet_{\rm d.R.}(X)$
do not depend on $\omega$ but only on the isomorphism class of the $\n{G}$-bundle $\bb{P}$. For an exhaustive treatment of this topic we refer to \cite[vol. II, Chapt. XII]{kobayashi-nomizu} or  \cite[Appendix C]{milnor-stasheff-74}.

\medskip

Let $V$ be a finite dimensional vector space and $\s{S}^k(V^*)$, $k\geqslant 1$ the space of  $k$-linear functions $P:V\times\ldots\times V\to \R$ which are \emph{symmetric} in the sense that $P(v_1,\ldots,v_k)=P(v_{\sigma(1)},\ldots,v_{\sigma(k)})$ for all $v_1,\ldots,v_k\in V$ and every permutation $\sigma$ of the indices $1,\ldots,k$.
One can also define a product $\circ:\s{S}^k(V^*)\times \s{S}^l(V^*)\to \s{S}^{k+l}(V^*)$ given by
$$
\big(P\;\circ\; Q\big)(v_1,\ldots,v_{k+l})\;:=\;\frac{1}{(k+l)!}\; \sum_{\sigma}P(v_{\sigma(1)},\ldots,v_{\sigma(k)})\, Q(v_{\sigma(k+1)},\ldots,v_{\sigma(k+l)})\;
$$
where $P\in \s{S}^k(V^*)$, $Q\in \s{S}^l(V^*)$
and $\sigma$ runs over all permutations of $1,\ldots,k+l$. It is  useful to introduce also $\s{S}^0(V^*):=\R$ and the graded space $\s{S}^\bullet(V^*):=\bigoplus_{k=0}^{+\infty} \s{S}^k(V^*)$. One can verify 
that $\s{S}^\bullet(V^*)$ is a ring with unit $1\in\R=\s{S}^0(V^*)$. This ring have a nice representation.
Let $n$ be the dimension of $V$ and $\{{\rm e}_1,\ldots,{\rm e}_n\}$ a basis. Consider the \emph{polynomial ring} $\R[x_1,\ldots,x_n]$ made by 
polynomials with real coefficients of every degree in the variables $x_1,\ldots,x_n$. Let $\R[x_1,\ldots,x_n]^k\subset\R[x_1,\ldots,x_n]$ be the subset of homogeneous polynomials of degree $k$. The map $v=\sum_{i}^nx_i{\rm e}_i\mapsto(x_1,\ldots,x_n)$ induces a map
$\s{S}^k(V^*)\ni P\mapsto \tilde{P}\in\R[x_1,\ldots,x_n]^k$ given by $\tilde{P}(x_1,\ldots,x_n):=P(v,\ldots,v)$. This map turns out to be an isomorphism of vector spaces $\s{S}^k(V^*)\simeq\R[x_1,\ldots,x_n]^k$ which extends to an isomorphism of rings $\s{S}^\bullet(V^*)\simeq\R[x_1,\ldots,x_n]$ as proved in \cite[vol. II, Chapt. XII, Proposition 2.1]{kobayashi-nomizu}.

\medskip

Let $\rr{g}$ be the Lie algebra of a Lie group $\n{G}$. The adjoint representation of $\n{G}$ on $\rr{g}$ induces an action of $\n{G}$ on  $\s{S}^k(\rr{g}^*)$ given by
$$
(g\cdot P)(\xi_1,\ldots,\xi_k)\;=\;P\left(g\cdot\xi_1\cdot g^{-1},\ldots,g\cdot\xi_k\cdot g^{-1}\right)
$$
for all $k\geqslant0$, $g\in\n{G}$ and $\xi_1,\ldots,\xi_k\in\rr{g}$. We say thet $P\in \s{S}^k(\rr{g}^*)$ is \emph{invariant} if $g\cdot P=P$ for all $g\in\n{G}$.
The (graded) subring of the invariant elements of $\s{S}^\bullet(\rr{g}^*)$ will be denoted by $\s{I}^\bullet(\rr{g}^*)$. We refer to the elements in 
$\s{I}^\bullet(\rr{g}^*)$ as \emph{invariant polynomials}.

\medskip

Now, let $\pi:\bb{P}\to X$ be a principal $\n{G}$-bundle with a connection 1-form $\omega$ and associated curvature $F_\omega\in\Omega^2(\bb{P},\rr{g})$ given by the structural equation \eqref{eq:struct_eq}.
For each $k\geqslant 1$ one has $F_\omega^k=F_\omega\wedge\ldots\wedge F_\omega\in \Omega^2(\bb{P},\rr{g}\otimes\ldots\otimes\rr{g})$. Every invariant polynomial $P\in \s{I}^k(\rr{g}^*)$
identifies a unique  linear map $P:\rr{g}\otimes\ldots\otimes\rr{g}\to\R$ (still denoted with the same symbol) just by the identification $P(\xi_1,\ldots,\xi_k)=P(\xi_1\otimes\ldots\otimes\xi_k)$. Then $P(F_\omega^k)$ is a $2k$-form, namely an element of $\Omega^{2k}(\bb{P})$. Since $F_\omega$ is ${\rm Ad}$-equivariant and horizontal and $P$ is invariant, it follows that $P(F_\omega^k)$ is invariant and horizontal, namely it is a \emph{basic} $2k$-form. Hence by Proposition \ref{prep:basic_conn} (ii) there is a unique $2k$-form  $\hat{P}(F_\omega^k)\in\Omega^{2k}(X)$  which pulls back to $P(F_\omega^k)$ by $\pi^*:\Omega^{2k}(X)\to \Omega^{2k}(\bb{P})$. This is  called the \emph{characteristic form} corresponding to $P$ and $\omega$. The most important facts about the
characteristic forms are summarized below:
\begin{proposition}
Let $\pi:\bb{P}\to X$ be a principal $\n{G}$-bundle, $\omega\in \Omega^1(\bb{P},\rr{g})$ a connection 1-form and $P\in \s{I}^k(\rr{g}^*)$, $Q\in \s{I}^l(\rr{g}^*)$ invariant polynomials. Then:
\begin{enumerate}
\item[(i)] $\hat{P}(F_\omega^k)\in\Omega^{2k}(X)$
is a closed form, that is ${\rm d}\hat{P}(F_\omega^k)=0$
.\vspace{1.3 mm}

\item[(ii)] $\widehat{(P\circ Q)}(F_\omega^k)=\hat{P}(F_\omega^k)\wedge\hat{Q}(F_\omega^k)$.\vspace{1.3 mm}

\item[(iii)] Let $(\varphi,\hat{\varphi})$ be a bundle map between $\bb{P}\to X$ and $\bb{P}'\to X'$ and $\omega':=\varphi^*\omega$ the induced connection on $\bb{P}'$. Then $\hat{P}(F_{\omega'}^k)=\hat{\varphi}^*\hat{P}(F_{\omega}^k)$. In particular \emph{gauge equivalent} connections  have the same characteristic forms (in this case $X=X'$, $\hat{\varphi}={\rm Id}_X$ and $\varphi$ reduces to an isomorphism).
\vspace{1.3 mm}
\end{enumerate}
\end{proposition}

\medskip

Let us denote with $cw(\bb{P},P):=[\hat{P}(F_{\omega'}^k)]\in H^{2k}_{\rm d.R.}(X)$  the de Rham class of the $2k$-form $\hat{P}(F_{\omega'}^k)$. The mapping
\begin{equation}
cw(\bb{P},\cdot)\;:\; \s{I}^k(\rr{g}^*)\;\longrightarrow H^{2k}_{\rm d.R.}(X)
\end{equation}
is called the \emph{Chern-Weil homomorphism} and 
$cw(\bb{P},P)$ is the \emph{characteristic class} of $\bb{P}$ corresponding to $P$. One has the following result \cite[vol. II, Chapt. XII, Theorem 1.1]{kobayashi-nomizu}:

\begin{theorem}\label{thero:Cclass0}
Let $\pi:\bb{P}\to X$ be a principal $\n{G}$-bundle. Then:
\begin{enumerate}
\item[(i)] For every $P\in \s{I}^k(\rr{g}^*)$ the cohomology class $cw(\bb{P},P)\in H^{2k}_{\rm d.R.}(X)$ does not depend on the choice of the connection $\omega$.\vspace{1.3 mm}

\item[(ii)] $cw(\bb{P},\cdot)$ is a homomorphism between the rings $(\s{I}^\bullet(\rr{g}^*),\circ)$ and $(H^{\bullet}_{\rm d.R.}(X),\wedge)$
.\vspace{1.3 mm}

\item[(iii)] Let $(\varphi,\hat{\varphi})$ be a bundle map between $\bb{P}\to X$ and $\bb{P}'\to X'$ Then $cw(\bb{P}',\cdot)=\hat{\varphi}^*cw(\bb{P},\cdot)$.
In particular for isomorphic principal $\n{G}$-bundles one has $cw(\bb{P}',\cdot)=cw(\bb{P},\cdot)$ (in this case $X=X'$, $\hat{\varphi}={\rm Id}_X$ and $\varphi$ reduces to an isomorphism).
\vspace{1.3 mm}
\end{enumerate}
\end{theorem}

\medskip
\medskip

\noindent
{\bf Chern classes.}
A \emph{characteristic class} $c$ (with coefficient in $\n{F}=\Z,\Z_2,\R,\C,\ldots$) for  principal $\n{G}$-bundles associates to every
principal $\n{G}$-bundle $\pi:\bb{P}\to X$  a cohomology class $c(\bb{P})\in H^\bullet(X,\n{F})$ such that 
$c(\bb{P})=c(\bb{P}')$ if $\bb{P}\simeq\bb{P}'$ and $f^*c(\bb{P})=c(f^*\bb{P})$ for each map $f:X\to X'$ between base spaces. The set of characteristic classes for principal $\n{G}$-bundles is a ring denoted with $H^\bullet_{\n{G}}(\n{F})$. The de Rham isomorphism \cite[Chapter II, Theorem 8.9]{bott-tu-82}
$$
H^{\bullet}_{\rm d.R.}(X)\;\stackrel{\imath}{\simeq}\; H^\bullet(X,\n{R})
$$
which is certainly true if $X$ is a smooth manifold and Theorem \ref{thero:Cclass0} provide the following:
\begin{corollary}
For every $P\in\s{I}^k(\rr{g}^*)$ the element $c_P(\cdot):= \imath\circ cw(\cdot, P)$ is a characteristic classes for principal $\n{G}$-bundles.
Moreover the mapping $\s{I}^k(\rr{g}^*)\ni P\mapsto c_P\in H^\bullet_{\n{G}}(\n{R})$ is a ring homomorphism.
\end{corollary}

\medskip

Let us focus on the case $\n{G}=\n{U}(m)$ with Lie algebra $\rr{u}(m)$ given by the $m\times m$ anti-hermitian matrices. The (real) polynomials $C_k$  given by
$$
{\rm det}\left(t\n{1}\;-\;\frac{1}{\ii\; 2\pi}\xi\right)\;=\;\sum_{k=0}^mC_k(\xi,\ldots,\xi)\; t^{m-k}\;,\qquad\quad\qquad t\in\R\;,\;\ \  \xi\in\rr{u}(m)
$$
are homogeneous of degree $k$ and $\rm{Ad}$-invariant. For a given principal $\n{G}$-bundle $\pi:\bb{P}\to X$
the characteristic class $c_k(\bb{P}):= \imath\circ cw(\bb{P}, C_k)\in H^{2k}(X,\n{R})$ is called the \emph{$k$-th Chern class} of $\bb{P}$. Without any risk of confusion we can also think to $c_k(\bb{P})$ in terms of its representative in the de Rham complex $H^{\bullet}_{\rm d.R.}(X)$.

\medskip

Let $\bb{E}\to X$ be a rank $m$ hermitian vector bundle and $\bb{F}(\bb{E})\to X$ the associated principal $\n{U}(m)$-bundle. We define
$$
c_k(\bb{E})\;:=\;c_k\big(\bb{F}(\bb{E})\big)
$$ 
to be the $k$-th Chern class of the vector bundle $\bb{E}$.

\medskip
\medskip

\noindent
{\bf Relation between topological and differential  Chern classes.}
Hermitian vector bundles (resp. principal $\n{U}(m)$-bundle) can be obtained, up to isomorphisms, as pullbacks of a \emph{universal} classifying vector bundle (resp. principal $\n{U}(m)$-bundle).
A model for this universal object is provided by the \emph{tautological} vector bundle (resp. principal $\n{U}(m)$-bundle) over the \emph{Grassmann manifold}  
$$
G_m(\C^\infty)\;:=\;\bigcup_{n=m}^{\infty}\;G_m(\C^n)\;,
$$
where,
for each pair $m\leqslant n$, $G_m(\C^n)\simeq\n{U}(n)/\big(\n{U}(m)\times \n{U}(n-m)\big)$ is the set of $m$-dimensional (complex) subspaces of $\C^n$. 
Let $\pi:\bb{T}_m^\infty\to G_m(\C^\infty)$ be the universal classifying bundle, meaning that each rank $m$ hermitian vector bundle $\bb{E}\to X$ (resp. principal $\n{U}(m)$-bundle $\bb{P}\to X$) can be realized, up to isomorphisms, as the pullback of  $\bb{T}_m^\infty$ with respect to a \emph{classifying map} $\varphi : X \to G_m(\C^\infty)$, that is $\bb{E}\simeq \varphi^\ast \bb{T}_m^\infty$ (resp. $\bb{P}\simeq \varphi^\ast \bb{T}_m^\infty$).
Since pullbacks of homotopic maps yield isomorphic  bundles (\emph{homotopy property}), the classification   only depends on the homotopy class  of $\varphi$.  This leads to the fundamental result \cite{steenrod-51,milnor-stasheff-74}
\beql{eq:class_compl_VB1}
{\rm Vec}^m_\C(X)\; \simeq\;{\rm Prin}_{\n{U}(m)}(X)\;\simeq\;[X , G_m(\C^\infty)]
\eeq
where in the right-hand side  there is the set of the equivalence classes of homotopic maps between $X$ and $G_m(\C^\infty)$.
An important result in the theory of fiber bundles is the computation of 
the cohomology ring of the {Grassmann manifold}  \cite[Theorem 14.5]{milnor-stasheff-74}:
\begin{equation}\label{eq:univ_chern_class}
H^\bullet\big(G_m(\C^\infty),\Z\big)\;\simeq\;\Z[\rr{c}_1,\ldots,\rr{c}_m]
\end{equation}
is the ring of polynomials with integer coefficients and $m$ generators $\rr{c}_k\in H^{2k}\big(G_m(\C^\infty),\Z\big)$. 
These generators $\rr{c}_k$ are called \emph{universal} Chern classes and there are no polynomial relationships between them.
The Chern classes of a general Hermitian vector bundle $\bb{E}$  (res. principal $\n{U}(m)$-bundle $\bb{P}$)  are constructed as follows: let $\varphi\in[X,G_m(\C^\infty)]$ be the map which represents $\bb{E}$ (resp. $\bb{P}$) according to \eqref{eq:class_compl_VB1}, then 
$\varphi^\ast:H^k(G_m(\C^\infty),\Z)\to H^k(X,\Z)$ is a homomorphism of cohomology groups for all $k$.
The \emph{$k$-th} \emph{topological} Chern class of $\bb{P}$  is by definition
$$
\tilde{c}_k(\bb{P})\;:=\;\varphi^\ast(\rr{c}_k)\;\in\;H^{2k}(X,\Z)\qquad\quad k=1,2,3,\ldots\;.
$$
and similarly for $\bb{E}$.
Since the homomorphism $\varphi^\ast$ only depends on the homotopy class of $\varphi$,  isomorphic  bundles possess
 the same family of topological Chern classes. Moreover, the $\tilde{c}_k$'s are functorial
 hence they define \emph{integer}
\emph{characteristic class} $\tilde{c}_k\in H^\bullet_{\n{U}(m)}(\n{Z})$.

\medskip

The inclusion $\Z\hookrightarrow\R$ induces a homomorphism 
$$
H^{k}(X,\Z)\;\stackrel{r}{\longrightarrow}\; H^{k}(X,\R)
$$
which is injective (\ie a monomorphism) if and only if $X$ has no torsion (\cf equation \eqref{eq:cohom_coef2}). The  \emph{differential} Chern classes $c_k(\bb{P})\in H^{2k}_{\rm d.R.}(X)\simeq H^{2k}(X,\n{R})$
are related to the topological Chern classes $\tilde{c}_k(\bb{P})\in H^{2k}(X,\Z)$ by the relation
$c_k(\bb{P})=r\big(\tilde{c}_k(\bb{P})\big)$. This shows that in the passage from the topological to the differential Chern classes possible information about torsion is lost.

\section{A reminder about spectral sequences}\label{spectral}

In the computation of the equivariant cohomology, a basic tool is the 
\emph{Leray-Serre-(Atiyah-Hirzebruch) spectral sequence} associated to the canonical fibration
\begin{equation}\label{eq:can_fib}
X\;\hookrightarrow\;X_{\sim\tau}\;\rightarrow\;\R P^\infty
\end{equation}
of the homotopy quotient \eqref{eq:homot_quot}.

\medskip

\noindent
{\bf Cohomology spectral sequence.}
We start by recalling some fundamental facts about the notion of spectral sequence. For a more accurate treatment we refer to \cite[Chapter 9]{spanier-66} or \cite[Chapter III]{fomenko-fuchs-gutenmacher-86} or \cite[Chapter 9]{davis-kirk-01}.

\medskip

A \emph{bigraded} abelian group (resp. $\s{R}$-module) is  a collection of the type $E=\{E^{p,q}\}_{p,q\in\Z}$ where $E^{p,q}$
are abelian groups (resp. $\s{R}$-modules). Sometimes 
 the notation $E=\bigoplus_{p,q\in\Z}E^{p,q}$ is also used.
 A \emph{differential} 
 on the bigraded abelian group (resp. $\s{R}$-module) $E$ is a homogeneous morphism $\delta:E\to E$ such that $\delta\circ\delta =0$. We say that $\delta_r$ is a differential of cohomological type if it has bidegree $(r,1-r)$ for some positive integer $r>0$, meaning that  
$\delta_r:E^{p,q}\to E^{p+r,q-r+1}$. 
A cohomology spectral sequence is a collection $\{E_r=\bigoplus_{p,q\in\Z}E_r^{p,q},\delta_r\}_{r>0}$ of {bigraded} abelian groups (resp. $\s{R}$-modules), endowed with a cohomological differential with bidegree $(r,1-r)$,
such that 
$$
E_{r+1}^{p,q}\;\simeq\;H^{p,q}(E_r,\delta_r)\;:=\;\frac{{\rm Ker}\left(\delta_r:E_{r}^{p,q}\to E_{r}^{p+r,q-r+1}\right)}{{\rm Im}
\left(\delta_r:E_{r}^{p-r,q+r-1}\to E_{r}^{p,q}\right)}\;.
$$
Intuitively we think of spectral sequences as \emph{pages} of abelian groups (resp. $\s{R}$-modules) ($E_r$ is called the $r$-th page of the
spectral sequence) such that the $(r + 1)$-th page is isomorphic to the homology of the $r$-th page.
Because the homology of a module is  a quotient of a submodule we would
expect that the modules on the $(r + 1)$-th page to be \virg{smaller} than the modules on the $r$-th page.
If we fix a grid position $(p, q)\in \Z^2$ and look at the sequence of modules $E^{p,q}_r$ as $r$ increases, we 
hope that either $E^{p,q}_r$ eventually becomes zero, or \emph{stabilizes} in the sense that $E^{p,q}_r=E^{p,q}_{r+1}$ for all $r$ greater than some $r_0$. If this is the case we write  $E^{p,q}_\infty=E^{p,q}_r$ for $r>r_0$.
This leads us to wanting to define in certain situations a  \virg{limit} page, called the $E_\infty$ page. To explain this concept  more precisely
let us drop the bigrading  for a moment on our modules $E_r$. Let $Z_1:={\rm Ker}(\delta_1)$ and $B_1:={\rm Im}(\delta_1)$
 be the cycles and boundaries so that $E_2\simeq Z_1/B_1$. By construction $B_1\subseteq Z_1\subseteq E_1$ are submodules. It is possible to show that to a spectral sequence $\{E_r,\delta_r\}_{r>0}$ is associated a 
tower of submodules
 \begin{equation}\label{eq:exa_seq1}
 0\;\subseteq\;B_1\;\subseteq\;B_2\subseteq\;\ldots\; \;\subseteq\;B_r\;\;\subseteq\;\ldots\;\subseteq Z_r\;\subseteq\;\ldots \;\;\subseteq Z_2\;\subseteq Z_1\subseteq\;E_1
\end{equation}
 such that $E_{r+1}\simeq Z_r/B_r$ and the differential $\delta_{r+1}:E_{r+1}\to E_{r+1}$ induces maps $\delta_{r+1}:Z_r/B_r\to Z_r/B_r$
such that
$$
{\rm Ker}(\delta_{r+1})\;=\;{Z_{r+1}}/{B_r}\;,\qquad\quad {\rm Im}(\delta_{r+1})\;=\; {B_{r+1}}/{B_r}\;.
$$
Notice that the above relations lead to an exact sequence
$$
0\;\longrightarrow\;{Z_{r+1}}/{B_r}\;\longrightarrow\;{Z_{r}}/{B_r}\;\stackrel{\delta_{r}}{\longrightarrow}\;{B_{r+1}}/{B_r}\;\longrightarrow\;0
$$
which imply the set of isomorphisms
\begin{equation}\label{eq:exa_seq2}
Z_{r}/{Z_{r+1}}\;\simeq\; B_{r+1}/{B_r}\;.
\end{equation}
We say that $Z_r$ is the set of elements that \virg{survived} to the $r$-th page, and $B_r$ is the set of
elements that are boundaries by the $r$-th page. Seeing as how the $B_r$'s form an increasing tower
and the $Z_r$'s form a decreasing tower one defines $Z_\infty: =\bigcap_r Z_r$ (set of elements
that  \virg{survive forever}) and $B_\infty: =\bigcup_r B_r$ (the set of elements that \virg{eventually bound}) and the  limiting page of the spectral sequence is given by $E_\infty: = Z_\infty/B_\infty$.

\medskip

The important aspect of the above construction is that it can be reversed. More precisely, it can be shown that given just a tower as in \eqref{eq:exa_seq1} and isomorphisms as in \eqref{eq:exa_seq2}, one can
work backwards and construct a spectral sequence. Let $A$ be an abelian group (resp. $\s{R}$-module). A \emph{(cohomology) filtration} for $A$  is an increasing tower
 \begin{equation}\label{eq:exa_seq3}
 0\;\subseteq\;\ldots\;\subseteq\;F^p\;\subseteq\;\ldots\;\subseteq\;F^1\subseteq\;F^0\;\subseteq\;F^{-1}\;\subseteq\;\ldots\;\subseteq F^{-p}\;\subseteq\;\ldots \;\;\subseteq A
\end{equation}
of submodules $A$. The filtration is \emph{convergent} if the union of the $F^p$'s is $A$ and their intersection is $0$. 
In the case $A=\bigoplus_{q\in\Z} A^q$  is a graded module, then the filtration is assumed to \emph{preserve} the grading, \ie
$F^p\cap A^q\subset F^{p-1}\cap A^q$ and this leads to a  bigraded filtration just by setting
$F^{p,q}:=  F^{p}\cap A^{p+q}$. Given a cohomology filtration ${F^p}$ of an $\s{R}$-module $A$ we can realize a graded $\s{R}$-module denoted by $Gr(A,F):=\bigoplus_{p\in\Z} Gr^p(A,F)$ by
$$
Gr^p(A,F)\;:=\;F^p/F^{p+1}\;.
$$
If $A$ is graded the associated module is automatically bigraded $Gr(A,F):=\bigoplus_{p,q\in\Z} Gr^{p,q}(A,F)$ by
$$
Gr^{p,q}(A,F)\;:=\;\big(F^{p}\cap A^{p+q}\big)/\big(F^{p+1}\cap A^{p+q}\big)\;.
$$
Now suppose that our filtered, graded module $(A,F)$ also had a differential $\delta$ \emph{compatible} with the filtration in the sense that $\delta:F^p\cap A^q\to F^p\cap A^{q+1}$. 
Then $\delta$ descends to a map on the associated bigraded module $\delta:Gr^{p,q}(A,F)\to Gr^{p,q+1}(A,F)$ such that $\delta\circ\delta=0$.

\medskip

Given a bigraded cohomology spectral sequence  $\{E_r=\bigoplus_{p,q\in\Z}E_r^{p,q},\delta_r\}_{r>0}$
and a graded $\s{R}$-module $A=\bigoplus_{q\in\Z} A^q$ we say that the spectral sequence \emph{converges} to $A$ and write
$$
E_2^{p,q}\;\Rightarrow A^{p+q}
$$
if: (i) for each $(p, q)$ there exists an $r_0$ so that $\delta_r:E_r^{p-r,q+r-1}\to E_r^{p,q}$ is zero for all $r\geqslant r_0$ and in particular 
this implies that there is an injection $E_{r+1}^{p,q}\hookrightarrow E_{r}^{p,q}$ for all $r\geqslant r_0$; (ii) there is a convergent
(cohomology) filtration $F$ for $A$ so that  the limit $E^{p,q}_\infty:=\bigcap_{r\geqslant r_0} E^{p,q}_r$ is isomorphic to the  associated graded module $Gr^{p,q}(A,F)$.

\medskip

We can now state the fundamental theorem for the construction of spectral sequences:

\begin{theorem}[Cohomology spectral sequences]
Let $(A=\bigoplus_{q\in\Z} A^q,\delta)$  be a cochain complex of abelian groups (resp. $\s{R}$-modules) and $F$ a
(cohomology) filtration for $A$ such that the filtration respects the grading and the differential $\delta$ is compatible with the filtration. Then there is a  cohomology spectral sequence  $\{E_r=\bigoplus_{p,q\in\Z}E_r^{p,q},\delta_r\}_{r>0}$ such that $E_0:=Gr(A,F)$ and 
$$
E_1^{p,q}\;=\;H^{p+q}\Big(F^p/F^{p+1},\delta\Big)\;:=\;\frac{{\rm Ker}\left(\delta:Gr^{p,q}(A,F)\to Gr^{p,q+1}(A,F)\right)}{{\rm Im}
\left(\delta:Gr^{p,q-1}(A,F)\to Gr^{p,q}(A,F)\right)}\;. 
$$
If the filtration is convergent then
$$
E_2^{p,q}\;\Rightarrow H^{p+q}(A,\delta)\;:=\; \frac{{\rm Ker}\left(\delta:A^{p+q}\to A^{p+q+1}\right)}{{\rm Im}
\left(\delta:A^{p+q-1}\to A^{p+q}\right)}
$$
namely the spectral sequence converges to cohomology of $H^\bullet(A,\delta)$ of the cochain complex $(A,\delta)$. More in detail, this means that
\begin{equation}
E_\infty^{p,q}\;\simeq\;\;\frac{{\rm Ker}\Big(H^{p+q}(A,\delta)\to H^{p+q}(F^{p-1}\cap A,\delta)\Big)}{{\rm Ker}
\Big(H^{p+q}(A,\delta)\to H^{p+q}(F^{p}\cap A,\delta)\Big)}\;
\end{equation}
where the maps are induced by the inclusions $F^{p}\cap A\subset A$.
\end{theorem}

\noindent
For the proof of this theorem see, fore instance, \cite[Chapter 9, Section 1]{spanier-66}.

\medskip

\noindent
{\bf Spectral sequences associated to fibrations.}
Let 
\begin{equation}
F\;\hookrightarrow\;M\;\stackrel{f}{\rightarrow}\;B
\end{equation}
be a (Serre) fibration with \emph{fiber} $F$ and base space $B$ which is  \emph{path connected} and has a CW-complex structure. Let $\rr{h}^\bullet$ be a \emph{generalized} cohomology theory\footnote{A generalized cohomology theory is a family of contravariant functors from the category of pairs of topological spaces and continuous functions (or some subcategory of them such as the category of CW-complexes) to the category of Abelian groups and group homomorphisms that satisfies the \emph{Eilenberg-Steenrod axioms} \cite[Section 1.5]{davis-kirk-01}.}. The fundamental group $\pi_1(B)$ acts as transformation group of the fiber $F$ and therefore it defines an action $\rho$ of $\pi_1(B)$
on the generalized cohomology groups $\rr{h}^q(F)$. The action $\rho$ endows the 
 groups  $\rr{h}^q(F)$ with the structure of a module over the group ring $\Z\pi(B)$ and one can define the cohomology with local coefficients  $H^\bullet(B,\rr{h}^q(F))$ as the cohomology of the cochain complex ${\rm Hom}_{\Z\pi(B)}(C_\bullet,\rr{h}^q(F))$ where $\{\partial_k:C_{k}\to C_{k-1}\}$ is the \emph{singular} chain complex of the 
 universal covering $\tilde{B}$ of $B$. 
 
 \medskip
 
 The CW-complex structure $\emptyset=B_{-1}\subset B_0\subset B_1\subset\ldots\subset B_p\subset\ldots\subset B$ induces a filtration $\emptyset=M_{-1}\subset M_0\subset M_1\subset\ldots\subset M_p\subset\ldots\subset M$ on the total space where $M_p:=f^{-1}(B_p)$
and the   inclusions $M_{p-1}\hookrightarrow M_p\hookrightarrow M$ induce maps in the cochains which define the cohomology theory $\rr{h}^\bullet$. More precisely one has 
\begin{equation}\label{eq:filt1} 
0\;\subseteq\;\ldots\;\subseteq\;^{(p)}C^q(M)\;\subseteq\;\ldots\;\subseteq\;^{(1)}C^q(M)\;\subseteq\; ^{(0)}C^q(M)\;\subseteq\; ^{(-1)}C^q(M)\;:=\;C^q(M)
\end{equation}
where $C^q(M)$ is the abelian group of the $q$-cochains of $M$ and $^{(p)}C^q(M)$ are the $q$-cochains which vanish on the $q$-chains $C^q(M_p)$ of  $M_p\subset M$. Let us set the graded group (resp. module) $A:=\bigoplus_{q\in\Z}C^q(M)$. The  filtration induced on $M$ by the skeleton structure of $B$ provides by \eqref{eq:filt1}  a filtration of the graded complex $A$  which preserves the grading and it is compatible with the differential. This allows us to define a bigraded 
group (resp. module) $E_0:=\bigoplus_{p,q\in\Z}E^{p,q}_0$ with 
$$
E^{p,q}_0\;:=\;^{(p-1)}C^{p+q}(M)\;/\; ^{(p)}C^{p+q}(M)\;=\;C^{p+q}(M_{p}|M_{p-1})\;.
$$
We want to think to $E_0$ as the zero page of a spectral sequence. The one page is then given by
$$
E_1^{p,q}\;=\; \frac{{\rm Ker}\left(\delta:C^{p+q}(M_{p}|M_{p-1})\to C^{p+q+1}(M_{p}|M_{p-1})\right)}{{\rm Im}
\left(\delta:C^{p+q-1}(M_{p}|M_{p-1})\to C^{p+q}(M_{p}|M_{p-1})\right)}\;=\;\rr{h}^{p+q}(M_{p}|M_{p-1})\;.
$$
In order to describe the  page $2$ of the spectral sequence we need the important isomorphism
$$
\rr{h}^{p+q}(M_{p}|M_{p-1})\;\simeq\; C^{p}(B,\rr{h}^{q}(F))
$$
where on the right-hand side we have the group of the $p$-cochains of $B$ with local system of coefficients given by the 
$\Z\pi(B)$-module $\rr{h}^q(F)$. This implies that
$$
E_2^{p,q}\;=\; \frac{{\rm Ker}\Big(\delta:C^{p}(B,\rr{h}^{q}(F))\to C^{p+1}(B,\rr{h}^{q}(F))\Big)}{{\rm Im}
\Big(\delta:C^{p-1}(B,\rr{h}^{q}(F))\to C^{p}(B,\rr{h}^{q}(F))\Big)}\;=\;H^{p}(B, \rr{h}^{q}(F))\;.
$$

\begin{theorem}[Cohomology Leray-Serre  spectral sequence]\label{theo:Leray-Serre}
Let 
$F\hookrightarrow M\stackrel{f}{\rightarrow}B$
be a (Serre) fibration with \emph{fiber} $F$ and base space $B$ which is  \emph{path connected} and has a CW-complex structure. Let $\rr{h}^\bullet$ be a \emph{generalized} cohomology theory. Assume also that there exists a $q_0$ such that $\rr{h}^q(F)=0$ for all $q<q_0$. Then, there exists a spectral sequence $\{E_r,\delta_r\}$ with
$$
E_1^{p,q}\;=\;C^{p}(B,\rr{h}^{q}(F))\;,\qquad\quad E_2^{p,q}\;=\;H^{p}(B, \rr{h}^{q}(F))
$$ 
that converges $E_2^{p,q}\Rightarrow\rr{h}^{p+q}(M)$, namely
\begin{equation}\label{eq:sdref}
E_\infty^{p,q}\;\simeq\;\frac{{\rm Ker}\Big(\rr{h}^{p+q}(M)\to {\rr{h}^{p+q}(M_{p-1}))}\Big)}{{\rm Ker}
\Big(\rr{h}^{p+q}(M)\to {\rr{h}^{p+q}(M_p)}\Big)}
\end{equation}
where the maps are induced by the inclusions $M_p\subset M$.
\end{theorem}

\noindent
For the proof of this theorem see, fore instance, \cite[Chapter 9, Section 2]{spanier-66}.

\medskip

\noindent
{\bf Group cohomology.}
We briefly recall the notion of  \emph{group cohomology} for the particular case of $\Z_2$.
For more details we refer to  \cite{brown-82}. As a  system of coefficients we fix an abelian group $\s{A}$ which has a (left) $\Z_2$-module structure given by a representation $\rho:\Z_2\to{\rm Aut}(\s{A})$. This allows us to  use  the notation $\epsilon\cdot a:=\rho(\epsilon)(a)$ for all $\epsilon\in\Z_2$ and $a\in\s{A}$.
As typical example, when $\Z_2$ acts on a topological involutive space $(X,\tau)$, then $H^k(X,\s{R})$ has a $\Z_2$-module structure given by the pullback induced by the map $\tau$.
Tensoring by $\Z(m)$ over $\Z$, we get also the  
$\Z_2$-module 
\begin{equation}\label{eq:Z2-mod}
H^k(X,\s{R}(m))\;:=\;H^k(X,\s{R})\otimes_\Z\Z(m)
\end{equation}
where the $\Z_2$-action on $\Z(m)$ is given by the multiplication by $(-1)^m$. We point out that, differently from the case of the 
{equivariant Borel cohomology} (\cf Section \ref{subsec:borel_cohom}) the coefficients 
 $\s{R}(m)$ in \eqref{eq:Z2-mod} are not local systems. 

\medskip

The cochains of the group cohomology with coefficients in $\s{A}$ are defined by
$$
C^0_{\rm group}(\Z_2,\s{A})\;:=\;\s{A}\;,\qquad\quad C^k_{\rm group}(\Z_2,\s{A})\;:=\;\left\{f:(\Z_2)^k\to\s{A}\ |\ \text{maps}\right\}\;,\quad k\geqslant 1
$$
The sets $C^k_{\rm group}(\Z_2,\s{A})$ inherit from $\s{A}$ the structure of abelian groups. If $\s{A}$ is further a vector space, then so are $C^k_{\rm group}(\Z_2,\s{A})$. The differential $\delta:C^k_{\rm group}(\Z_2,\s{A})\to C^{k+1}_{\rm group}(\Z_2,\s{A})$
is given by 
$$
\begin{aligned}
(\delta f)(\epsilon_1,\ldots,\epsilon_{k+1})\;:=\;&f(\epsilon_2,\ldots,\epsilon_{k+1})\;+\;(-1)^{k+1}\epsilon_{k+1}\cdot f(\epsilon_1,\ldots,\epsilon_{k})\\
&+\sum_{i=1}^k(-1)^i\; f\big(\epsilon_1,\ldots,\epsilon_{i-1},(\epsilon_{i}\epsilon_{i+1}),\epsilon_{i+2},\ldots,\epsilon_{k+1}\big)
\end{aligned}
$$
where a multiplicative notation for element in $\Z_2$ has been used. Since one directly checks that $\delta^2=0$, one gets a cochain complex $\big(C^k_{\rm group}(\Z_2,\s{A}),\delta\big)$ with related cocycle groups $Z^k_{\rm group}(\Z_2,\s{A})$ and
coboundary groups $B^k_{\rm group}(\Z_2,\s{A})$. Then, the $k$-th  group cohomology is defined as
$$
H^k_{\rm group}(\Z_2,\s{A})\;:=\; {Z^k_{\rm group}(\Z_2,\s{A})}\;/\;{B^k_{\rm group}(\Z_2,\s{A})}\;.
$$
The following result will be used in the following:
\begin{lemma}\label{lemma:vanish}
If the abelian group underlying the $\Z_2$-module $\s{A}$ is a vector space
(over $\R$), then
$$
H^k_{\rm group}(\Z_2,\s{A})\;=\;
\left\{
\begin{aligned}
&\big\{a\in \s{A}\ |\ \epsilon\cdot a=a\;,\quad \epsilon\in\Z_2\big\}&&\qquad\text{if}\ \ k=0\\
&0&&\qquad\text{if}\ \ k>0\;.
\end{aligned}
\right.
$$
Moreover
\begin{equation}\label{eq:Z2-mod_bis}
H^k_{\rm group}(\Z_2,\s{A}\otimes_\Z\Z(m))\;=\;\left\{
\begin{aligned}
&\big\{a\in \s{A}\ |\ \epsilon\cdot a=(-1)^m a\;,\quad \epsilon\ \ \text{\upshape generator of}\ \ \Z_2\big\}&&\qquad\text{if}\ \ k=0\\
&0&&\qquad\text{if}\ \ k>0\;.
\end{aligned}
\right.
\end{equation}
where the  $\Z_2$-action on $\Z(m)$ is given by the multiplication by $(-1)^m$.
\end{lemma}
\proof
In the case $k=0$ one has $H^0_{\rm group}(\Z_2,\s{A})=Z^0_{\rm group}(\Z_2,\s{A})$ and a $0$-cochain $a\in\s{A}$
is a $0$-cocycle if $(\delta a)(\epsilon)=a-\epsilon\cdot a=0$ for all $\epsilon\in\Z_2$.

To prove that $H^k_{\rm group}(\Z_2,\s{A})=0$ if $k>0$ it is enough to
show that for each 	$k$-cocycle $f\in Z^k_{\rm group}(\Z_2,\s{A})$ there is a $(k-1)$-cochain $h\in C^{k-1}_{\rm group}(\Z_2,\s{A})$
such that $\delta h=f$. Such a $(k-1)$-cochain $h$ is explicitly constructed from $f$ by means of an \virg{average}, \ie
\begin{equation}\label{eq:h_average}
h(\epsilon_2,\ldots,\epsilon_{k})\;:=\;\frac{1}{2}\sum_{\epsilon_1\in\Z_2}f(\epsilon_1,\epsilon_2,\ldots,\epsilon_{k})\;.
\end{equation}
The cocycle condition $(\delta f)(\epsilon_1,\epsilon_2,\ldots,\epsilon_{k},\epsilon_{k+1})=0$ implies that
$$
\begin{aligned}
f(\epsilon_2,\ldots,\epsilon_{k+1})\;=\;& f\big((\epsilon_1\epsilon_2),\epsilon_3\ldots,\epsilon_{k+1}\big)\;+\;(-1)^{k}\epsilon_{k+1}\cdot f(\epsilon_1,\ldots,\epsilon_{k})\\
&+\sum_{i=2}^k(-1)^{i+1}\; f\big(\epsilon_1,\ldots,\epsilon_{i-1},(\epsilon_{i}\epsilon_{i+1}),\epsilon_{i+2},\ldots,\epsilon_{k+1}\big)\;.
\end{aligned}
$$
By averaging the right-hand side on $\epsilon_1$ (the left-hand side does not depend on this variable) one gets
$$
\begin{aligned}
f(\epsilon_2,\ldots,\epsilon_{k+1})\;=\;& h\big(\epsilon_3\ldots,\epsilon_{k+1}\big)\;+\;(-1)^{(k-1)+1}\epsilon_{k+1}\cdot h(\epsilon_2,\ldots,\epsilon_{k})\\
&+\sum_{i=2}^k(-1)^{i+1}\; h\big(\epsilon_2,\ldots,\epsilon_{i-1},(\epsilon_{i}\epsilon_{i+1}),\epsilon_{i+2},\ldots,\epsilon_{k+1}\big)\;
\end{aligned}
$$
and, after suitably reshuffling the indices, one recognizes the desired relation $f=\delta h$.

Finally, the isomorphism \eqref{eq:Z2-mod_bis} is certainly true for $k>0$. For $k=0$ the isomorphism follows from 
the isomorphism of $\Z_2$-modules
$\s{A}\otimes_\Z\Z(m)\simeq\s{A}$. At level of abelian groups this isomorphism is given by the identification $a\otimes_\Z n\equiv na\otimes_\Z 1$. However, in order to have an  isomorphism of $\Z_2$-modules we need to endow 
$\s{A}$ with a new $\Z_2$-action \virg{twisted} by $\Z(m)$. More precisely the $\Z_2$-action on $\s{A}\otimes_\Z\Z(m)$  induced by the non trivial element $\epsilon\in\Z_2$ by $a\otimes n\mapsto\epsilon\cdot a\otimes (-1)^mn$ 
and the \virg{twisted} $\Z_2$-action on $\s{A}$ given by $a\mapsto(-1)^m\epsilon\cdot a$ are compatible with respect to the identification $a\otimes_\Z n\equiv na\otimes_\Z 1$  and provides the desired isomorphism. This fact concludes the proof of \ref{eq:Z2-mod_bis}.
\qed

\medskip

\noindent
We point out that the possibility to define the \emph{average} \eqref{eq:h_average} depends on the vector space nature of $\s{A}$ which allows the multiplication by the scalar $\frac{1}{2}$. At this point it is easy to understand that the proof generalizes to the case of a $\Z_2$-module $\s{A}$
whose underlying abelian group is a module over a ring in which $2$ is invertible.

\medskip

\noindent
{\bf Application to the Borel equivariant cohomology.}
The Leray-Serre  spectral sequence is an extremely useful tool for the computation of the  Borel equivariant cohomology. The application of Theorem \ref{theo:Leray-Serre} to the canonical fibration \eqref{eq:can_fib} for the generalized  cohomology theory $H^\bullet(\cdot\ ,\s{R}(m))$ defined in \eqref{eq:Z2-mod}
provides for the page 2 of the spectral sequence
$$
E_2^{p,q}\;=\;H^p(\R P^{\infty},H^q(X,\s{R}(m)))\;=\; H^p_{\rm group}(\Z_2,H^q(X,\s{R}(m)))
$$
where  the last equality is provided by the standard identification of {the group cohomology} for the group $\Z_2$ with coefficients in the $\Z$-module $H^q(X,\s{R}(m))$ \cite{brown-82}.	
Moreover, the spectral sequence converges to
$$
E_\infty^{p,q}\;=\;H^{p+q}(X_{\sim} ,\s{R}(m))\;=:\;H^{p+q}_{\Z_2}(X ,\s{R}(m))\;.
$$
The page 2 of $E^{p,q}$ is concentrated in the \emph{first quadrant}, meaning that $E_2^{p,q}=0$ if $p<0$ or $q<0$. This gives us important information. For example from $E_2^{-2,1}=E^{2,-1}_2=0$ we deduce that
$$
E_3^{0,0}\;=\;\frac{{\rm Ker}\left(\delta:E_{2}^{0,0}\to E_{2}^{2,-1}\right)}{{\rm Im}
\left(\delta:E_2^{-2,1}\to E_{2}^{0,0}\right)}\;=\; E_2^{0,0}
$$
and similarly
$$
E_2^{0,0}\;=\;E_3^{0,0}\;=\;\ldots\;=E_\infty^{0,0}\;=\;H^{0}_{\Z_2}(X ,\s{R}(m))\;.
$$
In particular, all the spectral sequence degenerates at the page 2 in the case of a coefficient system $\s{R}(m)$ which is a vector space over $\R$ (in particular in this case Lemma \ref{lemma:vanish} applies and one has immediately $E_2^{p,q}=0$ for all $q>0$). For instance, in the case of coefficients $\R(m)$ the isomorphism \eqref{eq:sdref}
provides
$$
E_2^{p,q}\;=\;E_3^{p,q}\;=\;\ldots\;=E_\infty^{p,q}\;\simeq\;\frac{{\rm Ker}\Big(H^{p+q}_{\Z_2}(X ,{\R}(m))\to {H^{p+q}_{\Z_2}(X_{p-1} ,{\R}(m))}\Big)}{{\rm Ker}
\Big(H^{p+q}_{\Z_2}(X ,{\R}(m))\to {H^{p+q}_{\Z_2}(X_p ,{\R}(m))}\Big)}
$$
and one obtains
 the  isomorphism
\begin{equation}\label{eq:iso_imp}
H^{q}_{\Z_2}(X ,{\R}(m))\;\simeq\;E_\infty^{0,q}\;\simeq E_2^{0,q}\simeq\;H^0_{\rm group}(\Z_2,H^q(X,{\R}(m)))\;.
\end{equation}
%

\section{Some properties of the cohomology $H^\bullet_{\Z_2}\big(X,\R/\Z(1)\big)$}\label{sec:new_cohom}
Let $H^k_{\Z_2}(X,\R/\Z(1))$ be  cohomology group defined by \eqref{new_cohom_coeffi}.
In analogy to the case of coefficients $\Z(1)$ investigated in \cite{gomi-13} also $H^k_{\Z_2}(X,\R/\Z(1))$ can be described as a Borel equivariant cohomology with local system of coefficients. Moreover the exact sequences of  \cite[Proposition 2.3]{gomi-13} can be generalized in the following way:
\begin{align}
&\ldots\; \stackrel{}{\longrightarrow}\;H^k_{\Z_2}\big(X,\R/\Z\big)\;\stackrel{\jmath}{\longrightarrow}\;H^k\big(X,\R/\Z\big)\;\stackrel{}{\longrightarrow}\;H^k_{\Z_2}\big(X,\R/\Z(1)\big)\;\stackrel{}{\longrightarrow}\;H^{k+1}_{\Z_2}\big(X,\R/\Z\big)\; \stackrel{}{\longrightarrow}\;\ldots\label{EXT_E1}\\
&\ldots\;\stackrel{}{\longrightarrow}\;H^k_{\Z_2}\big(X,\R/\Z(1)\big)\;\stackrel{\jmath}{\longrightarrow}\;H^k\big(X,\R/\Z\big)\;\stackrel{}{\longrightarrow}\;H^k_{\Z_2}\big(X,\R/\Z\big)\;\stackrel{}{\longrightarrow}\;H^{k+1}_{\Z_2}\big(X,\R/\Z(1)\big)\; \stackrel{}{\longrightarrow}\;\ldots\label{EXT_E2}
\end{align}
where $\jmath$ is the map which forgets the $\Z_2$-action. One has also the exact sequence
\eqref{eq:flat_exact_seq}
associated to the short exact sequence $0\to\Z(m)\to\R(m)\to\R/\Z(m)\to0$ of coefficients both for $m=0,1$. Let $X=\{\ast\}$ be a single point. From  \cite[Proposition 2.4]{gomi-13} and the isomorphism \eqref{eq:cohom_coef1} one has $H^k_{\Z_2}(\{\ast\},\R(m))=0$ for all $k=\N$,  $H^0_{\Z_2}(\{\ast\},\R(1))=0$ and $H^0_{\Z_2}(\{\ast\},\R)=\R$. Then, by exploiting the exact sequence
\eqref{eq:flat_exact_seq} one obtains
\begin{equation}\label{eq:coho_R/Z1}
H^k_{\Z_2}\big(\{\ast\},\R/\Z(1)\big)\;\simeq\; H^{k+1}_{\Z_2}\big(\{\ast\},\Z(1)\big)\;\simeq\;
\left\{
\begin{aligned}
&\Z_2&&\text{if}\ \ k\ \ \text{is even or}\ \ k=0\\
&0&&\text{if}\ \ k\ \ \text{is odd}\\
\end{aligned}
\right.
\end{equation}
and similarly
\begin{equation}\label{eq:coho_R/Z0}
 H^0_{\Z_2}\big(\{\ast\},\R/\Z\big)\;\simeq\;\R/\Z\;,\qquad\quad H^k_{\Z_2}\big(\{\ast\},\R/\Z\big)\;\simeq\; H^{k+1}_{\Z_2}\big(\{\ast\},\Z\big)\;\simeq\;
\left\{
\begin{aligned}
&0&&\text{if}\ \ k>0\ \ \text{is even}\\
&\Z_2&&\text{if}\ \ k\ \ \text{is odd}\\
\end{aligned}
\right.\;.
\end{equation}

\begin{proposition}\label{prop:inj_foget_R/Z}
Let $(X,\tau)$ be an involutive space. If $X$ is connected and there is at least one fixed point then:
\begin{enumerate}
\item[(i)] $H^0_{\Z_2}\big(X,\R/\Z\big)\simeq\R/\Z$ and $H^0_{\Z_2}\big(X,\R/\Z(1)\big)\simeq\Z_2$;\vspace{1.3 mm}

\item[(ii)] The map $\jmath:H^1_{\Z_2}(X,\R/\Z(1))\to H^1(X,\R/\Z)$ which forgets the $\Z_2$-action is injective.\vspace{1.3 mm}
\end{enumerate}
\end{proposition}
\proof
(i)  If $Y$ is a connected space then $H_0(Y)\simeq\Z$ and $H^0(Y,\s{R}):={\rm Hom}_\Z(\Z,\s{R})$. If $X$ is connected also the homotopy quotient $X_{\sim\tau}$ turns out to be connected, hence by the very definition of the Borel cohomology one has $H^0_{\Z_2}(X,\R/\Z)={\rm Hom}_\Z(\Z,\R/\Z)\simeq \R/\Z$. By definition 
of the reduced cohomology one has $H^0_{\Z_2}(X,\R/\Z)=\tilde{H}^0_{\Z_2}(X,\R/\Z)\oplus H^0_{\Z_2}(\{\ast\},\R/\Z)$ where $\{\ast\} \subset X$ is any fixed point. A comparison with \eqref{eq:coho_R/Z0} shows that 
$\tilde{H}^0_{\Z_2}(X,\R/\Z)=0$. A similar argument also proves that $\tilde{H}^0(X,\R/\Z)=0$. The exact sequence \eqref{EXT_E2} leads to the following exact sequence for the reduced theory:
$$
0\;\stackrel{}{\longrightarrow}\;\tilde{H}^0_{\Z_2}\big(X,\R/\Z(1)\big)\;\stackrel{\jmath}{\longrightarrow}\;\tilde{H}^0\big(X,\R/\Z\big)=0\;\stackrel{}{\longrightarrow}\;\tilde{H}^0_{\Z_2}\big(X,\R/\Z\big)=0
$$
proving that also $\tilde{H}^0_{\Z_2} (X,\R/\Z(1))=0$. Then $H^0_{\Z_2}(X,\R/\Z(1))\simeq H^0_{\Z_2}(\{\ast\},\R/\Z(1))\simeq\Z_2$.

(ii) Since the map $H^0(X,\R/\Z)\to H^0_{\Z_2}(X,\R/\Z)$ is surjective (the full contribution to both groups is given by a fixed point) one deduce from \eqref{EXT_E2} the injectivity of $\jmath$.
\qed


\medskip
\medskip

\end{document}